\documentclass[a4paper,fleqn,usenatbib]{mnras}
\pdfoutput=1
\usepackage{newtxtext,newtxmath}

\usepackage[T1]{fontenc}
\usepackage{ae,aecompl}

\usepackage{graphicx}	
\usepackage{amsmath}	
\usepackage{amssymb}	

\newcommand   {\sgra}  {Sgr A$\textrm{*}$}

\newcommand   {\kms}   {\mbox{km\,s$^{-1}$}}
\renewcommand {\ga}    {\mbox{\rlap{\hbox{\lower5pt\hbox{$\sim$}}}\hbox{$>$}}}
\renewcommand {\la}    {\mbox{\rlap{\hbox{\lower5pt\hbox{$\sim$}}}\hbox{$<$}}}

\newcommand{\e}[1]{\times 10^{#1}}

\def\msol{\hbox{$\hbox{M}_\odot$}}

\def\kms{km s$^{-1}$}

\def\pdeg           {$.\kern-.25em ^{^\circ}$}
\def\degree{\ifmmode{^\circ} \else{$^\circ$}\fi}
\def\ut #1 #2 { \, \textrm{#1}^{#2}} 
\def\u #1 { \, \textrm{#1}}          

\def\msol   {\hbox{$M_\odot$}}                  
\def\v      {\hbox{\it V}}                      

\title[]{Evidence for a jet and  outflow from Sgr A*:\\
a continuum and spectral line study}

\author[Yusef-Zadeh et al.]{
F. Yusef-Zadeh$^{1}$\thanks{E-mail: zadeh@northwestern.edu}
M. Royster$^1$
M. Wardle$^2$
W. Cotton$^3$
D. Kunneriath$^3$ 
\newauthor I. Heywood$^{4,5,6}$ \& J. Michail${^1}$
\\
$^{1}$CIERA, Department of Physics and Astronomy Northwestern University, Evanston, IL 60208, USA\\
$^{2}$Dept of Physics and Astronomy,  Research Centre for Astronomy, Astrophysics
and Astrophotonics, Macquarie University, Sydney NSW 2109, Australia\\
$^{3}$National Radio Astronomy Observatory, Charlottesville, VA, 22903, USA\\
$^{4}$Astrophysics, Department of Physics, University of Oxford, Keble Road, Oxford, OX1 3RH, UK\\
$^{5}$Department of Physics and Electronics, Rhodes University, PO Box 94, Makhanda, 6140, South Africa\\
$^{6}$South African Radio Astronomical Observatory, 2 Fir Street, Black River Park, Observatory, Cape Town, 7925, South Africa
}

\date{Accepted XXX. Received YYY; in original form ZZZ}

\pubyear{2020}

\begin{document}
\label{firstpage}
\pagerange{\pageref{firstpage}--\pageref{lastpage}}
\maketitle

\begin{abstract}
We study the environment of Sgr A* using spectral and continuum observations with the ALMA and VLA.
Our analysis of sub-arcsecond H30$\alpha$, H39$\alpha$, H52$\alpha$ and H56$\alpha$ line emission
towards Sgr A* confirm the recently published broad peak $\sim500$ \kms~spectrum toward Sgr~A*. We also
detect emission at more extreme radial velocities peaking near $-2500$ and $4000$ \kms\, within 0.2$''$.
We then present broad band radio continuum images at multiple frequencies on scales from arcseconds to
arcminutes. A number of elongated continuum structures lie parallel to the Galactic plane, extending from
$\sim0.4''$ to $\sim10'$. We note a nonthermal elongated structure on an arcminute scale emanating from
Sgr A* at low frequencies between 1 and 1.4 GHz where thermal emission from the mini-spiral is depressed
by optical depth effects. The position angle of this elongated structure and the sense of motion of
ionized features with respect to Sgr A* suggest a symmetric, collimated jet emerging from Sgr A* with an
opening angle of $\sim30^\circ$ and a position angle of $\sim$60$^\circ$ punching through the medium
before accelerating a significant fraction of the orbiting ionized gas to high velocities.  The jet with
estimated  mass flow rate  $\sim1.4\times10^{-5}$ \msol\, yr$^{-1}$ emerges perpendicular to the equatorial plane of
the accretion flow near the event horizon of Sgr A* and runs along the Galactic plane. To explain a
number of  east-west features near Sgr A*, we also consider the possibility of an outflow component with a wider-angle launched from
the accretion flow at larger radii.
\end{abstract}



\section{Introduction}

Our relative proximity to the supermassive black hole, Sgr A*, at the center of the Galaxy provides a 
laboratory in which to investigate the processes that lead to accretion and ejection flows in the low 
luminosity nuclei of galaxies.The bolometric luminosity of Sgr~A* powered by the magnetized accretion flow is 
several orders of magnitude below the luminosity estimated from the accretion of ionized stellar winds 
feeding Sgr A*.  Two broad classes of models have addressed the low luminosity of Sgr A*.  The first 
describes a radiatively inefficient accretion flow (RIAF) which argues that only a fraction of the initially 
infalling material from stellar winds fall onto Sgr A* and the rest is driven off as an outflow from Sgr A*
\citep[e.g.][]{yuan04,quataert04,wang13}.
 Alternatively, most of the gas approaching Sgr A* may be pushed 
away as part of a jet or an outflow \citep[e.g.][]{falcke00,becker11,paper2}. In either case, the accretion 
rate is well below the Bondi accretion rate. The presence of a jet constrains the accretion flow model and 
its interaction with the surrounding gas provides constraints on the jet power. Thus it is important to 
establish if there is indeed a jet arising from Sgr A*.

Recent H30$\alpha$ millimeter recombination line (mmRL) observations using Atacama Large Millimeter/submillimeter Array (ALMA)  detected  high velocity ionized gas associated with Sgr A*. A double peak mmRL was detected  in the spectrum of Sgr A* showing evidence of blue and red-shifted ionized gas from the inner 0.23$''$ (9 milli-pc) of Sgr~A* \citep{elena19}. The full linewidth of the velocity profile was  estimated to be $2200$ \kms\, with peak velocity $\pm 500$ \kms.  The ionized features have a position angle (PA) of $\sim 64^\circ$, directed roughly along the Galactic plane.  This double-lobed ionized structure $\sim3600R_s$ from Sgr A* is interpreted as a tracer of a cool ($\sim10^4$ K) accretion flow \citep{elena19}.

Another recent study detected high velocity ionized gas with velocities ranging between --480 and  --300 \kms\,
SW of Sgr A* and within
2$''-3''$ (0.08--0.12 pc) \citep{paper1}. These had been
 the highest radial velocities of diffuse ionized gas beyond the inner 1$''$ of Sgr A*.
To the NE of Sgr A*, a number of red-shifted isolated features  associated with  cometary stellar sources
have  also  been reported \citep{paper1}.
The  blue- and red-shifted features
were  interpreted  in terms of the interaction of a collimated
outflow with a position angle  $\sim60^\circ$ East of North  with respect to Sgr A*.  An outflow rate
of 2$\times10^{-7}$ or 4$\times10^{-5}$ \msol\, yr$^{-1}$ was  estimated for  a relativistic jet-driven outflow
from Sgr A* or collimated stellar winds, respectively.  Similar to this picture, several past studies have also  argued for the presence of an outflow either due to a wind or a jet from Sgr A* in the direction either along the Galactic plane or other directions, details of which are provided below. There is currently no direct evidence for a jet emanating from Sgr A* in the highly confused region of the Galactic center containing: the mini-spiral feature tracing both ionized and cool dense material, the Sgr A East SNR tracing nonthermal emission and high density of evolved and young mass-losing stars orbiting Sgr A*. However,  morphological and kinematical studies  infer sites where the outflowing material emerging from Sgr A* interact  with orbiting gas  and  stars. A brief summary of the results of these investigations indicate a collimated outflow arising from Sgr A*, as we describe them next from small to large scales.

$\,\,\,\,\,\,\,\bullet$ 
Polarimetric near-IR measurements of flare emission from Sgr A* indicate  a preferred direction of the polarization angle
\cite{eckart06,meyer07,zaman11}. 
 These measurements give a statistically significant mean  polarization angle of 60$^\circ(240^\circ) \pm20^\circ$ \citep{eckart06}. 
 An ejection  model of plasma blobs has been considered to explain flaring activity of 
Sgr A* \citep{fyz06}. If this arises in a jet, the  PA of synchrotron-emitting ejected material 
should be in the direction of the preferred polarization angle, thus directed on both sides along the Galactic plane. 

$\,\,\,\,\,\,\,\bullet$ 
A chain of ionized blobs seen in  radio
continuum  leads
from Sgr A* to  a hole in the distribution of
orbiting ionized gas, the ``mini-cavity'' \cite{paper0}.
The  mini-cavity is
kinematically disturbed  and
is located about $\sim3''$ SW of Sgr A*.
The blobs, connected by a ridge of emission to Sgr A* could be produced 
by an outflow from Sgr A*.

$\,\,\,\,\,\,\,\bullet$ 
\cite{muzic07,muzic10,peisker19}
reported the discovery of 
head-tail cometary stellar sources 
 within a  few arcseconds to the SW of  Sgr A*. These sources 
display bow-shock structures pointing  toward  Sgr A*. 
The PAs of these four stellar sources,   X1, X3, X7 and X8,  
with respect to Sgr A*
are  greater than 45$^\circ$. There are also cometary radio sources, f1 to f3, found to the NE of Sgr A* with similar PAs \citep{paper2}.
\cite{muzic10} argue for  a collimated wind-driven  outflow arising roughly parallel to the Galactic plane to account for the cometary features X3 and X7. 


$\,\,\,\,\,\,\,\bullet$ 
\cite{paper3} reported  a 2.5$''\times1.5''$ (0.1$\times0.06$ pc) mm halo surrounding Sgr A*. 
We interpreted this diffuse halo as being generated by synchrotron emission from relativistic
electrons. The halo has an X-ray counterpart \citep{wang13} and coincides with a dust cavity with
the low infrared extinction of $\sim$2.4 magnitudes. Two dust cavities within 2$''$ are elongated and 
run along the Galactic plane \citep{schodel10,paper3}.  
The spatial anti-correlation of the halo  emission  and the low near-IR extinction  
suggests evidence of an outflow sweeping up the interstellar material, 
creating a dust cavity within 2 arcsec of Sgr A* along the Galactic plane \citep{paper3}. 

$\,\,\,\,\,\,\,\bullet$ 
\cite{peisker20a} 
reported a number of dusty stellar sources with near-infrared excess and 
a  large 
 [FeIII] to Br$\gamma$  line ratio. 
This is indicative of 
 a spatial distribution of  sources with a high abundance to ionization parameter by 
 the interaction of  a non-spherical outflow with the atmosphere of dusty stellar sources. 
The line ratio suggests an excitation by  an external non-spherical wind \citep{peisker20a}. 

$\,\,\,\,\,\,\,\bullet$ 
Narrow and long cometary tails (fibrils) at millimeter   are identified with two mass-losing young stars AF and AFNW.  
These fibrils are  pointing in the direction of Sgr A* along the Galactic plane\citep{paper3}. 
These characteristics suggest an outflow is
responsible for creating: a synchrotron halo,  dust cavities and cometary tails \citep{paper3}.

$\,\,\,\,\,\,\,\bullet$ 
Morphological studies of
high resolution radio continuum observations
of  the inner few pc of Sgr A*  indicate a faint
continuous linear structure centered on Sgr A*
with a PA$\sim60^\circ$ \cite{paper4}. 
 Several blobs of radio emission  were also identified 
along the continuous linear structure. 
The extension
of this feature terminated symmetrically by two linearly
polarized structures  $\sim75''$ (3 pc) from Sgr A*.  
\cite{paper4}  
considered a
mildly relativistic jet from Sgr A*  with
an outflow rate of 10$^{-6}$ \msol\, yr$^{-1}$ to
explain disturbed
kinematics and  enhanced FeII/III line emission
from the minicavity \citep{lutz93}. 

$\,\,\,\,\,\,\,\bullet$ 
Large-scale morphological studies of radio continuum emission
showed   a striking tower of nonthermal
radio emission
150$''$ (6 pc) extending away from Sgr A* at PA$\sim50^\circ - 60^\circ$ \citep{paper2}.
The tower structure
was   argued to be the result of
an interaction of  a jet from Sgr A*
with the Sgr A East SNR which itself  lies behind the mini-spiral \citep{fyz86b,pedlar89}. 
The jet appears to drag   a portion of the
shell to the NE along the Galactic plane.
A jet outflow rate of $10^{-7}$ \msol\, yr$^{-1}$ is needed to explain the origin of the tower \citep{paper2}.

All the above studies generally indicate  a collimated jet from Sgr A* 
directed roughly along the Galactic plane.  
There are other claims of a jet emanating from Sgr A* on different 
scales with 
discrepant values 
of the inclination and position angle of the jet. 
Most of these measurements indicate an   outflow with an axis
projected on the sky  perpendicular to the Galactic plane or  within 30$^\circ$ of  the Galactic plane
(\cite{fyz86a,markoff,broderick11,zaman11,muno08,li13}. 
On a milli-arcsecond (mas) scale, a number of VLBI
measurements  have also characterized 
the size and  PA of millimeter emission from  Sgr A*. 
Recent 86 GHz continuum measurements of Sgr A* using ALMA and 
VLBI indicate a size scale of of $\sim0.2$ mas and appears to 
be elliptical with major axis PA$\sim84^\circ$ \citep{sara19}. 
Other  mas-resolution 86 GHz measurements of Sgr A* 
also indicate PAs ranging between 75$^\circ$ and  83$^\circ$ 
\citep{shen05,lu09,johnson18,ortiz16,brink19}. 


Discrepant values of the PA of the jet are seen on small scales which may result from 
the fact  that 
data reduction is challenging for a strong,  time variable source  on an hourly time scale.
 In addition, Sgr A* has an inverted spectrum, thus continuum imaging 
over a broad bandwidth has to account for the spectral index of Sgr A*.  Confusing sources 
add another challenge in searching  for a large-scale 
jet with an established PA using continuum data.
Lastly, the {\it uv} converge of interferometric measurements will provide additional limitations 
 in imaging this source where there is emission on a wide range of angular scales. 
However, 
detection of high velocity ionized gas associated with Sgr A* is significant in opening a new window to study 
processes that lead to accretion or ejection. The limiting factor is proper continuum subtraction due to 
frequency-dependent time variability and continuum flux of Sgr A* as well as a lack of sufficient line free channels 
for continuum subtraction.
 
Unlike previous measurements, this study combines spectroscopic and radio continuum data at multiple frequencies with different spatial resolutions. In particular, to detect recombination line emission from Sgr A*, we use a different technique, multiple transitions of hydrogen atom and a wider spectral coverage. We use a technique in which the continuum subtraction is applied in the image plane toward Sgr A* unlike the uv plane where the subtraction is done over the entire field with contributions from multiple sources. These new ALMA and VLA measurements  have broader velocity coverage than earlier measurements \citep{elena19}. Amazingly, the lines we detect are so broad that the measurements are still limited by not being able to conclusively identify line free channels from Sgr A*.  Thus, we make some assumptions and the results indicate that the 10 milli-pc (mpc) scale broad mmRL emission from Sgr A* is a small  extension of arcsecond and arc-minute scale structures inferred from past radio continuum studies.


We also present new multi-wavelength radio continuum images of Sgr A*. The time variability of Sgr A* due to flaring activity and spectral variation of Sgr A* across broad observing bands can also lead to radio artifacts that can easily be confused with real features. Given that high frequency features near Sgr A* are contaminated by phase and amplitude errors, we analyze numerous data sets and describe features that persist   and  appear at different frequencies, epochs, and VLA array configurations, as these are unlikely to be generated by imaging artifacts. 



The outline of this paper is as follows. In section 2, we first describe and analyze spectroscopic data reduction using both ALMA and VLA. In section 3, we focus on radio continuum data analysis of  high and 
low-frequency measurements. In section 4, we interpret the results of these measurements before we provide a summary.


\section{Spectroscopic Data Analysis and Results}
\subsection{Data Preparation}

We examined the spectrum of Sgr A* in  four  different   hydrogen recombination lines
utilizing datasets from both ALMA and the VLA.
The data reduction techniques used to make the final spectra of
H30$\alpha$, H39$\alpha$, H52$\alpha$  and H56$\alpha$ line emission are described below.

\subsubsection{H30$\alpha$ ALMA Data}

Our earlier ALMA H30$\alpha$ mmRL observations of the Galactic center were described in detail
by \cite{paper1}.  H30$\alpha$ is the same transition used by \cite{elena19}. Our
ALMA data provides an independent means to confirm the double-peak spectrum of Sgr A* because 
of  the different observational setup and  epochs.  
Observations were carried out as part of a long multi-wavelength monitoring campaign of Sgr A*,
thus achieving excellent sensitivity to  weak structures.

The calibrated Band 6 archival data (project code 2015.A.00021.S) was observed on 2016 July
12/13 in Cycle 3.  The dataset was first combined along all spectral windows prior to deriving
and applying phase self-calibration solutions three times with decreasing time intervals before
a final phase and amplitude  self-calibration solution was derived.  The image cube
with a $\sigma=0.4$ mJy per 3~\kms
channel was constructed after applying
the self-calibration solution. Continuum subtraction was not performed in the {\it{uv}} plane,
instead the image cube was constructed immediately after self-calibration.
The data cube is made up of 45 \kms~channels ranging from $-3770$ to $+700$ \kms~over two
spectral windows relative to the rest frequency of H30$\alpha$~at 231.9 GHz.

A spectrum (which includes the continuum) was obtained from the inner 0.23$''$ towards the position of Sgr A* with a synthesized beam of $0.30''\times0.33''$.
In order to equalize the flux between the two spectral windows that are used here, the five channels on the edge of each window were averaged to determine the
relative difference.  This resulted in the need to multiply the lower velocity window by a factor of 1.0052 before concatenating the spectra together.  Line
free channels were then chosen by inspection in order to subtract the continuum emission.

\subsubsection{H39$\alpha$ ALMA Data}

We used archival ALMA data (project code 2011.0.00887.S, PI Heino Falcke) observed on May 18,
2012 with 19 antennas using band 3. The calibrated dataset provided by the ALMA archive consists of
4 spectral windows, each with 2 GHz bandwidth and 15.625 MHz channel width.
Titan and Neptune were used as flux calibrators, and J1924-292 and NRAO530 were used as the
bandpass and phase calibrator, respectively. We used CASA 5.4.0-74 to split out the Band 3
spectral windows (spws), self-calibrate the measurement set and image the continuum as well as
the H39$\alpha$ and H49$\beta$ line cubes at rest frequencies 106.737 GHz and 105.302 GHz,
respectively. We performed two rounds of phase  self-calibration, deriving
solutions per spectral window in the first round and then for all spectral windows  combined.

\subsubsection{H52$\alpha$ VLA Data}

A-array observations were carried out in the (7mm) band on 2019, July 18 (19A-229).
We used the 3-bit system and
the same calibration strategy as the  7mm (Q band) continuum  observations described in section 
3.1.1. Phase self-calibration was applied
three times to all the data using the bright radio source Sgr A*. The broad 8 GHz bandwidth is comprised of 64 spws,
each having 64 channels of 2 MHz width.  We imaged 12  individual spws close to the rest frequency of
H52$\alpha$ at 45.4538 GHz before continuum subtraction was applied.
These spectral windows correspond to a velocity range of$\sim$ --4200 \kms~to +6000 \kms.

Before a continuum was fitted and subtracted, it was necessary to manually equalize the flux of each spectral
window due to errors in flux calibration. We employed the following procedure beginning with the lowest frequency
spectral window $j$: (1) determine the ratio of the mean flux of the last ten channels of spectral window $j$
with the mean flux of the first ten channels of the next (higher frequency) spectral window $j+1$, (2) multiply
each channel of spectral window $j+1$ by this ratio to equalize it, (3) repeat the previous steps but now
starting with the newly equalized spectral window $j+1$. The derived ratios used to normalize the spws ranged
from $0.9832$ to $1.003$.  The spectral windows were then concatenated as shown in Figure 3a.

A linear continuum was fit to the concatenated data with channels with no strong line emission
chosen between --4200 to --3900 \kms and 5800 \kms~to 6000 \kms.  The resultant subtracted
spectra is shown in Figure 3b. Note that although the absolute continuum flux is lost in Figure
3a, the line emission flux density is recovered in Figure 3b.

\subsubsection{H56$\alpha$ VLA Data}

We carried out A-array observations (19A-229) in the (9mm) band on 2019, July 26.  We used the
3-bit system and a calibration strategy similar to that employed for the VLA H52$\alpha$~data. Phase self-calibration was applied three times to all the data using the bright radio source Sgr
A*. The broad 8 GHz bandwidth is comprised of 64 spws, each having 64 channels of 2 MHz width.
We imaged ten individual spws close to the rest frequency of H56$\alpha$ at 36.466 GHz before
continuum subtraction.  These spectral windows correspond to a velocity range between
$\sim$--4000 \kms~and +6300 \kms.

Continuum subtraction was performed with the same technique that was applied to the H52$\alpha$ line data.
A linear continuum was fit to the concatenated data with channels
chosen between --4400 to --4150 \kms and 5800 \kms~to 6000 \kms.  The resultant with and without continuum
subtracted spectra are shown in Figure 4a,b, respectively.


\subsection{Spectroscopic  Results}
To confirm the results presented by \cite{elena19}, we used the four  independent measurements of 
 H30$\alpha$, H39$\alpha$, H52$\alpha$ and H56$\alpha$ with different spectral setups. Despite having significantly broader spectral coverage than \cite{elena19}, it is still not clear where the true continuum is because of a limited set of line free channels. Clearly, a different spectral setup is required in the future to determine true line free channels for proper continuum subtraction.

\subsubsection{H30$\alpha$ Line Emission}
We extracted H$30\alpha$ spectrum using the same scale size and spectral resolution that were used by \cite{elena19}. However, our spectral coverage is wider than that shown by~\cite{elena19}.  As a result, we selected two different sets of line free channels. The first was chosen to confirm the double-peak spectrum of Sgr A* as seen by~\cite{elena19}.  Channels between velocities of --1300 \kms~and --1800 \kms~were used in a linear $\chi$-squared fit in the image plane to subtract the continuum.  This mimics a line free channel selection process if bandwidth was not available beyond --1800 \kms. Figure 1a,b~shows the spectra of Sgr A* before and after the continuum subtraction, respectively. The subtracted linear fit in Figure 1b shows a spectrum with similar shape to that presented by \cite{elena19}. This confirms the double-peak spectrum of Sgr A* at $\sim\pm$500 \kms\, utilizing a different technique in continuum subtraction. The only significant difference to \cite{elena19} is a higher estimated flux density by $\sim$10 mJy. This is likely the result of the amplitude and phase self-calibration applied to our data. Atmospheric phase errors can cause reduced flux densities of Sgr A* if proper self-calibration is not carried out.  Alternatively, 
it is  possible that the stimulated emission from ionized gas toward Sgr A* is time variable and is 
responsible for different flux  density measured at different epochs.


The second set of line free channels was selected at the extreme end of blue-shifted velocities between --3600 and --3800 \kms. Figure 1c,d shows the constructed spectra before and after applying continuum subtraction, respectively. There is a broad absorption line centered at --2600 \kms.  Although this absorption contaminates the emission from Sgr A*, it nevertheless shows emission that extends beyond --1500 \kms. The data presented by \cite{elena19}~did not contain velocity coverage beyond --2000 \kms. Thus, we tentatively show evidence of highly blue-shifted velocities extending to --3500 \kms\, and a third high-velocity peak near --2000 \kms, in addition to $\pm500$ \kms\, peaks.

Potential sources of the molecular absorption line found at roughly --2665 km/s, as noted in Figures 1 and 2, are 
identified in the H30$\alpha$ rest frame.  The absorption is centered at 233.963 GHz. There are three candidates 
that could be responsible for the absorption. One is CH$_3$OH(5(1,5)-4(1,4)) at center frequency 234.01158 GHz which 
would result in a red-shifted velocity of 62.19 \kms. This component has similar velocity to that of dense gas 
associated with the molecular ring orbiting Sgr A* (e.g., \cite{fyz17}), thus this red-shifted gas cloud close to 
Sgr A* may be detected in absorption against Sgr A*. The others are OS$^{18}$O(24(2,22)-24(1,23)) at 233.94997 GHz, 
blue-shifted by --16.75 \kms\, or OS$^{18}$O(52(7,46)-51(8,43)) centered at 233.9506176 GHz, blueshifted by --15.94 
km/s.\footnote{http://www.cv.nrao.edu/php/splat/sp-basic.html}

\subsubsection{H39$\alpha$ Line Emission}

Two spectra were obtained over the inner 0.75$''$ towards the position of Sgr A*.  With respect to the rest frequency of H39$\alpha$~at 106.737 GHz, one spectral window was available between --3050 \kms~to 1600 \kms~and the other ranged from 2575 \kms~to 7200 \kms, as shown in Figure 2. There is no data available between these two spectral windows because of the initial observing setup. A linear continuum was fit individually to each spectra.  For the more blue-shifted window, the range --3000 \kms~to --2300 \kms~was selected, while 5500~\kms~to 7000 \kms~was chosen for the red-shifted component. The resultant H39$\alpha$ emission with and without continuum subtraction is shown in Figure 2a,b. We note broad H39$\alpha$ line emission from Sgr A*.  The emission confirms the results of \cite{elena19} in that there are two high velocity ionized components peaking at $\sim\pm500$ \kms.  Figure 2c also shows excess emission detected at $\sim-2000$ \kms, confirming the reality of the H30$\alpha$ peak line emission shown in Figure 1c,d with the caveat that the true line free channels in H39$\alpha$ and H30$\alpha$ are unknown. Thus, we note tentative detection of H$39\alpha$ line emission extending to 5500 \kms\, reflecting broader velocity coverage towards positive velocities.

\subsubsection{H52$\alpha$ Line Emission}

We used twelve adjacent spectral windows covering radial velocities between $\sim-4000$ to 
 $\sim6300$ \kms\, with respect to the rest frequency of H52$\alpha$ at 45.4538 GHz. The 
 spectra shown in Figure 3a,b were obtained over the inner 0.20$''\times0.11''$ of Sgr A* with 
 a spatial resolution of 77 $\times$ 35 mas and PA$\sim-2^\circ$. We detect an excess flux of 
 H52$\alpha$ line emission at roughly $\sim500$ \kms~with a dip near 0 \kms\, consistent 
 with the earlier H$30\alpha$ measurements \citep{elena19}.  The $-500$ \kms\, feature is much weaker 
than the $500$ \kms\, feature but extends to more negative velocities. 
In addition, we note enhanced 
 line emission that extends to --1500 and +4000 \kms; similar trends are also noted in the 
 H39$\alpha$ spectrum, as shown in Figure 3c.

In order to determine where high velocity ionized features are located with respect to Sgr A*, we selected the spectral windows that showed enhanced --2500 and 4000 \kms~peak emission. Using two spectral windows,  we made moment maps of five channels centered near the low and high velocity channels (channels 5-10 and 50-55), determined the slope between the low and high velocity features for each spectral window   before subtracting them  from each other. The resulting continuum subtracted image of Figure 3c shows contours of 
blue- and red-shifted H56$\alpha$ line emission corresponding to $\sim$--2500 and $\sim$4000 \kms, respectively. These high velocity features show a diagonal NE-SW distribution along the Galactic plane similar to that found by \cite{elena19} 
at $\pm500$ \kms.

\subsubsection{H56$\alpha$ Line Emission}

We used ten adjacent spectral windows covering radial velocities between $\sim-4000$ to $\sim+6000$ \kms\, 
with respect to the rest frequency of H56$\alpha$ at 36.466 GHz.  The spectra shown in Figure 4a,b were 
obtained over a 0.22$''\times0.27''$ region centered on Sgr A* with a spatial resolution of 
$\sim95\times49$ mas and PA$\sim82^\circ$. We detect an excess flux of H56$\alpha$ line emission at 
roughly $\sim500$ \kms~with a dip near 0 \kms\, consistent with earlier H$30\alpha$ measurements 
\citep{elena19}. However, the red-shifted component at radio is much stronger than the blue-shifted component which is faint. 
It is likely that the $-500$ \kms\, feature that \cite{elena19} detects at 
higher frequencies is  faint at H56$\alpha$ line but extended to more negative velocities 
corresponding to $\sim$--2500 and $\sim$4000 \kms.  
These high velocity features show a 
diagonal NE-SW distribution along the Galactic plane 
similar to that found by 
The trend is similar to mm 
spectra except that enhanced radio recombination line emission at red-shifted velocities are 
3-4 times stronger than the emission at blue-sifted velocities. In addition, we note enhanced 
line emission that extends to --1500 and +4000 \kms; similar trends are also noted in the 
H39$\alpha$ spectrum, as shown in Figure 2c. We note two narrow and one broad absorption 
feature at velocities between --2500 and --4200 \kms. The two narrow absorption lines are 
detected at 36.796 and 36.947 GHz, and are shown in Figure 4b. The 36.947 absorption line is 
centered on a broad absorption ranging over a few hundred \kms. It is not clear what these 
lines are.  There is a family of methyl cyanide lines that range from 36.7937 - 36.7975 GHz as 
well as a methanol line at 36.7955 GHz. The latter  corresponds to 
 a  radial velocity of $\sim-4$ \kms with respect to 
the rest frequency. We also note that acetone lines between 36.952 and 
36.940 GHz  correspond  to 40 km/s and $-$57 \kms,  respectively. These radial velocities are in the 
range of velocities detected toward the molecular ring orbiting Sgr A*.  


In order to determine where the high velocity ionized features are located with respect to Sgr A*, we selected spectral windows that showed enhanced --2500 and 4000 \kms~peak emission. From these windows five channels were averaged near both the low and high velocity band edges (channels 5-10 and 50-55), before subtracting them from each other. The resulting continuum subtracted image of Figure 4c shows contours of blue- and red-shifted H56$\alpha$ line emission corresponding to $\sim$--2500 and $\sim$4000 \kms, respectively. These high velocity features show a similar distribution to that found for H52$\alpha$ line emission, i.e., diagonal NE-SW orientation.

\section{Continuum Data Analysis and Results}
\subsection{Data Preparation}


We carried out radio continuum  observations of Sgr A* at multiple frequencies, each of which is  described below.
All our  continuum observations are centered at the J2000 absolute position of Sgr A*, 
$\alpha, \delta=17^h 45^m 40^s.038,  -29^\circ 00' 28''.069$. 
We used 3C286 to calibrate the flux density scale, 3C286 and J1733-1304 (aka NRAO530) to 
calibrate the bandpass and  J1744-3116 to calibrate the complex gains in all of our observations.

\subsubsection{Broadband Q-band (40--48 GHz) and Ka-band (32--40 GHz)}
We observed \sgra\ on 2014 February 21 (14A-231) and  2018 April 3 (18A-091) 
in the most extended A-array configuration at 44 GHz. The Q-band (7mm band) was 
used in the 3-bit system, which provided full polarization correlations in 4 basebands, each 2 GHz wide, centered around 
44.5 GHz.  Each baseband was composed of 16 subbands (spws), each 128 MHz wide.  Each subband in turn was made up of 64 channels, each  2 MHz wide. 
This epoch of 44 GHz observation, had effective bandwidth of 8 GHz. 

In addition, A-array observations (14A-231, 16B138) in the Ka (9mm) band on 2014, March 9 and  2016 July 19 at 34.5 GHz were carried out.  We used the same 3-bit system and the same calibration strategy for Ka band as we did for the Q-band observations. 
A phase and amplitude 
self-calibration procedure was applied to all data using  Sgr A*. To construct spectra at Ka and Q bands, as shown in Figures 3 and 4,  
only phase calibration  was applied to the VLA data sets. 


\subsubsection{Broadband Ku band 13--15 GHz}
A-array observations (14A-231) were carried out on March 10, 2014 at  14 GHz utilizing 
 full polarization correlations in 2 basebands, each 1 GHz wide between 
13.18  and 14.95 GHz.  Each baseband was composed of eight 128 MHz wide subbands. 
 Each subband was made up of sixty-four 2 MHz channels for a total effective bandwidth of 2 GHz.
All of the subbands were self-calibrated in phase separately. We then combined all 14  
useful spectral windows with the task  VBGLU in AIPS before we applied additional  phase 
self-calibration. 

\subsubsection{Broadband X band 8--10 GHz}
A-array observations were carried out in two epochs on April 17, 2014, and August 17, 2019 at 8 GHz with identical system 
setup.  We used the same backend as that of Ku data for both data sets but with 
center frequency at 9 GHz and  total 
bandwidth of 2 GHz.  Calibration on both data sets was done independently similar to that of Ku data by phase and amplitude 
calibration of individual spectral windows. We combined both data sets followed by phase and amplitude self-calibration.
All Stokes parameters were calibrated in both data sets. 

\subsubsection{Broadband L Band 1.4--2 GHz}

We carried out broad band VLA A-array observations of the Galactic center at 1-2 GHz on 2014
April 02 (project 14A-231). The system we used provided full polarization in both basebands. Each 1.4 GHz
subband was made up of 64 channels of 1 MHz each.
Several phase and amplitude self-calibration procedures were applied to the data in OBIT \citep{cotton08} 
using the bright radio source Sgr A* before  MFImager was used  to construct a cube of 16 channels 
corresponding 
to each spectral window.



\subsection{Radio Continuum Results}

We begin first by describing high frequency features within 5$''-10''$ of Sgr A* followed by 
the   arcminute structures in the vicinity of Sgr A*. We then argue for a radio jet emanating from Sgr A*. A trace of this jet is revealed at high frequencies, though with low 
signal-to-noise ratio. Many high frequency features that are described in the vicinity of Sgr~A* are several thousand times 
fainter than the bright source Sgr A*, thus errors generated by the  variable, strong 
continuum from Sgr A* contaminate  structural details in its vicinity. 
To mitigate these difficulties, we focus  on persistent 
structures revealed from multiple observations with  different frequencies and baselines.

\subsubsection{High Frequency Radio Data}
{\it (a) NE-SW features}

Figure 5 shows a linear feature J1 with a PA$\sim23^\circ$ extending NE of Sgr A* 
for $\sim0.4''$, and having a typical surface brightness of 
$\sim300\,\mu$Jy beam$^{-1}$ at 44 GHz. As described below, this feature may be related to a more extended 
loop-like structure.  
The rms noise of J1 in the 
immediate vicinity of Sgr A* is $\sim 46\,\mu$Jy beam$^{-1}$.  
There is also a fainter feature,  J2, which is  1/2 of typical surface brightness of  J1,  
with PA$\sim 39^\circ$. The latter needs to be confirmed, 
though it is likely to be associated with large-scale features J5, as described below. 
The PAs of J1 and J5 range between 23$^\circ$ and 59$^\circ$  and are   
similar to the PAs of highly red-shifted RRL components,  as shown in Figures 3 and 4, 
and mmRL reported by {\cite{elena19}. 
The H39$\alpha$ and H56$\alpha$ spectra shown in Figures 2 and 4 
include the region that contains both J1 and J2. 
Thus, it is possible that the high velocity 
redshifted components are  related to elongated structures J3 and J5  near Sgr A*, as 
seen in Figure 5.

Figure 6 shows another 44 GHz image of the same region based on our second epoch  observation. 
 This image has a dynamic range of $\sim9000$ and rms noise of 17 $\mu$Jy beam$^{-1}$. We 
note a weak elongated feature J5 with a surface brightness of $\sim50 \mu$Jy beam$^{-1}$ to the NE of 
IRS 16C and a PA$\sim59^\circ$. Two faint linear features J3 and J4 are also found to the SW of Sgr A* 
with surface brightness of 20-100 $\mu$Jy beam$^{-1}$. The position angles of J3 and J4 are 51$^\circ$ 
and 57$^\circ$, respectively. In order to enhance the signal-to-noise ratio of J3, J4 and J5, Figures 
7a,b present the inner 5$''$ of Sgr A* with a convolved resolution of 76 milliarcsecond (mas). 
All three 
features have PAs within six degrees of each other. 
The longest linear feature 
is J4 with an extent of $\sim3.''5$.  
A 9 GHz image of this region in Figure 9d with lower spatial resolution 
than Figure 7 shows the same elongated structure emanating from Sgr A*. 
The PAs of these features appear to be in the direction where IRS 16C 
as well as cometary radio and near-IR stellar sources (F1, F2, F3, X3, X7, X8) lie 
\citep{muzic07,muzic10,paper2,peisker19}.

We notice that in numerous high frequency, high resolution images, the rms noise increases {\it 
only} within $\sim 1''$ of Sgr A*. For example, the rms noise in Figure 6 is about three times 
higher. 
 It is possible that phase and amplitude errors 
are responsible for the increased  noise. Alternatively, the excess noise is due to diffuse 
emission at 44 GHz and is intrinsic due to residual synchrotron emission detected at mm and 
X-rays. This is a region where diffuse mm and X-ray emission has 
previously been detected \citep{wang13,paper3}.
Thus, that the dynamic range may be  limited due to highly turbulent multi-phase medium. 

\

{\noindent\it(b) western minicavity}

Figure 8a shows a 36.8 GHz contour image of the inner 0.4$''$ of Sgr A* constructed from data taken on 
2016 July 19. The linear feature J1 is detected at a PA$\sim26^\circ$. 
Figure 8b shows contour emission 
from the inner 1$''$ of Sgr A* with a 
view of J1 based on  a 9 GHz observation carried out on 2014 April 17. Again we notice an extension of 
J1  with similar PA to  that at 36.8 GHz. 
Figure 8c shows the inner 4.5$''$ of Sgr A* at 15 GHz 
where we note two faint linear features directed to 
the NE of Sgr A*. Both features, though with low 
signal-to-noise ratio, appear to have a coherent structure with similar PAs to J1 and J5, detected at 
other frequencies. Figure 8d  shows a greyscale image of the inner 4$''$ of Sgr A* at 9 GHz. 
The region to the southwest of Sgr A* in Figure 8c,d 
 contains highly blue-shifted ionized gas associated with the minicavity and a string of highly blue-shifted 
compact velocity features which appear to bridge Sgr A* to the western edge of the mini-cavity.  Compact 
sources, labeled C1 to C4, are identified to have extremely blueshifted ionized gas within 0.2pc of Sgr A* 
from --480 to --300 \kms~\citep{paper1}. The highly blue-shifted ionized gas has been detected not only 
toward C1-C4, as labeled on Figure 8c,d, but also toward the minicavity (\cite{zhao10} and multiple compact 
sources that lie close to the western edge of the minicavity. Figure 8d shows a linear feature J4, as shown 
in Figure 6a, that appears to cross a string of HII regions, including C3. An excess of FeII/III line emission 
(Lutz et al. 1993) and disturbed kinematics from the western edge of the mini-cavity are unlikely to be due 
to orbital motion because enhanced FeII/III line emission is detected only in the mini-cavity.  An external 
source of outflow impacting the orbiting gas is likely to be responsible for producing shocked gas, shaping 
morphological, kinematic structures to the SW of Sgr A* \citep{paper2,paper1}. Lastly, Figures 8e,f show detailed 
images of the inner  4$''$ constructed by combining two epochs of 9 GHz data.  We note that J2 and J5 
appear to be parts of a coherent elongated 4$''$ structure that extends the NE of Sgr A*. 
A loop-like feature  is also noted extending from the vertical feature to the NE of Sgr A*. 
It is possible that J1 is part of 
this  loop feature within which  there are  additional diffuse features. A north-south feature 
appears to be the radio counterpart to a  Br$\gamma$ ionized bar  that has recently been studied 
\citep{peisker20b}. Figure 9c also shows contour representation of  the Br$\gamma$ feature  lying between
J1 and the vertical feature.

{\noindent \it(c) EW ridge}

Figure 8 shows an EW ridge of mm emission which extends to the west of  Sgr A* for $\sim2''$. This 
is also prominently detected at 224 GHz within which a number of radio sources, RS1-RS8, 
and a cluster of dusty infrared sources lie \citep{paper3,peisker20a}. The 
trajectories of dusty stellar 
sources detected at infrared 
indicate that they are in an elliptical orbital motion around Sgr A* \citep{peisker20a}. Unlike 
many NE-SW features running along the Galactic plane, 
as described earlier,  the EW ridge has a 
different orientation detected only on the western side of Sgr A* 
 but nevertheless appears to be associated with Sgr A*.  
The EW ridge is revealed best at high 
radio frequencies, millimeter and infrared, suggesting that it consists of a mixture of dust, gas and 
dusty stellar sources. 
The  kinematics of dusty stellar sources in the ridge indicate that they are 
in an elliptical counter-rotating orbit around Sgr A* \citep{peisker20a}, suggesting that they are  
impacted by a collimated  outflow from Sgr A*, as argued  below. 

To examine the morphology of the mini-cavity further, Figures 9a,b present 9 GHz images of the inner 20$''$ 
of Sgr A* with two different brightness contrasts to highlight the bright region of the mini-cavity.
 The mean flux density in the western edge of the minicavity is twice that of the eastern edge. Assuming 
that the enhanced emission and the disturbed kinematics are due to a collimated outflow from Sgr A*, the 
opening angle of the outflow must be  $\sim30^\circ$, as indicated  in Figure 9a, which covers not only 
the western half of the mini-cavity but also the EW ridge.  

{\noindent \it(d)  AF and AFNW stars}

The SW extension of the 
minicavity directed towards  the southern arm of the minicavity shows a number of 
elongated sources and a tail of ionized gas crossing the southern arm.  Two mass-losing stars AF and 
AFNW have previously been identified to have cometary tails directed  away from Sgr A* 
\citep{paper3}. These features are likely to reflect  the interaction of an outflow from the 
direction of Sgr A* with stellar sources and  orbiting gas. Both the AF and 
AFNW sources are located   within the opening angle of the jet beyond the minicavity; the  
white lines drawn on Figures 9a,b run parallel to a number of features that 
trace the direction of the outflow from Sgr A*. 
lie within the opening angle of the jet.   
In this picture, infrared stellar sources 
in the EW ridge are  also impacted  by  this outflow. 

{\noindent\it(e) bent filament}

We also note in Figure 9 a  bent filament  to the north of Sgr A* \citep{paper3,morris17}. Previous 
measurements have not been able to determine if this filament is connected to Sgr A* and perhaps produced by 
Sgr A* \citep{paper3,morris17}. There is no evidence for morphological association  with 
Sgr A* at 9 GHz. We note a compact source with a flux density of 
0.15$\pm0.5$ mJy beam$^{-1}$ coinciding  with the southern tip of the filament 
at $\alpha, \delta_{J2000}=17^h\, 45^m\, 40^s.101\pm0.002\, -29^\circ\,  00'\,  23^{''}.269\pm0.118$. 
To the NE of Sgr A*, J5 and J1  are  revealed with an average 
flux density of $\sim100$ $\mu$Jy beam$^{-1}$ and a PA$\sim56^\circ$ and 21$^\circ$, respectively.  
The extension of J5  appears to encounter the cometary tail of stellar source F1 to  within one degree. 
 
{\noindent\it(f) vertical feature}

We also note a vertical feature to the NW of Sgr A*. A close up of this region, shown in Figure 9c,  indicates 
a continuous structure from the eastern boundary of the minicavity. There is no kinematic information but it appears that the ionized gas associated with the eastern 
boundary of the minicavity is turning counterclockwise as it  orbits Sgr A*. 
A schematic diagram of  features noted at high frequencies  is  shown in Figure 9d. 
 
\subsubsection{Low Frequency Radio Data}

{\noindent \it(a) NE-SW radio jet}

We have constructed images based on 
low-frequency data taken with the VLA using its broad band capability. 
Figure 10a,b  show   grayscale and contours of 20 cm continuum emission from the inner $\sim35''$ of Sgr A*. 
These images are  constructed 
 by combining  spectral windows with  frequencies between 1 and 1.4 GHz. 
A  prominent linear structure at PA$\sim60^\circ$ is detected 
symmetrically only at the lowest frequencies of the broad  band. 
The arms of the mini-spiral are  seen in absorption against the strong nonthermal emission from the shell-type 
Sgr A East SNR, known to lie behind Sgr A* \citep{fyz86b,pedlar89}. 
The  bent filament with PA$\sim-30^\circ$ 
is seen immediately to the north of Sgr A* \citep{paper3,morris17}.

In order to show how features projected against a strong nonthermal source, i.e. Sgr A East, 
are identified, we present grayscale and contour   images  of all 16  spectral channels between 1 and 2 GHz, as seen in Figure 
11a,b. 
At frequencies $>1.5$ GHz, thermal emission from the
mini-spiral  is bright and  stronger than the background nonthermal emission from the Sgr A East SNR.
On the other hand at low
frequencies (1-1.4 GHz), the ionized gas of the mini-spiral  becomes  optically thick and 
is only detected  in absorption against the strong nonthermal 
background  emission. A  "sweet spot", 1.4 GHz, demarcates 
where  thermal and nonthermal   emission dominate at higher and lower frequencies, respectively.   


Figure 12a shows a composite image of nonthermal and thermal sources within 30$''$ of Sgr A*. Low 
frequency (1--1.4 GHz) and high frequency images (1.5--1.9 GHz) are shown in red and blue whereas green 
shows an image using the entire range (1--1.9 GHz). We note that the optically thick ionized gas 
associated with the mini-spiral is absent at low frequencies (red) and becomes optically thin at higher 
frequencies (blue). Figure 12b shows another color image constructed from combining a narrow 
low-frequency channel map (green) and a broad high frequency channel map (red). The linear nonthermal 
features are best seen at low frequencies where thermal emission is suppressed.  
These images confirm that the ionized mini-spiral features are in front of the nonthermal emission. More 
importantly, new linear features, a radio jet and the bent filament are detected at low frequencies below 
1.4 GHz, consistent with being nonthermal. Figure 12c shows a diagram of the prominent thermal and 
nonthermal components surrounding Sgr~A*.


Figure 13a,b show a larger view of the region on a scale of arcminutes where the mini-spiral, the Sgr A East 
SNR and a string of HII regions at the eastern edge of the Sgr A East shell lie. Unlike the broad band 20 cm 
image of Figure 13a, the narrow band image of Figure 13b shows the same region where the low frequency radio 
jet is detected. The position angle of the narrow jet runs $20^\circ$ above the Galactic plane at negative 
longitudes, centered on Sgr A*.

Figure 13c shows a close-up view of the southern extension of the radio jet and a number of Galactic 
center nonthermal radio filaments. The prominent shell of the Sgr A East SNR is known to be elongated 
along the Galactic plane. 
One possibility that could explain the shape of Sgr A East is that the 
expansion occurred in a medium that had 
become anisotropic by an earlier outflow driven by the jet activity of 
Sgr A* along the Galactic plane.  In this picture, 
the oval shape of the shell is produced by the material  being swept along the outflow axis  to the 
onset of SNR expansion. The  magnetic field along the Galactic plane 
could also contribute  to  the formation of  the elongated Sgr A East shell. 



A  327 MHz image of large scale structures in the Sgr A complex also shows extensions of arcminute 
jet feature   along the Galactic plane. 
The 327 MHz feature extends up to about 5$'$ away from Sgr A*  (see Figure 2 of  \cite{nord14}) having   
PAs similar to arcminute scale structures revealed at 1.4 GHz. 
Thus, it is possible that the jet to the NE of Sgr A*, extends to $\sim10'$ (24 pc). High resolution 
broad band observations are needed to determine the association of linear features detected at 327 MHz 
with those at 1.4 GHz.


\section{Discussion}

The Galactic center is a challenging site to search for a jet near Sgr A* because the dynamic range of 
high resolution images is limited by confusing sources, limited {\it uv} coverage, 
 and the intrinsic hourly time variability of Sgr 
A*, especially at high frequencies. It is  also  difficult to identify nonthermal sources
because
at low frequencies (325 MHz), the mini-spiral of ionized gas 
orbiting Sgr A* is seen in absorption due to optical depth effects. At higher frequencies (8 GHz), 
thermal features dominate the emission, making it difficult to identify nonthermal sources. 
One approach to identify both thermal and nonthermal sources is to 
employ  broad bandwidths at multiple frequencies. 
The upgraded VLA 1-2 GHz band, has changed our view of this region studied by 
traditional narrow band observations. 
New images are providing 
evidence for a linear feature with the appearance of a symmetrical jet originating from Sgr A*, pointing 
in the direction of the Galactic plane.

\subsubsection{Collimated Outflow and Interaction Sites}

The overall distribution of the PAs of the linear features 
described here range between $\sim30^\circ$ and $\sim60^\circ$.  As we discuss below, the inferred 
position angle of the jet is supported by sources within the opening angle of the jet showing 
kinematical  signatures of interactions with an anisotropic outflow.  In addition, the 
inferred jet and outflow properties are consistent with what is expected from models of the interaction 
of the stellar winds from the WR stars on the central parsec with Sgr A*. 

First, the  mini-cavity to the SW of Sgr A* shows the highest radial velocities and electron 
temperatures from radio recombination lines, as well as
 excess FeII/III and FeIII/Br$\gamma$ line ratios
\citep{lutz93,zhao10,paper4,peisker19,peisker20a}. 
Radio images presented here clearly indicate an increase in the 
surface brightness of the western edge of the cavity. 
There is also the EW ridge showing similar characteristics to the  mini-cavity, e.g., 
 excess FeII/III and FeIII/Br$\gamma$ line ratios.   

Recent spectroscopic studies suggest dust embedded pre-main sequence stars in the EW ridge \citep{peisker19}. 
All these indicate  
shock excitation due to violent disturbance.  
We previously used the implied rate of change in momentum of the disturbed gas and the angle that the
 western edge of the cavity subtends from Sgr A* to infer that the (two-sided) jet momentum and opening 
angle are approximately 0.1\,\msol\,yr$^{-1}$\,km\,s$^{-1}$ and 30\degr, respectively.  We assume that the 
jet is mildly relativistic, with $\beta\gamma=3$, implying a rest mass outflow rate
$\approx1\e{-7}$\,\msol\,yr$^{-1}$ \citep{paper3}.

Other observations that can be explained by the impact of the jet with the mini-cavity is
an X-ray source with a luminosity of 2$\times10^{33}$ erg\, s$^{-1}$ 
coincides with the cluster of stars, IRS 13E, \citep{wang20,zhu20} 
embedded within the western edge of the minicavity. Recent  analysis of X-ray data concludes that 
colliding winds of mass-losing stars in the cluster 
are responsible for production of X-rays \citep{wang20,zhu20}. 
Alternatively, we note  that X-ray emission from the minicavity 
can also be explained by the  violent collision   with the jet shocking the gas to high temperatures. 
I another  hard X-ray observations between 20-40 keV  over the inner few arcminutes of 
of  Sgr A* show extended symmetric emission with respect to Sgr A* 
 along the Galactic  plane \citep{perez15}. The diffuse hard X-ray emission is 
interpreted to be due to a population of compact objects  or particle  outflow from Sgr A*.
If the latter, the jet-picture discussed   here is likely to explain  the origin of diffuse X-ray emission.

Furthermore, gas compression can in principle increase the gas density to overcome tidal shear by Sgr A*, 
thus inducing star formation activity. Recent measurements Recent spectroscopic studies suggest dust embedded 
pre-main sequence stars in the EW ridge as as well as bipolar  CO outflows indicating  young star 
formation activity near Sgr A*\citep{fyz17,peisker19}. 

Second, a chain of radio blobs linking Sgr A* and the western edge of the minicavity \citep{paper0}, show 
the highest velocity ionized gas, $\sim-300$ to $-480$ \kms\, detected beyond the inner 1$''$ of Sgr 
A* \citep{paper1}. Along the axis defined by the detection of highly blue-shifted ionized gas, the 
cometary dusty stellar sources, X3, X7, and X8 are also found to the SW of Sgr A* along the extension of 
the jet feature \citep{muzic07,muzic10,peisker19}.  These sources show a high abundance of [FeIII] line 
emission, with similar characteristics to the western edge of the minicavity \citep{peisker19}.  To the 
NE of Sgr A*, there is evidence of an elongated X-ray structure, NE plume in \cite{paper4}, a number of 
cometary stellar sources pointing towards Sgr A* (F1, F2, F3) and red-shifted ionized gas 
\citep{muzic07,muzic10,paper1}.  \cite{muzic10} suggested that an external wind with density $\sim10$ 
cm$^{-3}$ and speed $\ga 1000$\,\kms\, emerging from the close vicinity of Sgr A* is responsible for 
shaping the bow shocks and orienting them towards Sgr A*.  The inferred ram pressure, $\rho\, 
v^2\sim2\times10^{-7}$ dyn cm$^{-2}$  is one twentieth of the 
jet ram pressure thought to be responsible for the formation of the minicavity, 
As a working 
hypothesis we assume that  the opening angle of the 
outflow is 60\degr, with the inner 30\degr occupied by a mildly  relativistic jet. 
In this picture, the EW ridge 
 at a PA $\sim$30\degr away from the direction of the relativistic jet 
is mainly produced by the interaction of the outflow with dusty  infrared 
stellar sources. 

Third, on scales of a few mpc, H30$\alpha$, H39$\alpha$ and H56$\alpha$ spectra exhibit broad blue- and 
 red-shifted line emission with a line width of thousands of \kms\, and oriented roughly along the Galactic 
 plane to the NE and SW, respectively.  These have previously been interpreted as the red and blue lobes of a 
 highly inclined Keplerian disk \citep{elena19}.  However, our observations have wider frequency coverage and 
 find significantly broader line widths, $\ga 4000$\kms consistent with Keplerian motion at about 1/5 the 
 observed angular scale of the emission.
Instead, this emission more likely arises from entrained material being accelerated outwards by the jet.  
The mass of accelerated material cannot easily be inferred from the recombination-line fluxes because of the 
dominant contribution of stimulated emission.  The upper limit on the Br$\gamma$ emission within 0.2$''$
of Sgr A* is $\sim 2\u Jy\, $\kms \citep{CCMD20}. 

The Br$\gamma$ emissivity for T$_e\sim10^4$ K and n$_e\sim10^4 \rm cm^{-3}$ is 
j$_\gamma\sim2.7\times10^{-28}$ erg s$^{-1}$ cm$^{-3}$ sr$^{-1}$ \citep{hummer87}. 
The Br$\gamma$ luminosity for a source with volume emission measure 
VEM=$\int n_e^2\, dV$, 
yielding a flux S$_\gamma=(\rm j_\gamma\, \rm VEM) / d^2 \sim 0.96 (\rm VEM/ (10^{58}\, \rm cm^{-3}))$  Jy\, \kms\, 
where a Galactic center distance d=8 kpc has been adopted. 
This implies that the volume emission measure  of 
the material is $\la\, 2\e{58}$\,cm$^{-3}$, and 
that stimulated emission 
amplifies the H30$\alpha$ line luminosity by at least a factor of 80 \citep{elena19}.  Our 
radio continuum observations have tentatively detected this material: the weak, extended features J1 and 
J3 that lie within $0.2''$ of Sgr A* have flat spectra between 9 and 44\,GHz, consistent with 
optically-thin thermal bremsstrahlung from the red- and blue-shifted components of this gas, respectively.  
J3 lies in an area confused by extended continuum emission, making it difficult to extract reliable 
fluxes, so we focus on J1, which has an integrated flux $\sim 0.2$\,mJy at 36\,GHz. 
The free-free emissivity at 36 GHz is 
j$_\nu/n_e^2\sim2.2\e{-40}\,  \textrm{erg}\,  \textrm{cm}^3\,  \textrm{Hz}^{-1}\,  \textrm{sr}^{-1}$,  
 yielding a flux S$_\nu= \textrm{j}_\nu\, \textrm{VEM}/\rm d^2 \sim 0.32\,  \textrm{VEM}/(10^{58}\, \rm 
\textrm{cm}^{-3})$ mJy. 
Thus,  we infer 
$\textrm{VEM}\approx6\e{57}\ut cm -3 $ for nominal temperature $10^4$\,K.  Adopting dimensions 
$0\farcs2\times0\farcs02\times0\farcs02$ consistent with Fig.~6(b) this implies $n_e\approx1\e5\ut cm -3 $ 
and mass $\approx6\e{-5}\msol$.  

One of  the implications of  the presence of 
ionized gas on such a small scale associated with Sgr A* is that it can also 
explain   the  large rotation measure detected toward Sgr A* (RM$\sim-5\e5$\, rad m$^{-2}$) \citep{bower18}.
Assuming that the parallel component of the  magnetic field is 1.5 mG, high velocity
ionized gas detected within 0.1$''$ of   Sgr A*  can produce  the observed Faraday rotation.


Fourth, on a scale of AU, the polarization angle of near-IR flaring activity of Sgr A* can be measured several times 
a day.  These measurements give a mean polarization angle of 60$^\circ\pm20^\circ$ consistent with the position 
angle of the larger scale jet features \citep{eckart06,meyer07}. In addition, VLBI measurements provide highest 
resolution images of millimeter emission from Sgr A* with mas size scales. The PA of millimeter emission from Sgr A* 
is roughly in agreement with a PA within 30$^\circ$ of the Galactic plane.  Thus, this emission could be associated 
with the large scale jet feature having similar PA.


Fifth, the linear pc-scale jet-like features that we have identified are stronger at 9\,GHz than at 
44\,GHz, suggestive of synchrotron emission from a jet emerging from the vicinity of Sgr A*.  This 
confirms our earlier narrow band observations reporting a tentative detection of continuous linear 
structure with an extent of $\sim$3\,pc and PA$\sim 60^\circ$ \citep{paper4}.  The extension of this 
feature appears to terminate symmetrically by linearly polarized structures at 8.4 GHz, $\sim75''$ from 
Sgr A* \citep{paper4}.  For the observed jet diameter ($\approx 2\arcsec$) and 1.6\,GHz surface brightness 
$\approx 0.2\u mJy \ut beam -1 $, we deduce an equipartition magnetic field $\sim 0.2$\,mG using standard 
(but uncertain) assumptions: an $E^{-2}$ electron spectrum running between 10\,MeV and 100\,GeV and equal 
energy density in electrons and protons.


Finally, the NE edge of  Sgr A East is  highly distorted along the Galactic place \citep{paper2}. In the jet picture, the kinetic luminosity of the jet is likely responsible for the distortion of Sgr A East.    
In addition, the puzzle of why the Sgr A East shell is elongated along the Galactic plane can also be explained by a medium that has been swept by a symmetric outflow from Sgr A* along the Galactic plane before the expansion of the remnant occurred. 

The six characteristics described above provide overwhelming evidence for a jet extending from AU to pc scales likely surrounded by a wider-angle outflow in the inner parsec.  

\subsection{Orientation of the Jet}

The orientation, and mass flow rate of the jet is consistent with jet-launching models, which predict that 
the jet emerges perpendicular to the equatorial plane of the accretion flow near the event horizon of Sgr A*.  
Here we define the jet orientation to be the 3D unit vector parallel to the red-shifted component, which we 
quantify by two angles, $\Omega_j$, the PA of the arm (measured E of N), and the angle $\theta_j$ that the 
arm makes to the line of sight (so $0\degr \leq \theta_j \leq 90 \degr$).  The inferred jet is aligned 
roughly NE-SW on the sky, and given that the disturbed material in the mini-cavity is blue shifted, we infer 
that the mini-cavity and the ionized bar lie in front of Sgr A* and that the SW arm of the jet is approaching 
us.  Thus the red shifted arm of the jet emerges to the NE of Sgr A* with PA $\Omega_j \approx 60\degr$ E of 
N and makes an angle $\theta_j \approx 45\degr $ to the line of sight (which is directed away from the 
observer).  The angular momentum of the inner accretion flow on to Sgr A* should then be parallel or anti 
parallel to the redshifted arm, implying that the orbital elements of the disk, i.e. PA of the ascending 
node, $\Omega$, and inclination to the plane of the sky, $i$, are either (i) $\Omega= \Omega_j-90\degr 
\approx 330\degr$, $i= 180\degr-\theta_j\approx120\degr$, or (ii) $\Omega= \Omega_j+90\degr \approx 
150\degr$, $i= \theta_j\approx120\degr$. Figure 14a,b show schematic diagrams  of the possible orientations of 
the jet  launched from the accretion disk. 


Sgr A* is fed by the combined winds of WR stars in the inner $\sim\,0.1$\,pc of the Galaxy 
\citep{CM97,CNSM06,CNM08,rsq19}.  The orbits of most of these stars lie in the ``clockwise disk'' , with 
orbital planes distributed around a mean PA of their ascending nodes $\Omega \approx 106\degr$ E of N and 
mean inclination $i\approx126\degr$ to the line of sight \citep{BMFG09,GETA09,YGLD14,GPES17}.  
Figure 14c shows the orientation of
clockwise stellar disk 
which is thought to be the source of the accreting material.
While the stellar disk orientation  is 
significantly misaligned with the average angular momentum of the inner accretion flow 
inferred from the jet orientation, simulations show that the material in the inner accretion flow was 
injected with low orbital angular momentum, i.e.~the wind material that is ejected antiparallel to the 
orbital motion of a small number of dominant WR stars \citep{CCSB20,GBDA20,rsq19,rsq20}.  The inherent 
granularity of this process introduces significant variations in the orientation of the low-angular 
momentum material on $\sim200$ year time scales as the stars move around their orbits \citep{rsq20}.  
Further, astrometric observations of infrared flares with the 
GRAVITY instrument on board the VLT suggest 
a hotspot source orbiting Sgr A* with inclination $i\approx130\degr$ and $\Omega$ most likely to be either 
$\approx130\degr$ -- consistent with the clockwise stellar disk, or $\approx300\degr$ -- consistent with 
our observed jet orientation \citep{GBDA20}.  Finally, we note also that the black hole spin, which 
reflects the accretion of angular momentum by the hole during its life time is likely to be significantly 
misaligned with the clockwise disk, reflecting instead the accumulation of material over cosmic time.  If 
the net spin is significant, one might expect the inner accretion flow to be forced into alignment with 
the equatorial plane of the Kerr space time on scales of a few gravitational radii. We conclude then that 
the misalignment of the jet and the angular momentum of the clockwise disk is not a significant issue.

\section{Summary}

We presented multi-wavelength spectroscopic and broad band continuum measurements over a wide range of 
angular scales toward Sgr A* and its local environment. These measurements indicate a jet driven outflow arising 
from Sgr A* on mpc to pc scales along the Galactic plane. The opening angle of the outflow is 60\degr, with the 
inner 30\degr occupied by the relativistic jet. Our millimeter recombination line data supported earlier 
studies that the red and blue-shifted ionized gas detected toward Sgr A* lie along the Galactic plane. We 
argued that the interaction of the collimated outflow with stellar atmospheres and gas clouds can be 
responsible for disturbed kinematics of ionized gas, enhanced FeII/III and [FeIII] to Br$\gamma$ line ratios, 
head-tail cometary stellar sources, X-ray emission from IRS 13 and high Faraday rotation observed toward Sgr 
A*. Shock heated gas resulting from the interaction has other implications such as compressing the gas to 
possibly overcome tidal shear, thus inducing star formation. 
The  outflow could also shape  the elongation of the Sgr A East SNR and be responsible for high 
cosmic ray ionization rate inferred from H$_3^+$ measurements within the molecular ring and the central 
molecular zone, e.g, \citep{goto14}. Furthermore, our radio continuum data demonstrates that lower frequency emission 
from 
Sgr A* at $\nu <1.4 $ GHz and its environment have the potential to determine the mass of ionized gas.  In the jet picture, a 
significant amount of the power comes out in the form of the kinetic luminosity carrying off energy and 
angular momentum.  This could explain why Sgr A* appears radiatively under luminous.

\section{Data Availability}

All the data that we used here are available online and are not proprietary. 
 We have reduced and calibrated these data and are available if  requested. 


\section*{Acknowledgments}

This work is partially supported by the grant AST-0807400 from the NSF. The 
National Radio Astronomy Observatory is a facility of the National Science Foundation operated under
cooperative agreement by Associated Universities, Inc.
This paper makes use of the following ALMA data: ADS/JAO.ALMA\#2015.0.00021.S. ALMA is a partnership of 
ESO (representing its member states), NSF (USA) and NINS (Japan), together with NRC (Canada), NSC and
ASIAA (Taiwan), and KASI (Republic of Korea), in cooperation with the Republic of Chile. The Joint ALMA
Observatory is operated by ESO, AUI/NRAO and NAOJ. 






\onecolumn


\begin{figure}
\centering
\includegraphics[scale=0.45]{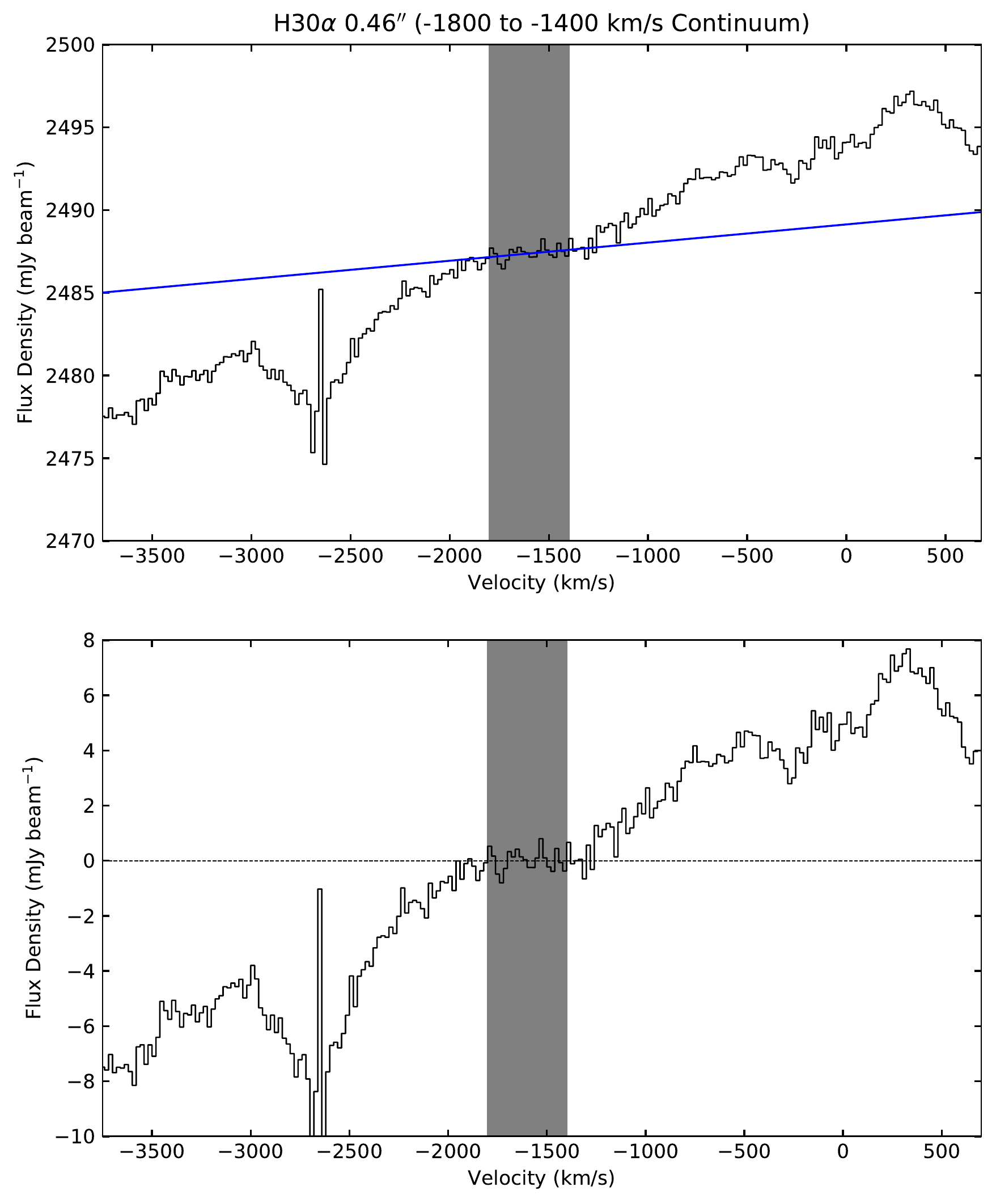}
\includegraphics[scale=0.45]{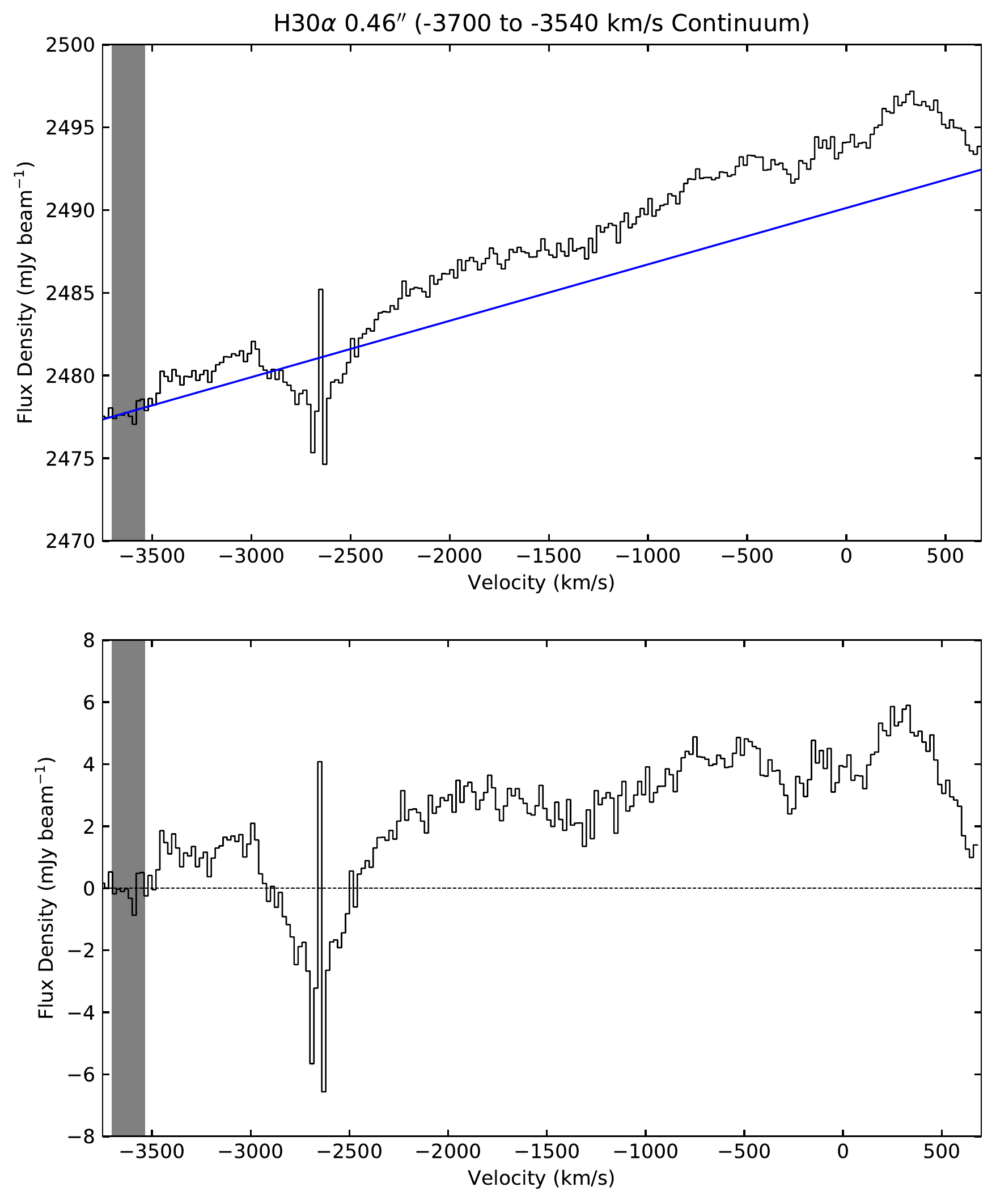}
\caption{
{\it (a Top Left)}
The spectrum of Sgr A* 
integrated over a beam with a diameter of 0.46$''$ is shown before  continuum subtraction. 
In order to confirm the double-peak spectrum of Sgr A*, we used similar parameters to those of 
\citet{elena19}, 
assuming  line free channels at velocities between --1300 and --1800 \kms\,  (the shaded region). 
The overlayed line is the 
least squares straight-line fit to the continuum in  the shaded region.
{\it (b Bottom Left)}
Identical to (a) except that
the continuum has now been subtracted from the spectrum.
The vertical gray bars show the selected
line free channels.
{\it (Top Right c, Bottom Right d)}
As for Figure 1 except line free channels are instead chosen from --3600 and --3800 \kms\, 
(the shaded region).  
As before the top panel is before continuum subtraction and the bottom panel is post continuum subtraction.
}
\end{figure}

\begin{figure}
\centering
\includegraphics[scale=0.45,angle=0]{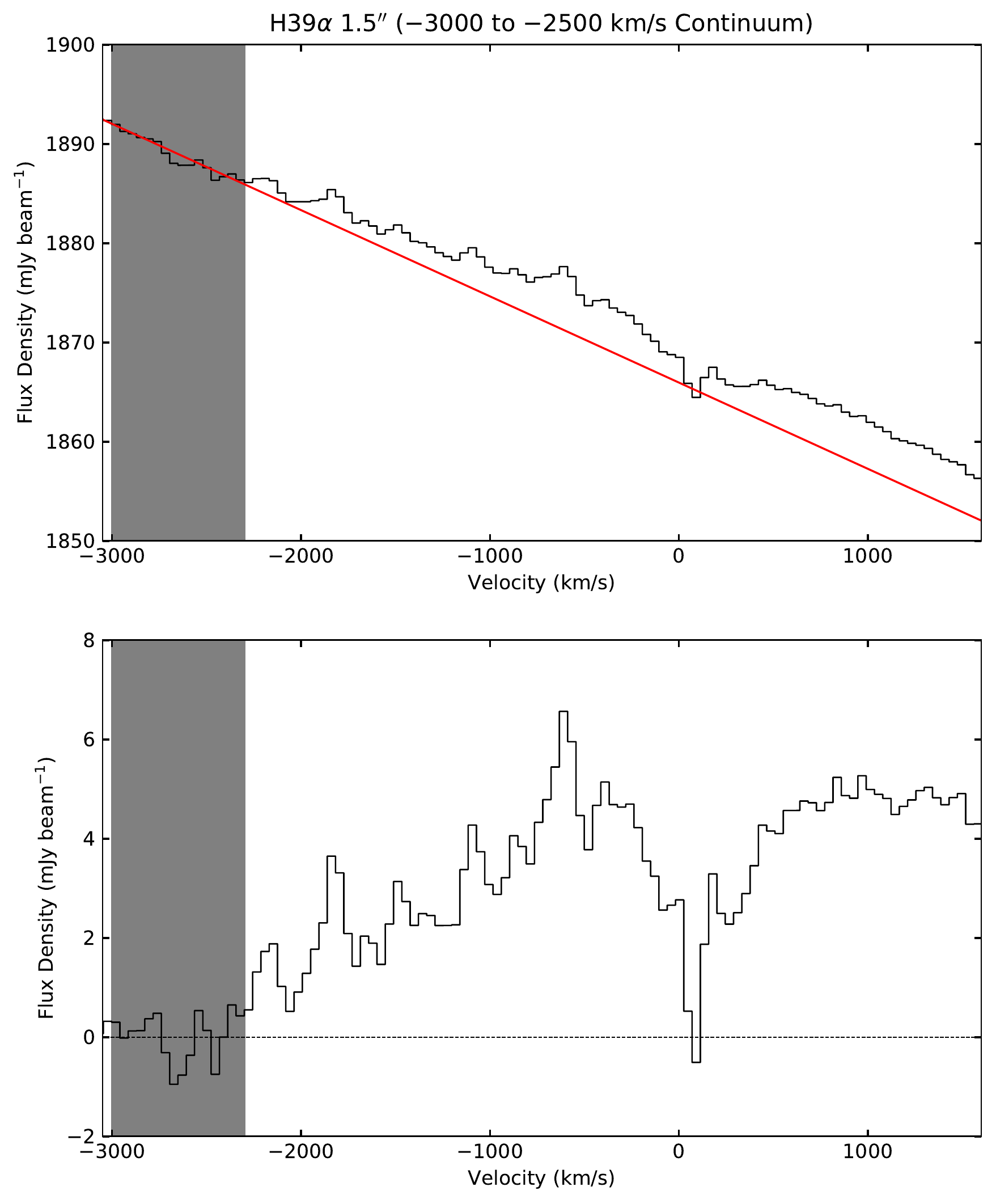}
\includegraphics[scale=0.45,angle=0]{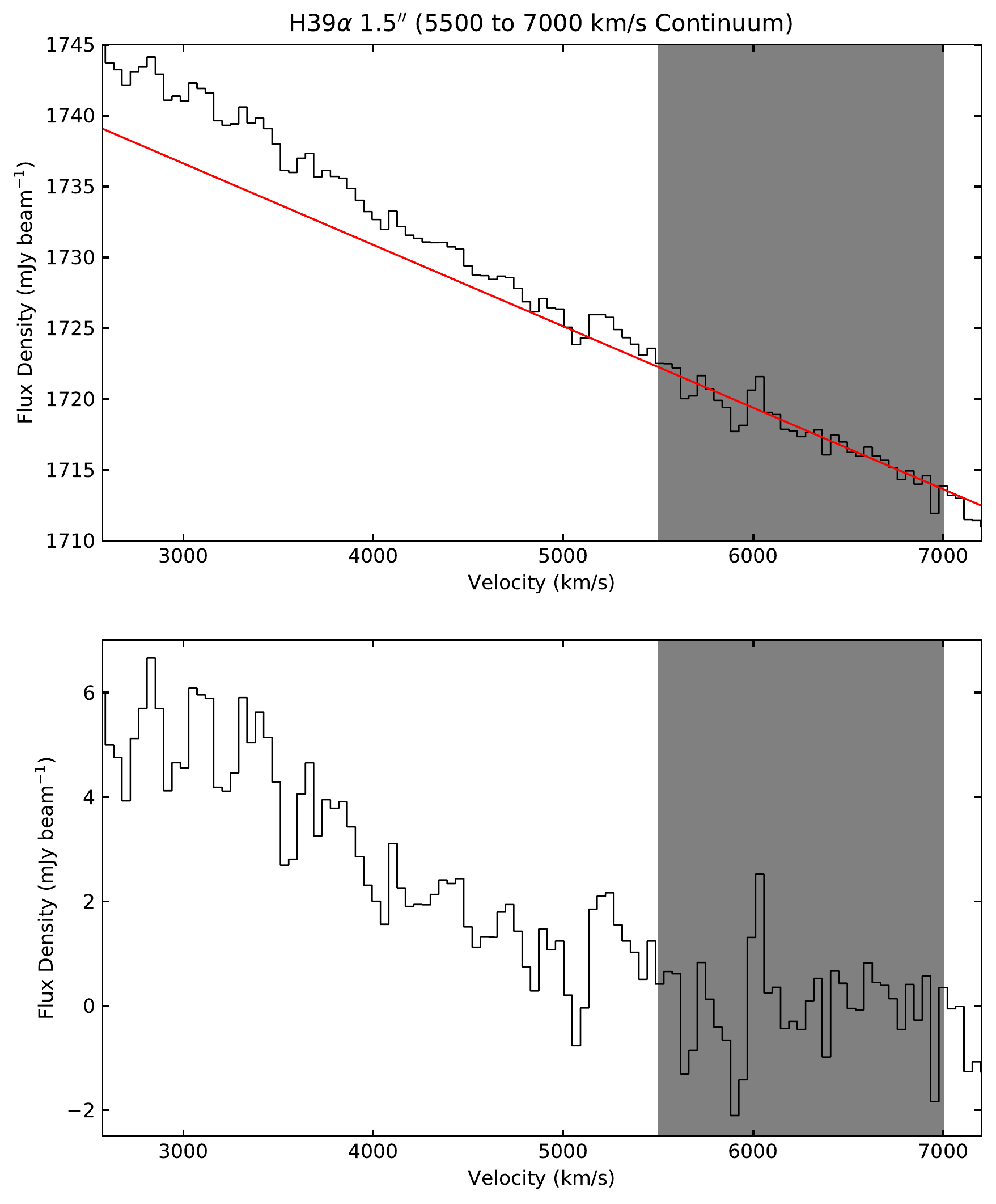}
\includegraphics[scale=0.55,angle=0]{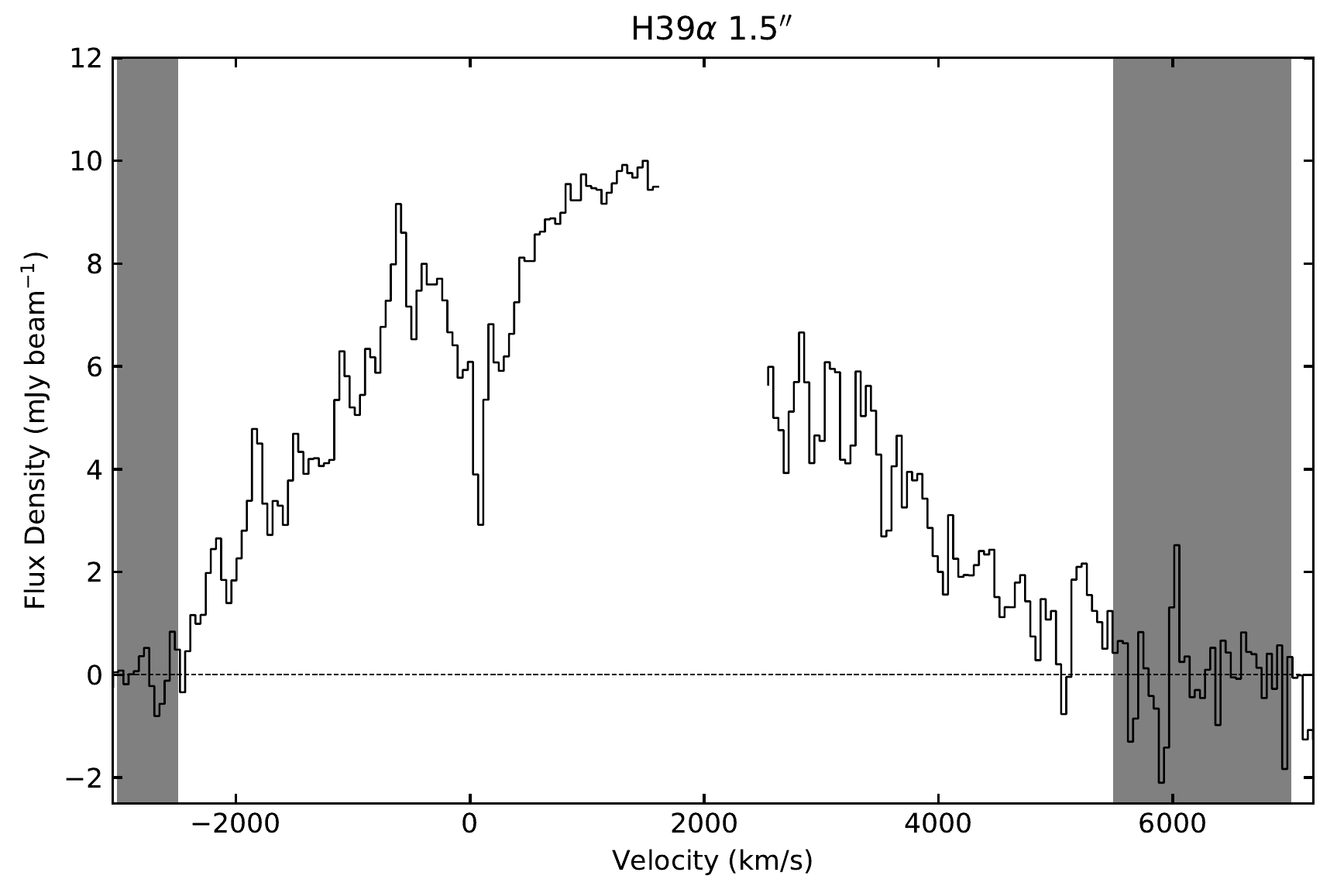}
\caption{
{\it (a Left)} 
Emission with or without continuum subtraction from  the inner 0.75$''$~at the peak position of Sgr A*~from --3050 \kms~to1600\kms. 
 A fitted continuum is shown as a result of fitting chosen line free channels from --3000\kms~to --2500\kms.  {\it (b Right)} 
Similar to {\it a} except the emission from 2575 \kms~to 7200 \kms~is shown.  The line free channels were chosen as 5500 \kms~to 7000 \kms~and the resultant fit is drawn.  {\it (c Bottom)}  
Excess H39$\alpha$ emission found from subtracting continuum in the respective top panels in (a,b).
}
\end{figure}

\begin{figure}
\centering
\includegraphics[scale=0.4,angle=0]{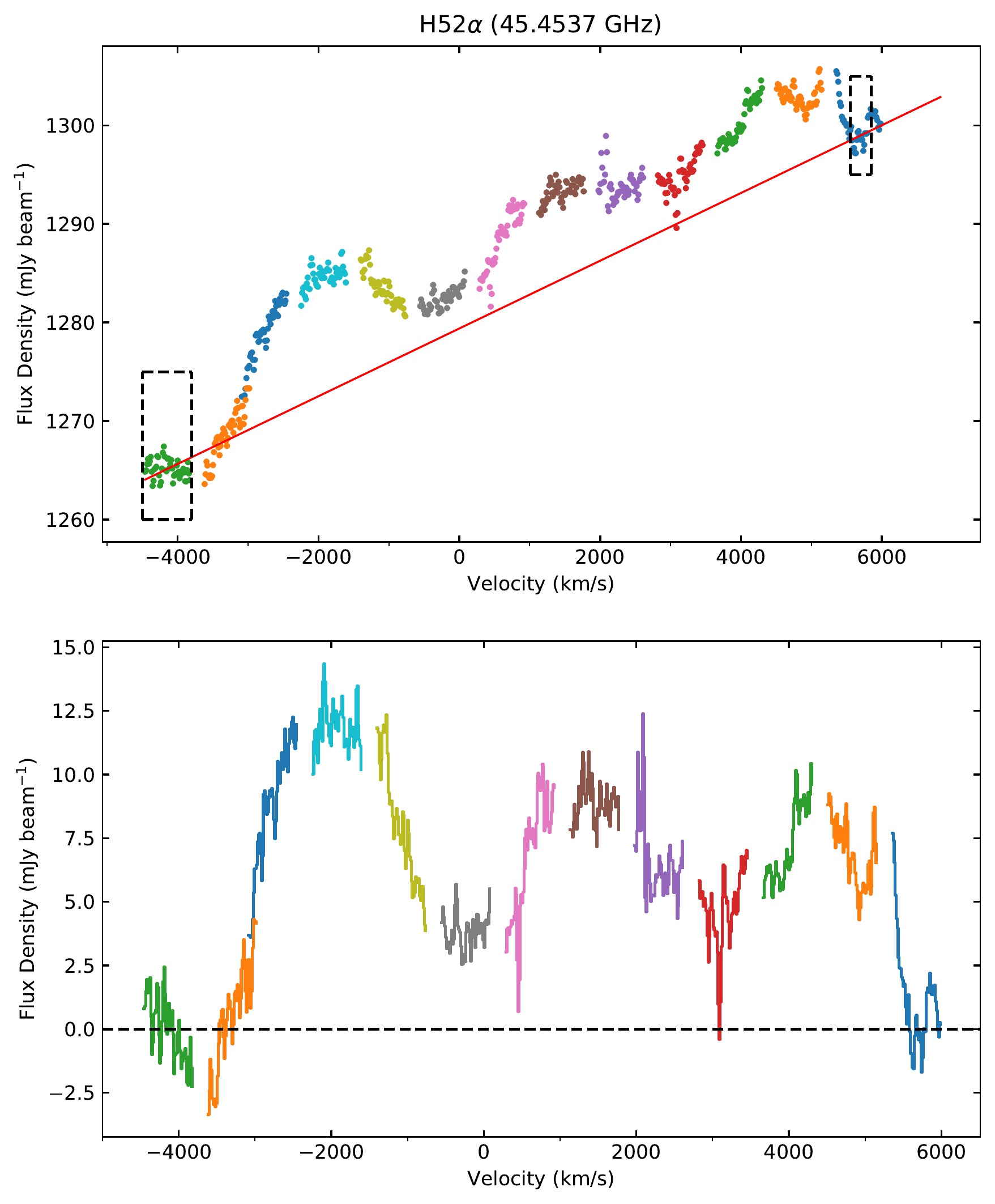}
\includegraphics[scale=0.5,angle=0]{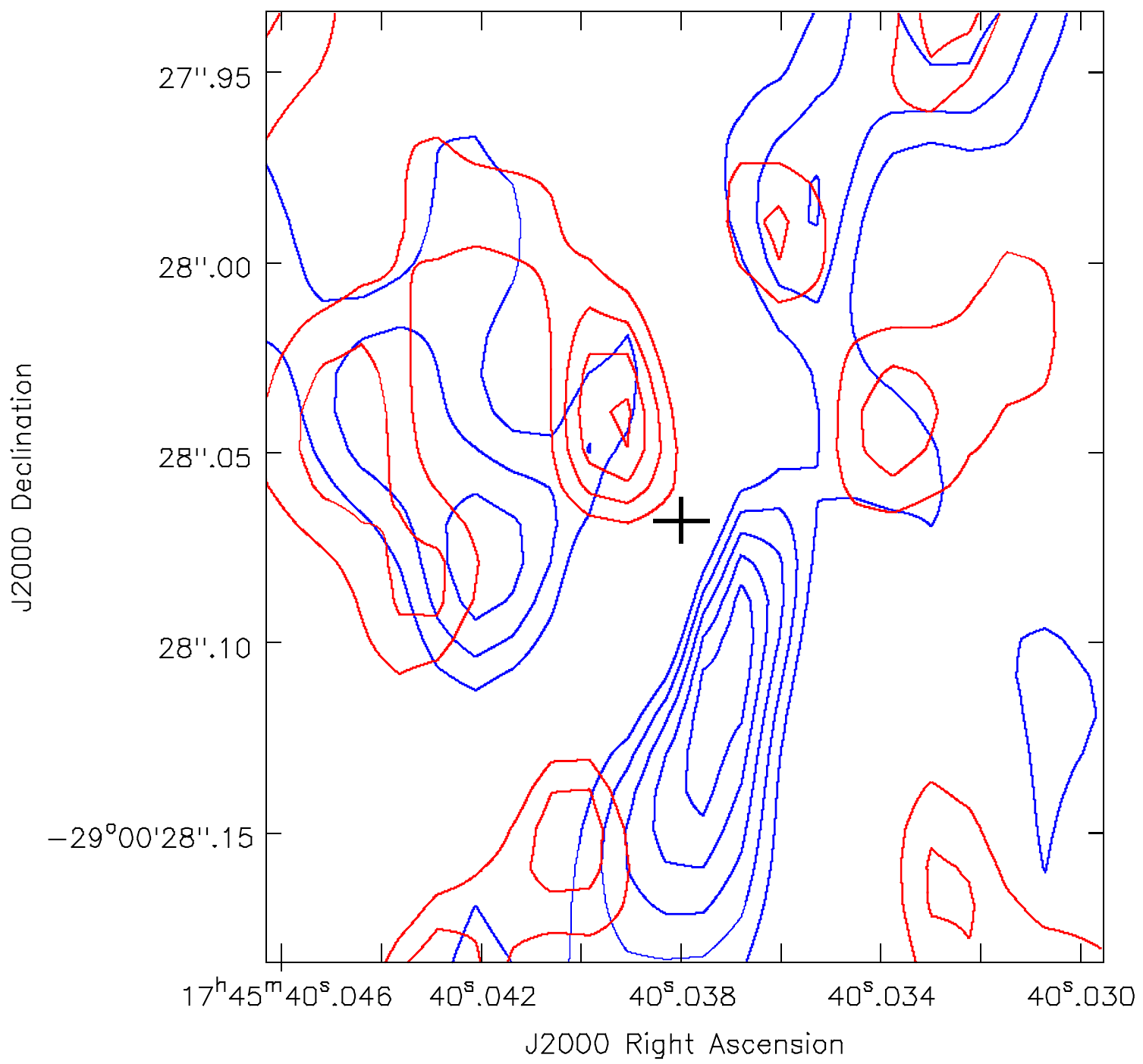}
\caption{
{\it (a Top Left)} 
A scatter plot of the spectrum of H52$\alpha$ emission without continuum subtraction from the inner 
0.077$''\times0.035''$ (PA$\sim-2^\circ$) of Sgr A* using twelve adjacent spectral windows. 
{\it (b Botton Left)} 
Continuum subtracted H52$\alpha$ emission from Sgr A* presented in histogram format. 
This  continuum was fit in the range  $-4000$  to $-3600$  and $6000$ to $6300$ \kms.
The spectrum of 
individual spectral window is scaled such that adjacent spws are continuous.  This scaling removes the flux 
scale across the band which varies at a level of few tenths of a percent.
{\it (c Left)} 
Contour levels of red- and blue-shifted components of 
H52$\alpha$ are set at 0.18,  0.36,  0.54,  0.72 Jy\, beam$^{-1}$\, \kms. 
The beam sizes for the blue- and red-shifted moment maps  are 
$0.075''\times0.034''$  and $0.077''\times0.035''$, respectively. 
The blue and red-shifted 
velocity ranges, in red, blue colors, correspond to $\sim4000$ and $-2000$ \kms, respectively. 
 The drawn 
cross coincides with the position of Sgr A*.
}
\end{figure}

\begin{figure}
\centering
\includegraphics[scale=0.4,angle=0]{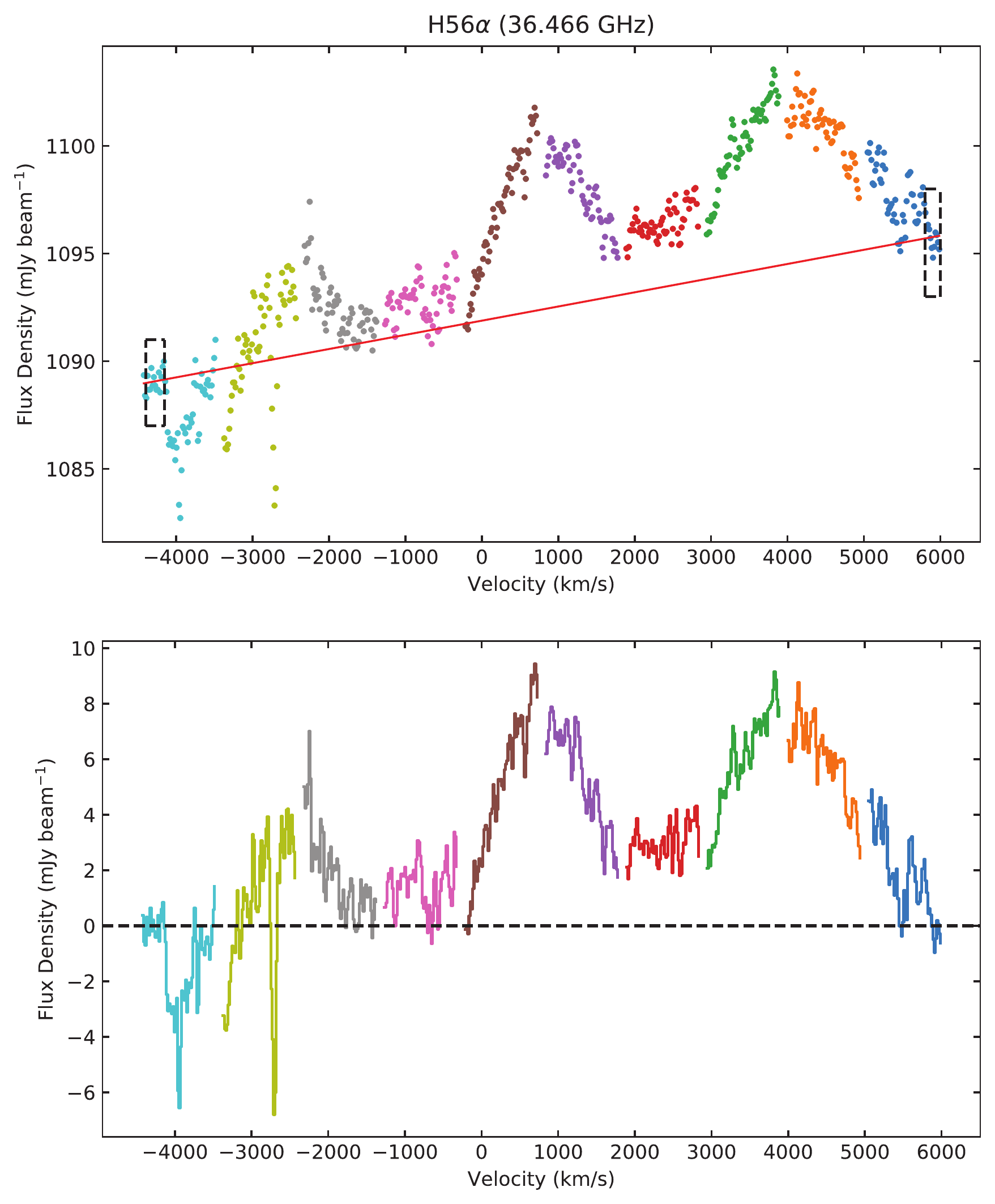}
\includegraphics[scale=0.5,angle=0]{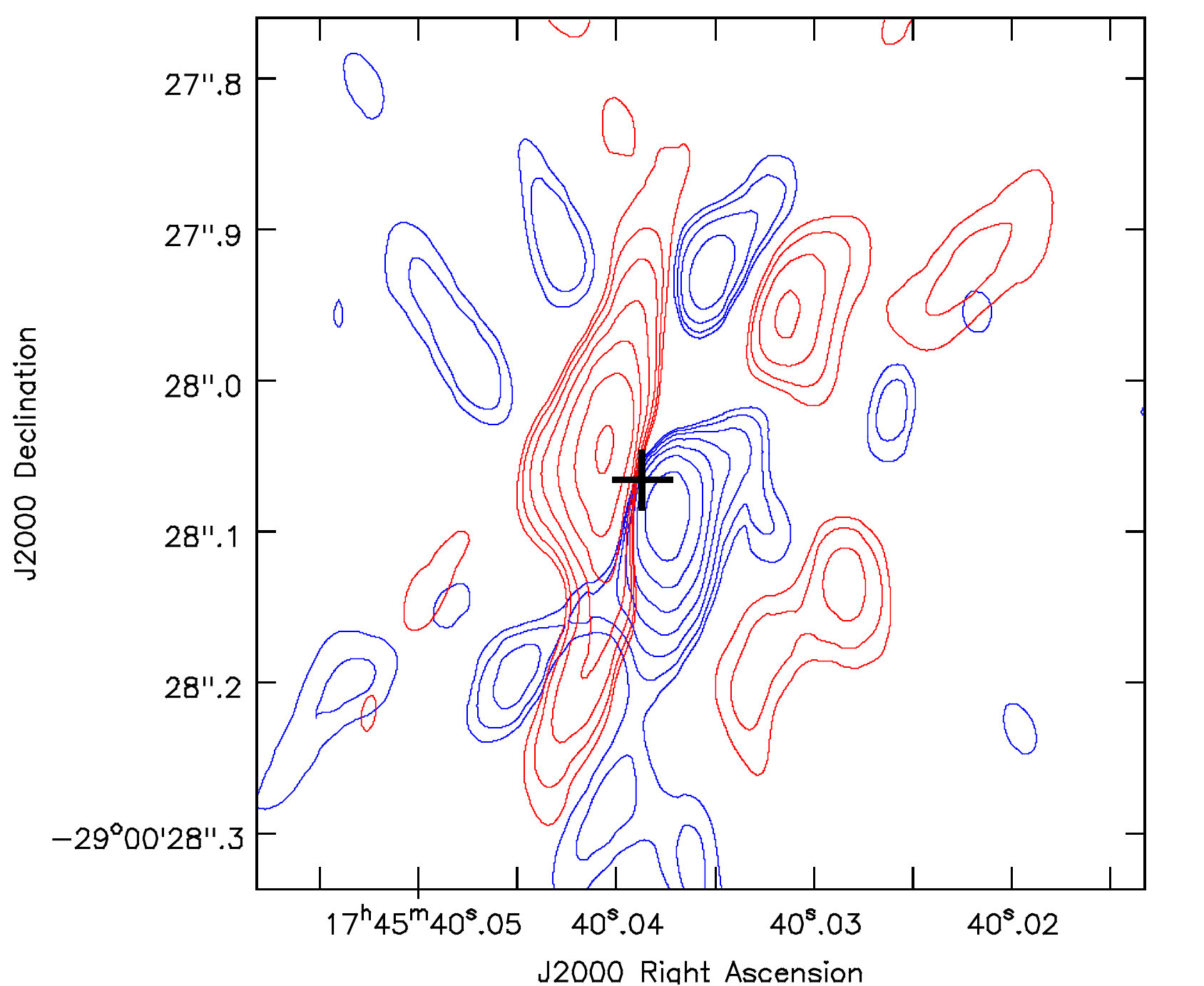}
\caption{
{\it (a Top Left)} 
A scatter plot of the spectrum of H56$\alpha$ emission without  continuum subtraction 
from the inner 0.22$''\times0.27''$ of Sgr A* using ten   adjacent spectral windows. 
The rest frequency of H56$\alpha$ transition is at 0 \kms. 
{\it (b Bottom Left)} 
Continuum subtracted H56$\alpha$ emission from Sgr A* presented in  histogram format. The spectrum of individual spectral window is scaled such that adjacent spws  are continuous.  This scaling removes 
the flux scale across the band 
which varies at a level of  few tenths of a percent. 
{\it (c Right)} 
Contour levels  of red-  and blue-shifted components of  H56$\alpha$ 
are set at 1, 1.5, 2.25, 3.38,  5.06,  7.6 and  11.4 mJy per $0.099''\times0.05''$ beam and $0.095''\times0.049''$ beam, 
respectively.  
The blue  
and red-shifted velocity ranges,  in red, blue colors, correspond  to $\sim4000$ and $-2000$ \kms, respectively.  
The drawn cross  coincides with the position of Sgr A*. 
}
\end{figure}

\begin{figure}
\centering
\includegraphics[scale=1.0,angle=0]{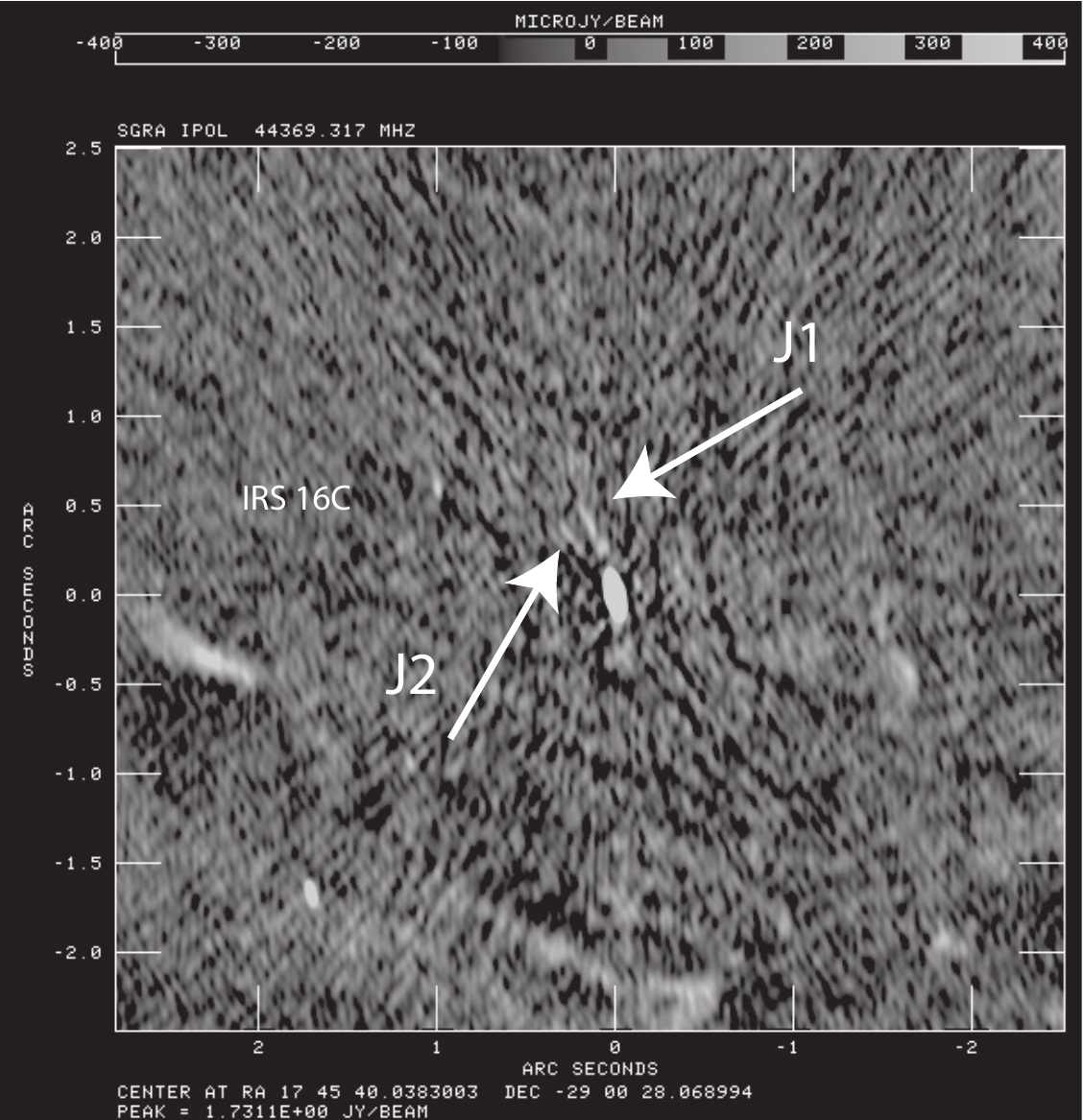}\\
\includegraphics[scale=1.0,angle=0]{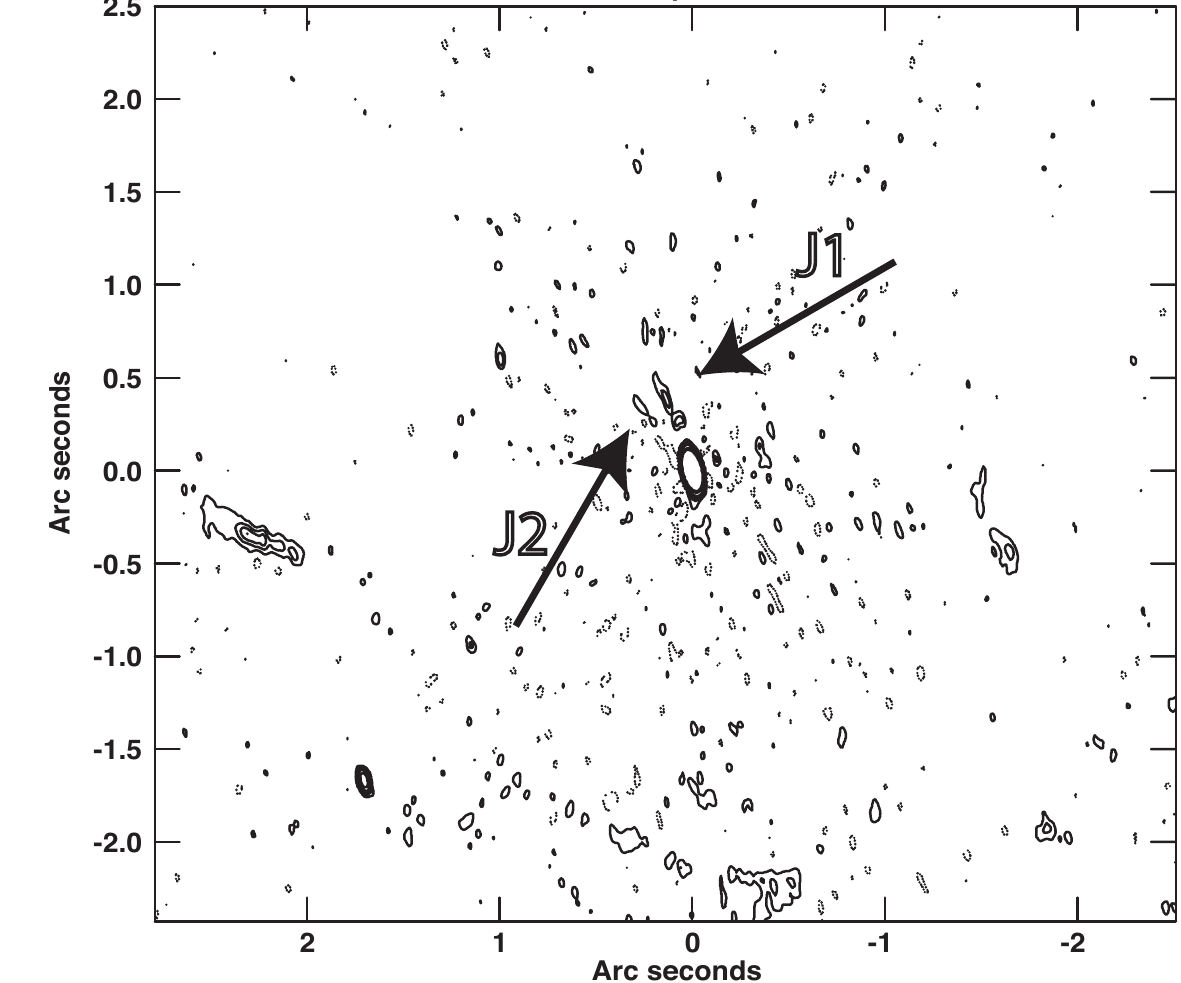}
\caption{
{\it (a)}
Grayscale continuum image of the inner 2.5$''$ of Sgr A*  at 44 GHz  with a resolution of 
$0.087''\times0.036''$ (PA=13.41$^\circ$).
The dynamic range of this image  with the peak flux  1.73 Jy beam$^{-1}$ is 3700. 
{\it (b)}
The same as (a) except that contours are displayed at levels 
-100, 100, 200, 300, 400, 500, 1000, 5000, 10000  $\mu$Jy beam$^{-1}$. 
This data set was  taken with the VLA on September 16,  2015. 
}
\end{figure}

\begin{figure}
\centering
\includegraphics[scale=.9,angle=0]{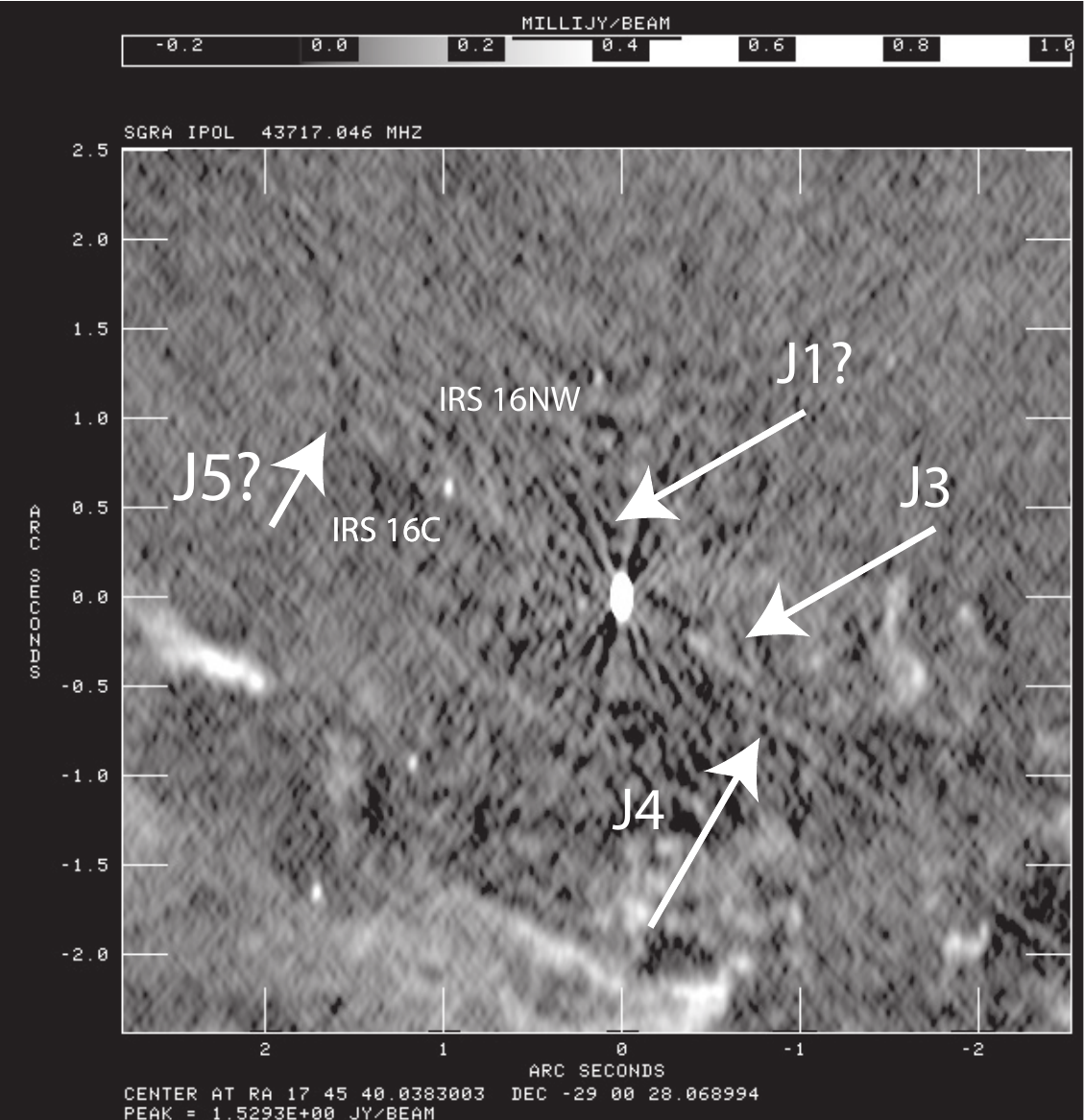}
\includegraphics[scale=0.65,angle=0]{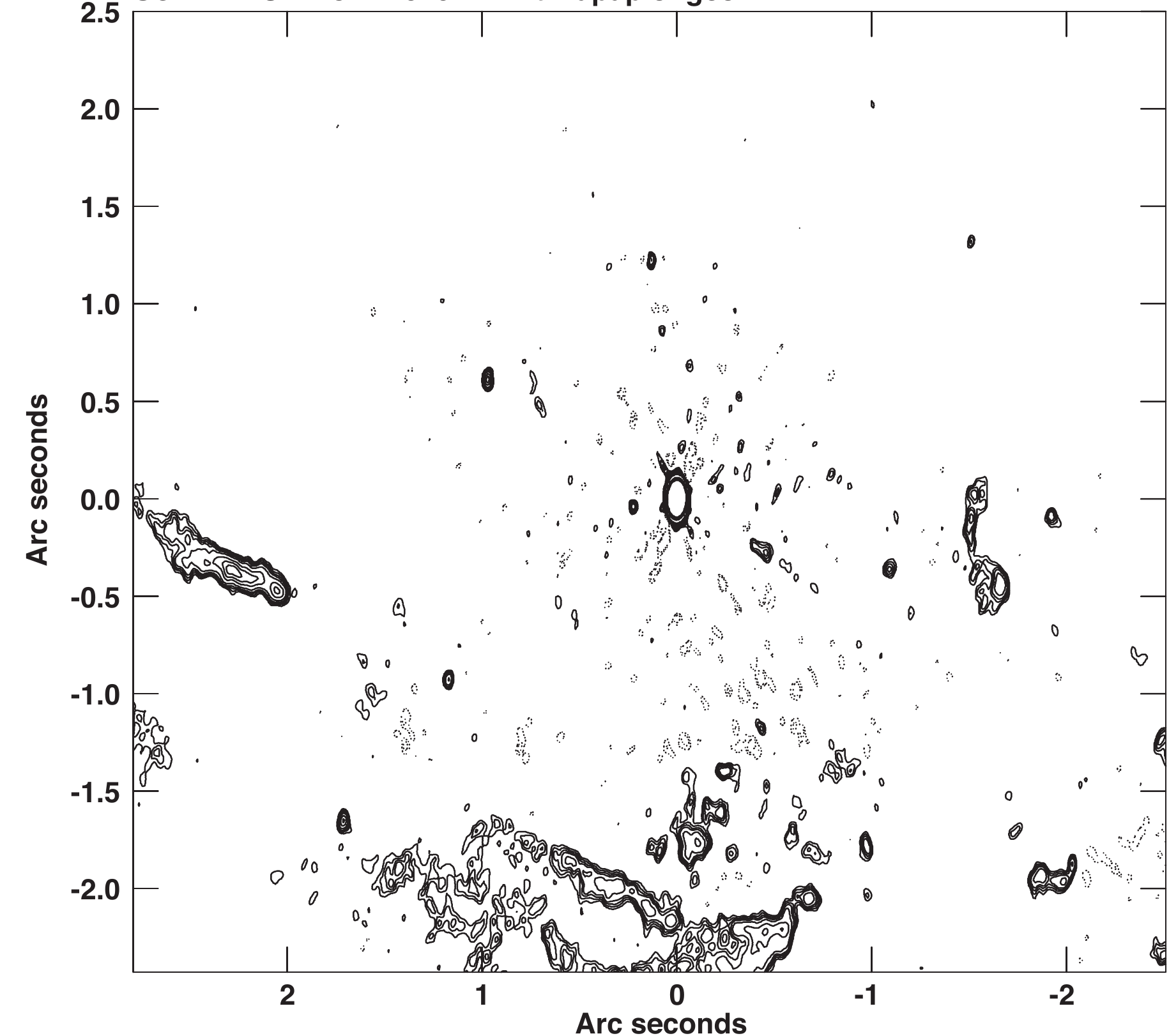}
\caption{
{\it (a)}
A grayscale image of the inner 2.5$''$ of Sgr A* at 44 GHz with a spatial resolution of  $0.075''\times0.037''$ (PA=$-0.86^\circ$)
taken on April 3, 2018.  The rms noise is  18$\mu$Jy. Prominent  features to the south are associated with the mini-spiral or Sgr A West.  
{\it (b)} The same as (a) except that contours are displayed at levels 
-100, -75, 75, 100, 125, 150, 200, 300, 400, 500, 750,  1000, 5000, 20000  $\mu$Jy beam$^{-1}$. 
}
\end{figure}

\begin{figure}
\centering
\includegraphics[scale=0.5,angle=0]{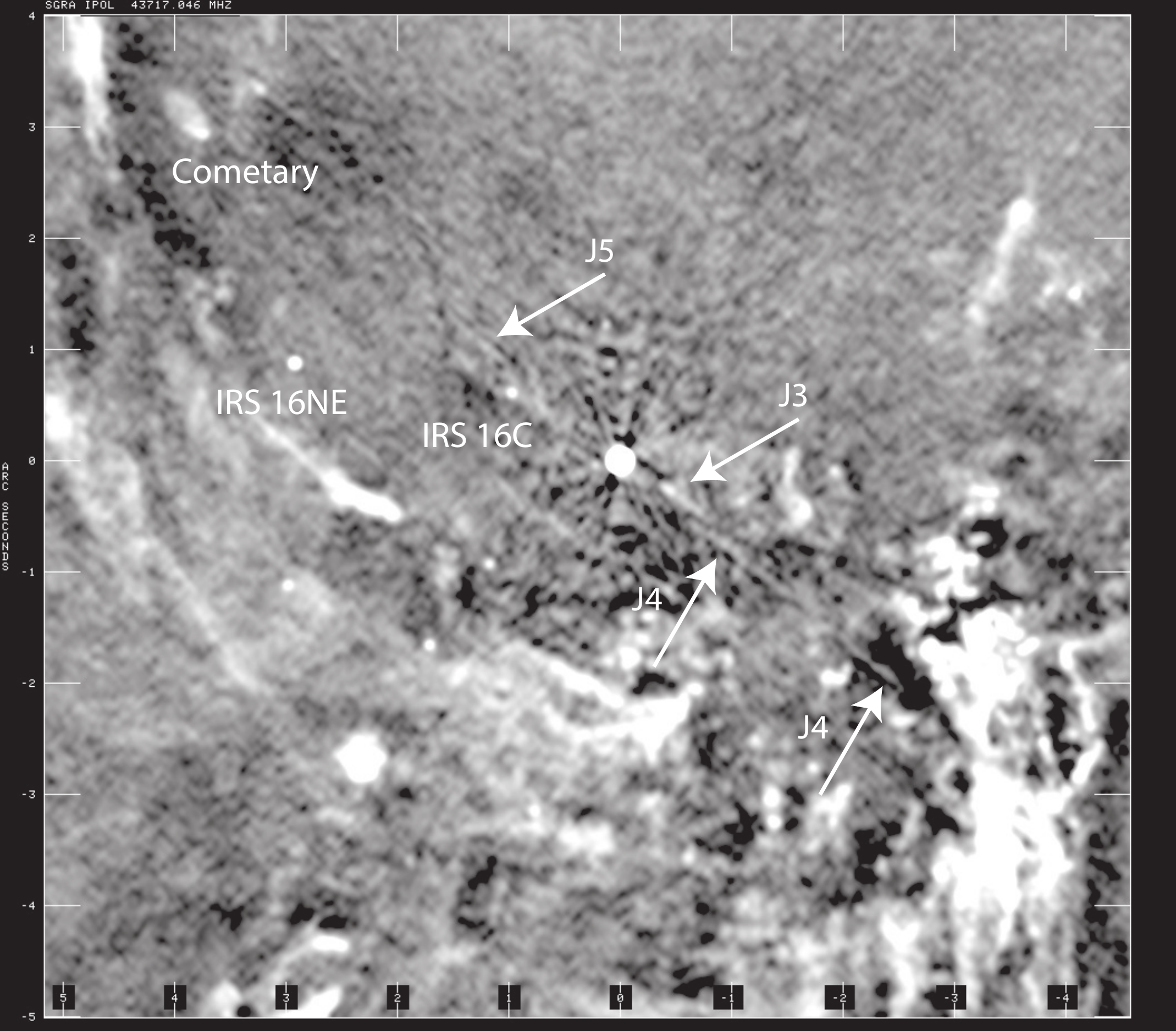}
\includegraphics[scale=0.5,angle=0]{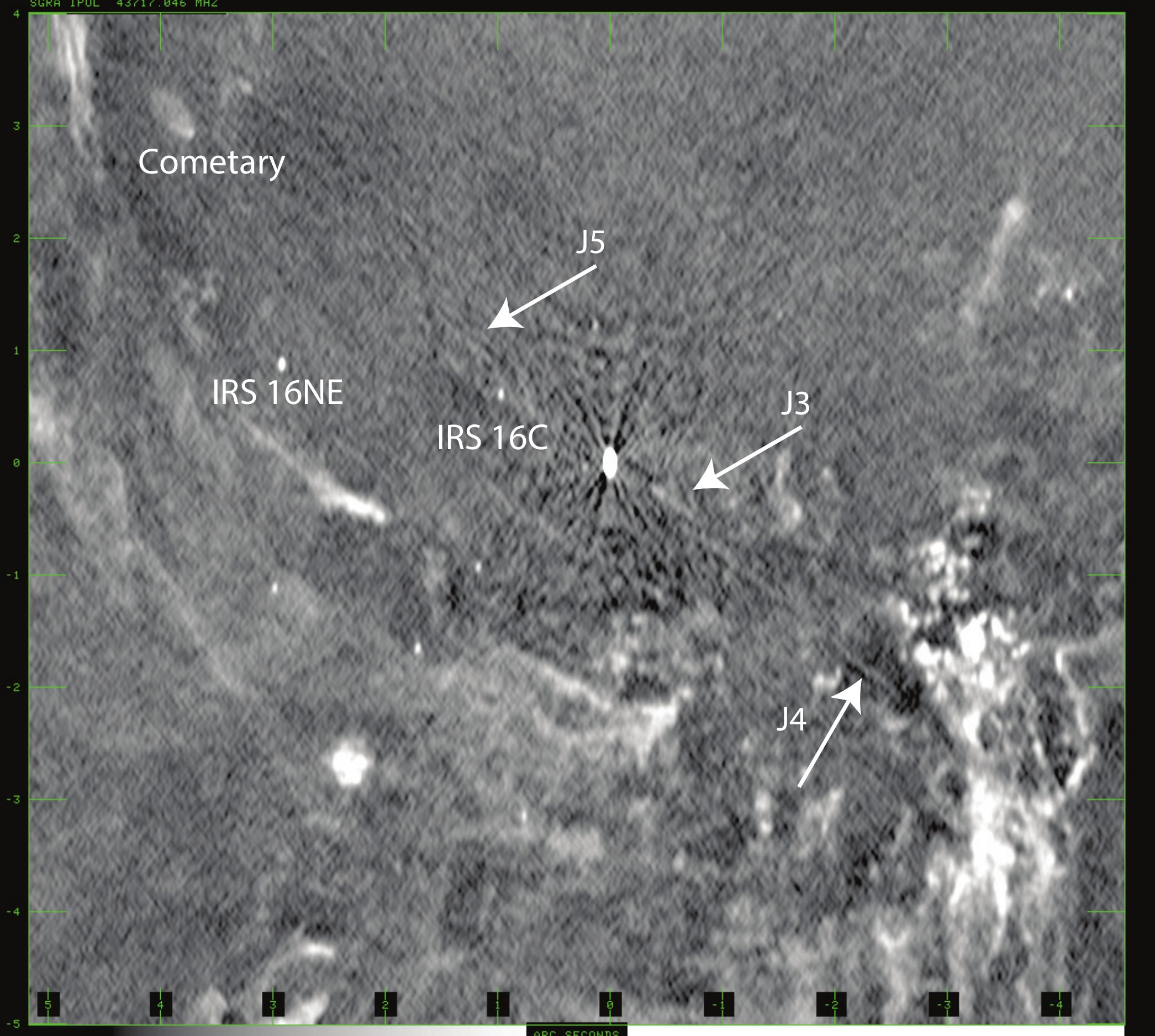}
\caption{
{\it (a, b)}
Two grayscale images of the inner 4$''$ of Sgr A* at 44 GHz with two different brightness contrasts. 
The images are similar to Figure 6 except they are convolved with a 0.076$''$ Gaussian beam. 
The rms noise is 21$\mu$Jy. 
The prominent feature to the west is the mini-cavity. 
}
\end{figure}

\begin{figure}
\centering
\includegraphics[scale=0.4,angle=0]{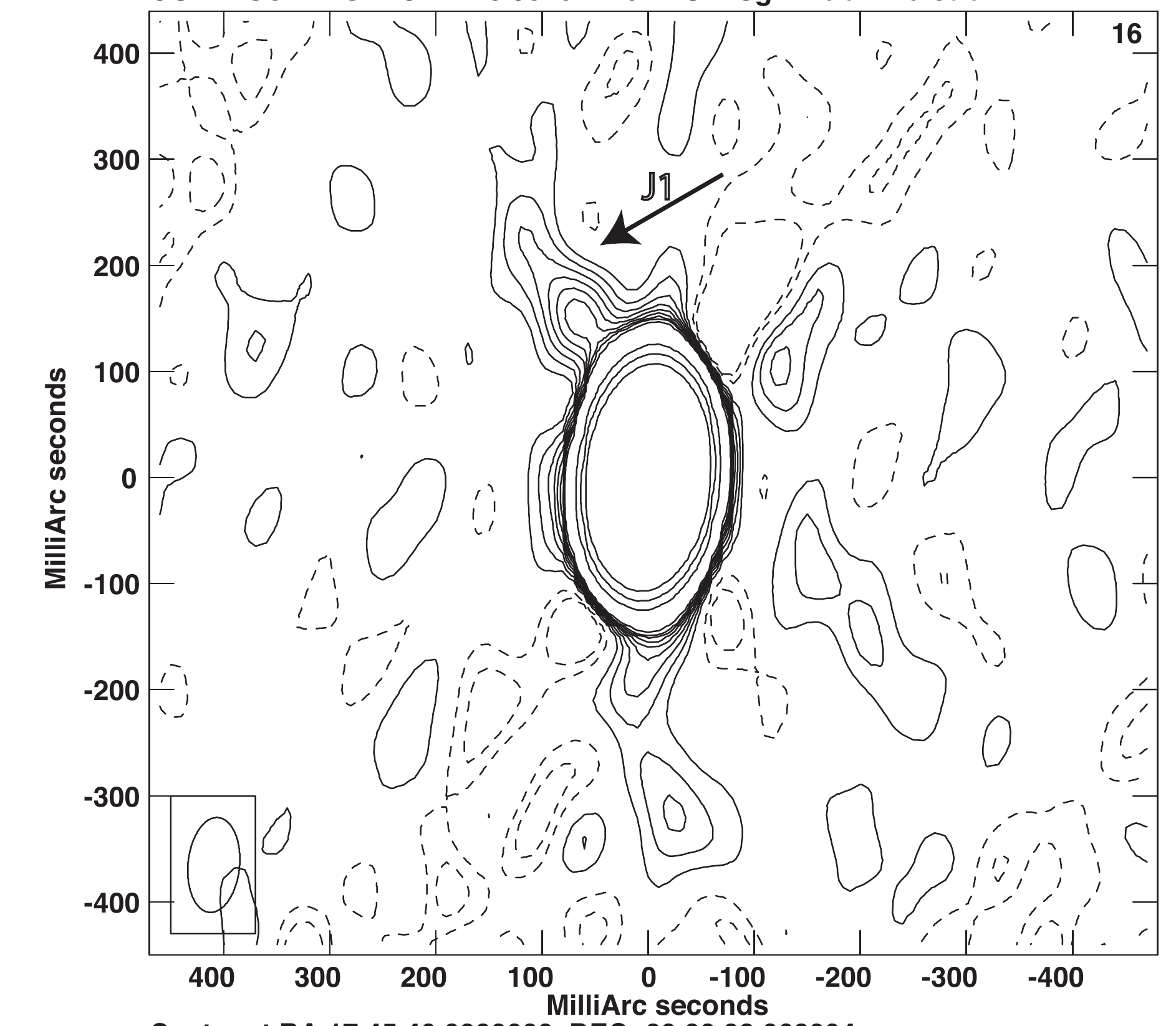}
\includegraphics[scale=0.4,angle=0]{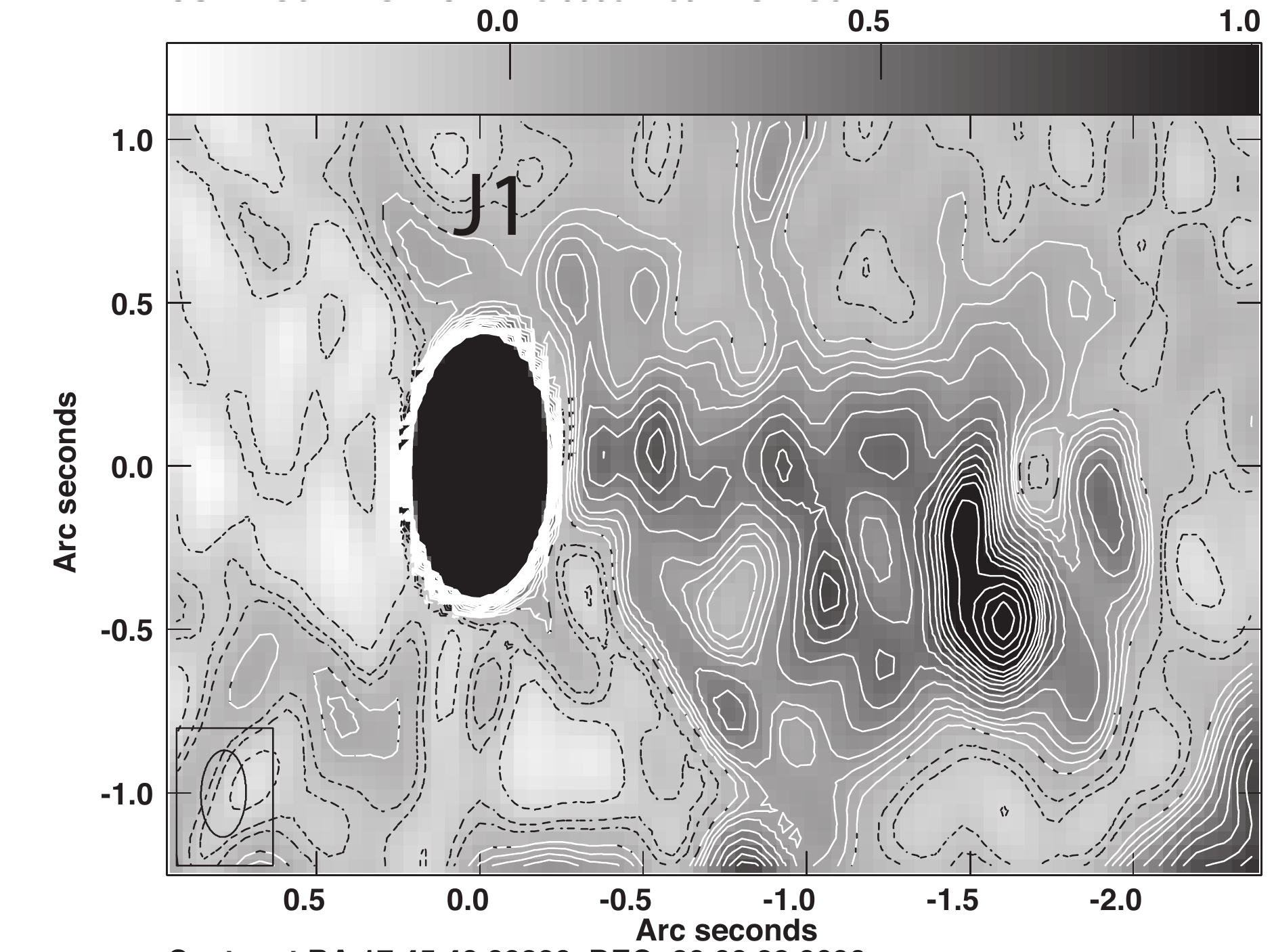}\\
\includegraphics[scale=0.35,angle=0]{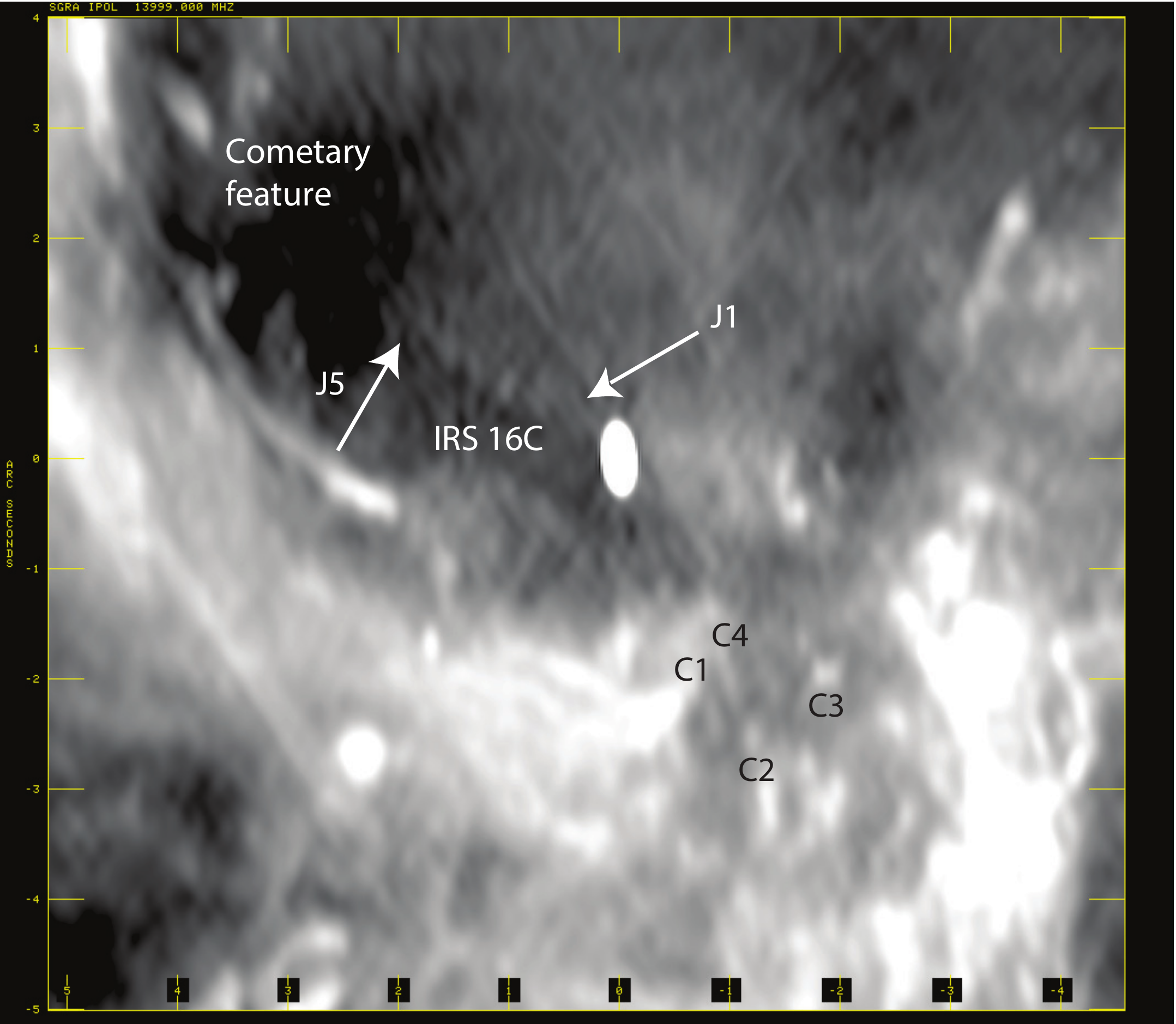}
\includegraphics[scale=0.5,angle=0]{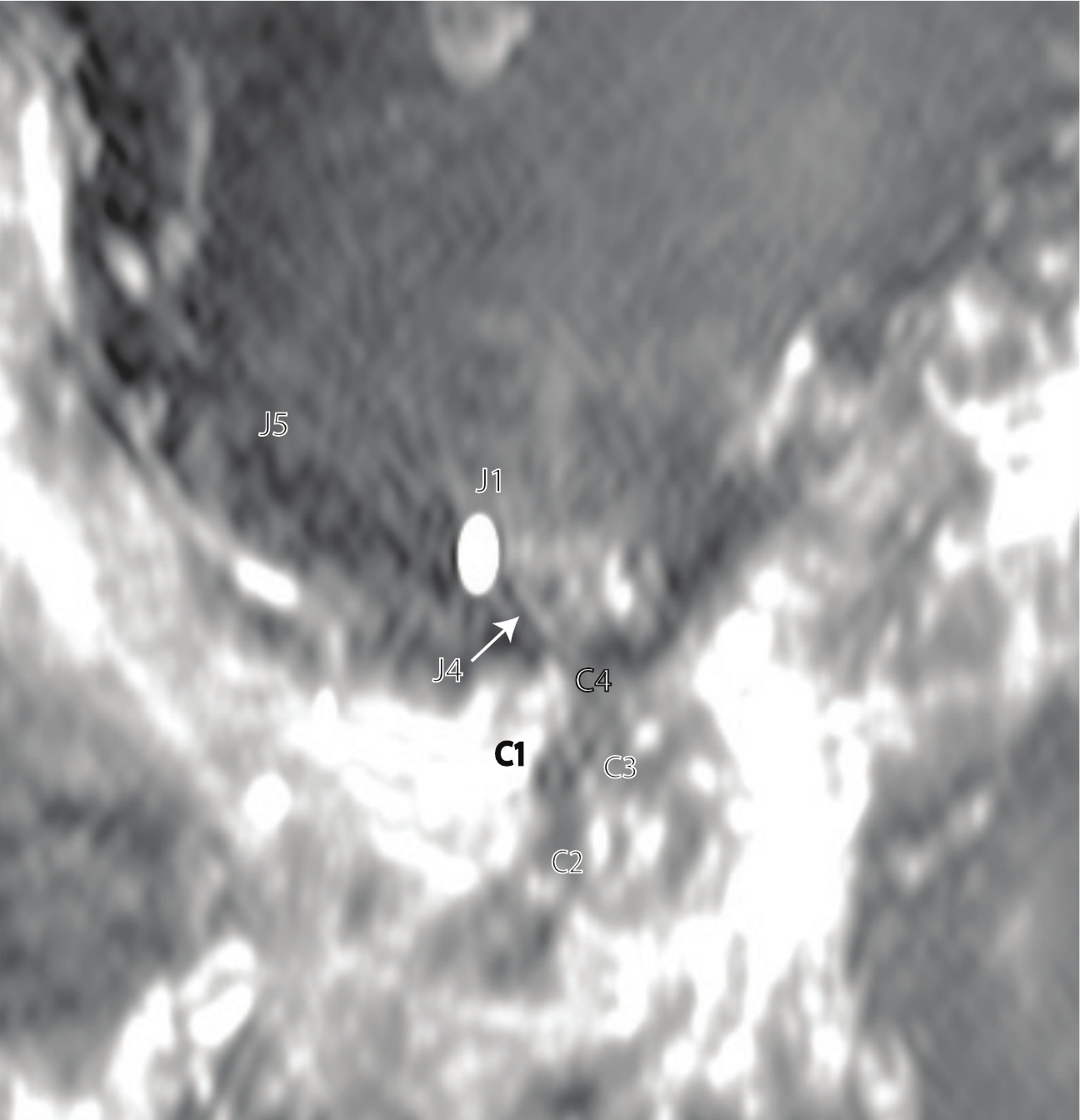}
\caption{
{\it (a) Top Left}
Contours of 35.72 GHz  emission from  the inner 0.45$''$  of Sgr A*
observed on March 9, 2014 at levels of  (-2, -1, 1, 2,..., 8, 10, 50, 100, 200) $\times 81\mu$Jy beam$^{-1}$. 
The 8-GHz bandwidth was imaged in 24 different frequency channels and this map corresponds to  channel 16. 
The resolution of the image is $0.090''\times0.049''$  (PA=-5.38$^\circ$). 
{\it (b) Top Right} Greyscale contours of 8.9 GHz emission  with levels at (-2, -1, 0..5, 0.5, 1, 1.5, 2, 3,..., 10, 12, 14,....20) $\
times$ 0.1 mJy beam$^{-1}$. of $0.26''\times0.14''$  (PA=$-1.5^\circ$). 
{\it (c) Bottom Left} 
A grayscale 15 GHz image of the inner 4$''$ of Sgr A* with a resolution of 
$0.22''\times0.11''$  (PA=5.25$^\circ$). 
The highly blue-shifted negative velocity features are labeled C1 to C4 \citep{paper1}.
{\it (d) Bottom Right} 
A 9  GHz image from data, as in (c), shows a grayscale image of the inner 4$''$ of Sgr A*. 
}
\end{figure}

\addtocounter{figure}{-1}
\begin{figure}
\centering
\includegraphics[scale=0.7,angle=0]{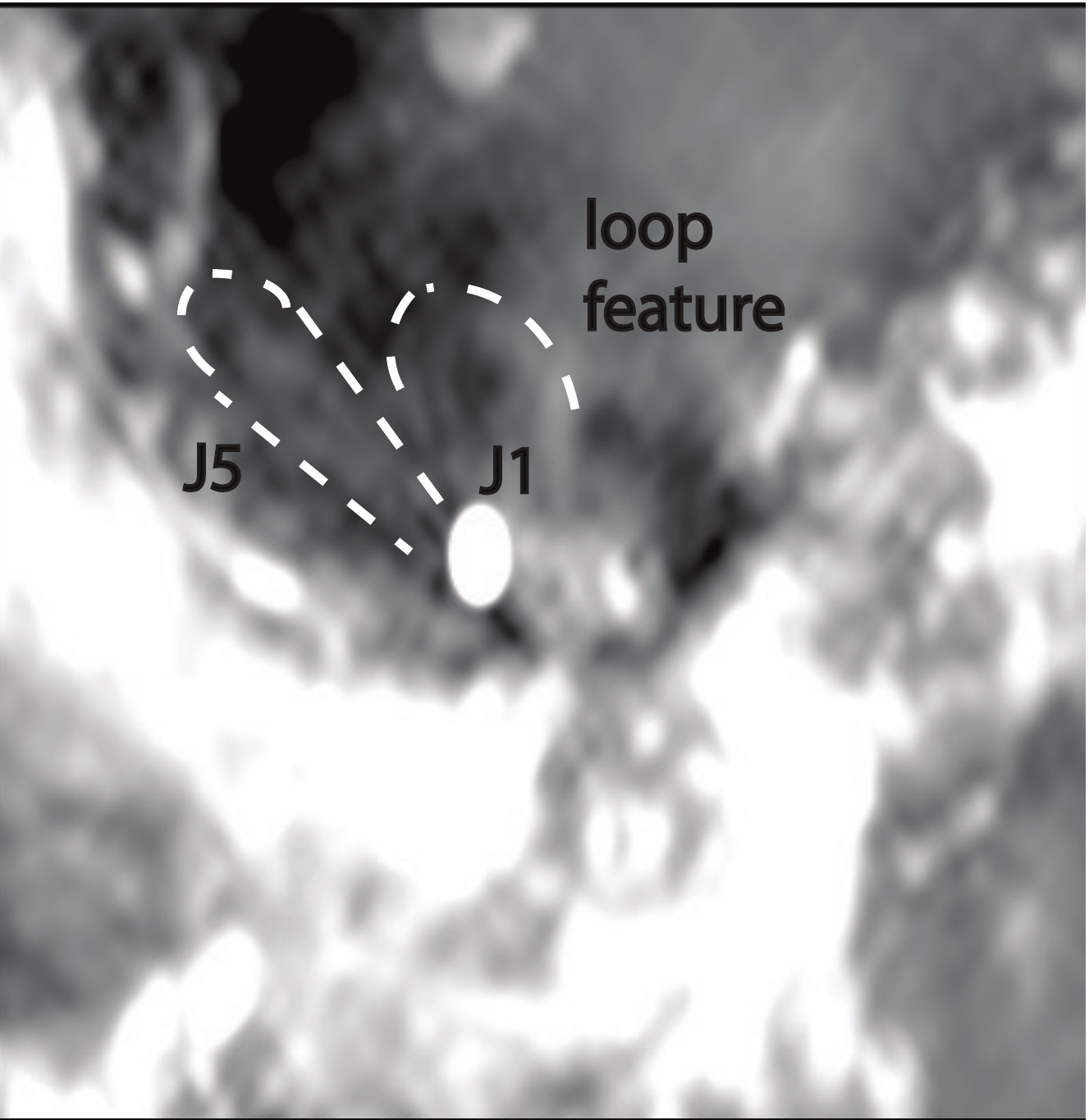}
\includegraphics[scale=0.7,angle=0]{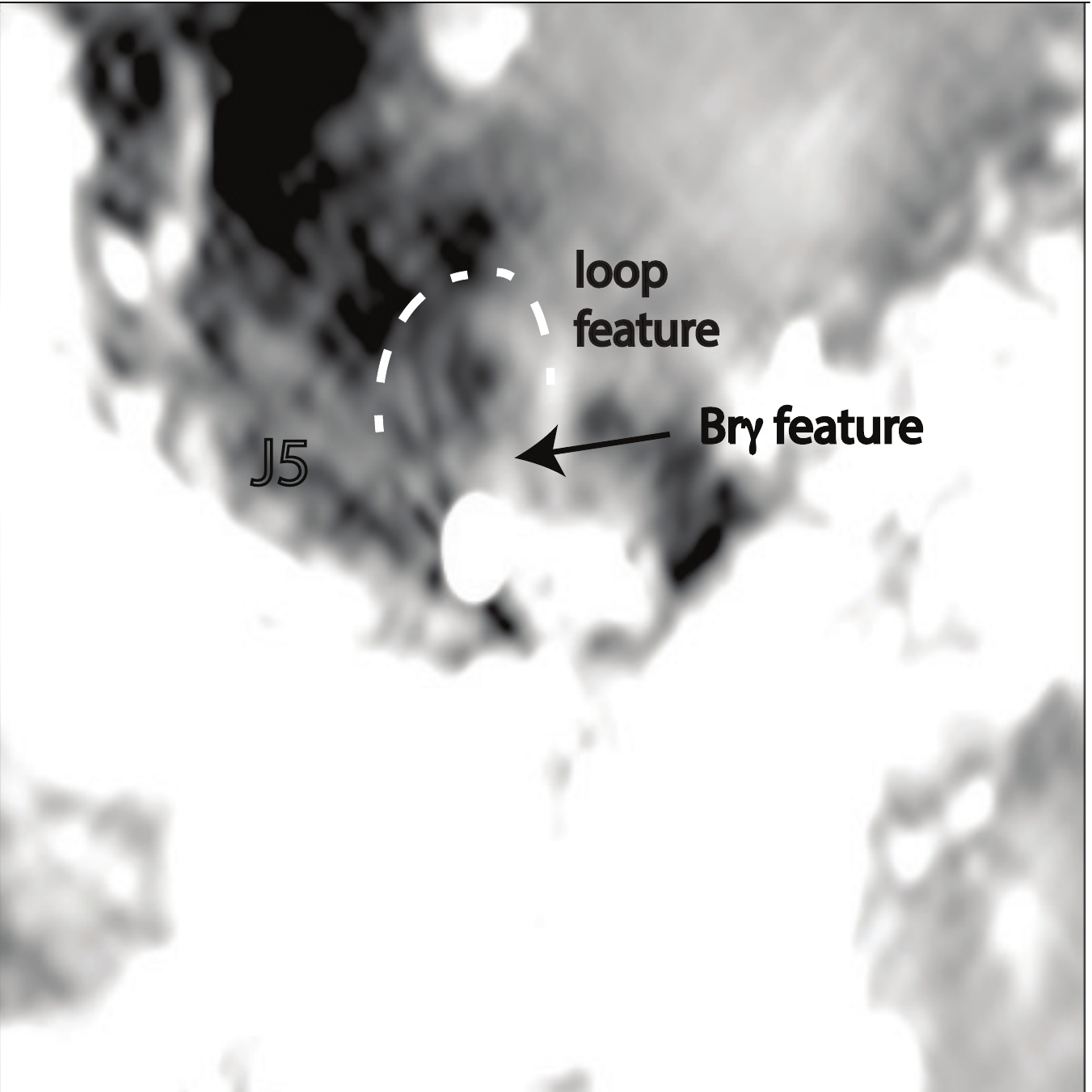}
\caption{
{\it (e) Top} A greyscale image of 9 GHz image using data taken in 17 April, 20014 
with a resolution of $0.36''\times0.22''$  (PA=-1.88$^\circ$). 
We used  OBIT by splitting the 2 GHz bandwidth of 9 GHz data  into 46 channels  to account 
for frequency response of Sgr A*  in each frequency channel before the final image was made.   
Similar imaging was done with  data taken on 17 August, 2019  and the features  described here 
were identified in both epochs. 
{\it (f) Bottom} 
Similar to (e) except with different contrasts to bring out faint features.
}
\end{figure}


\begin{figure}
\centering
\includegraphics[scale=0.55,angle=0]{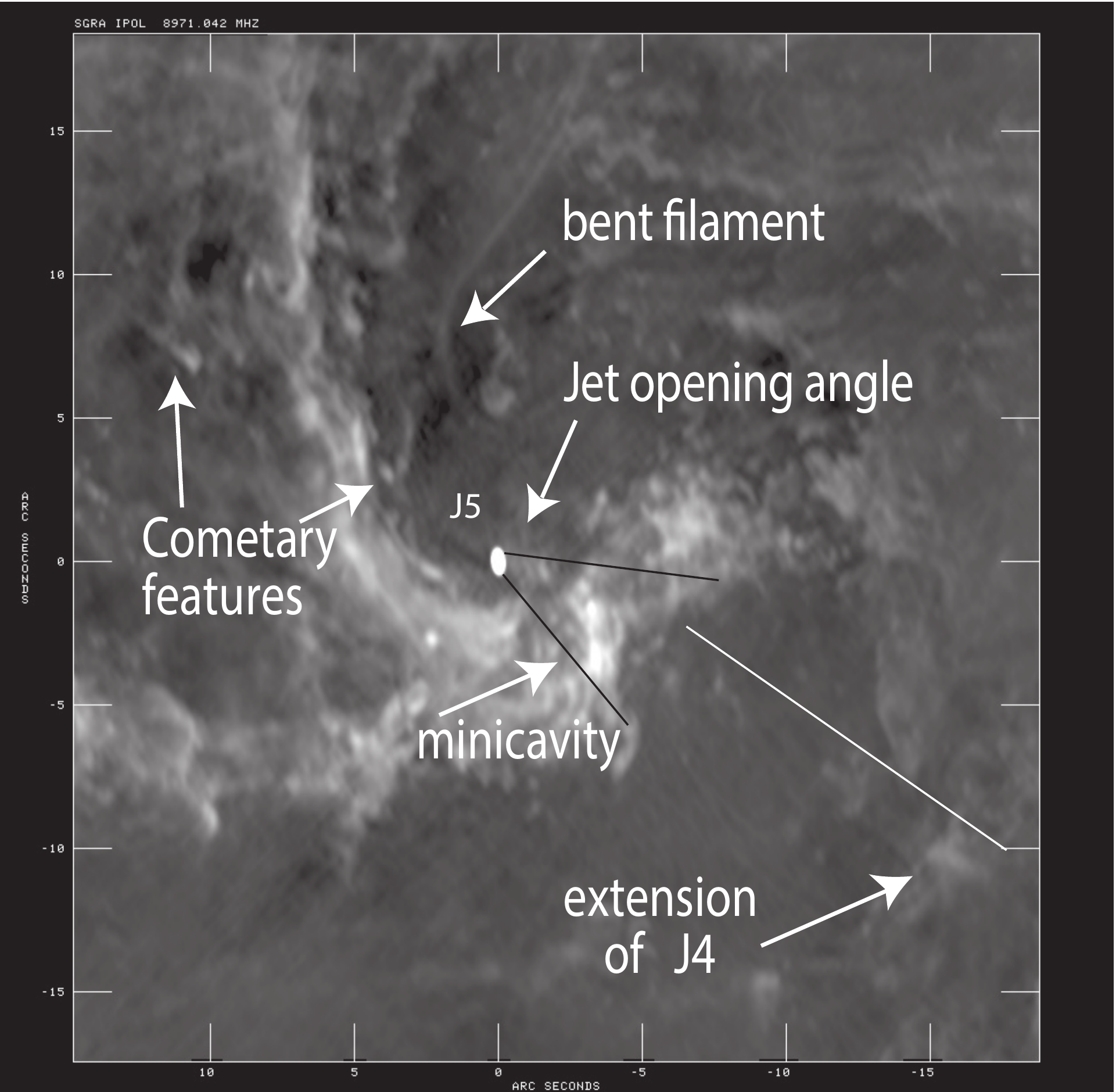}
\includegraphics[scale=0.55,angle=0]{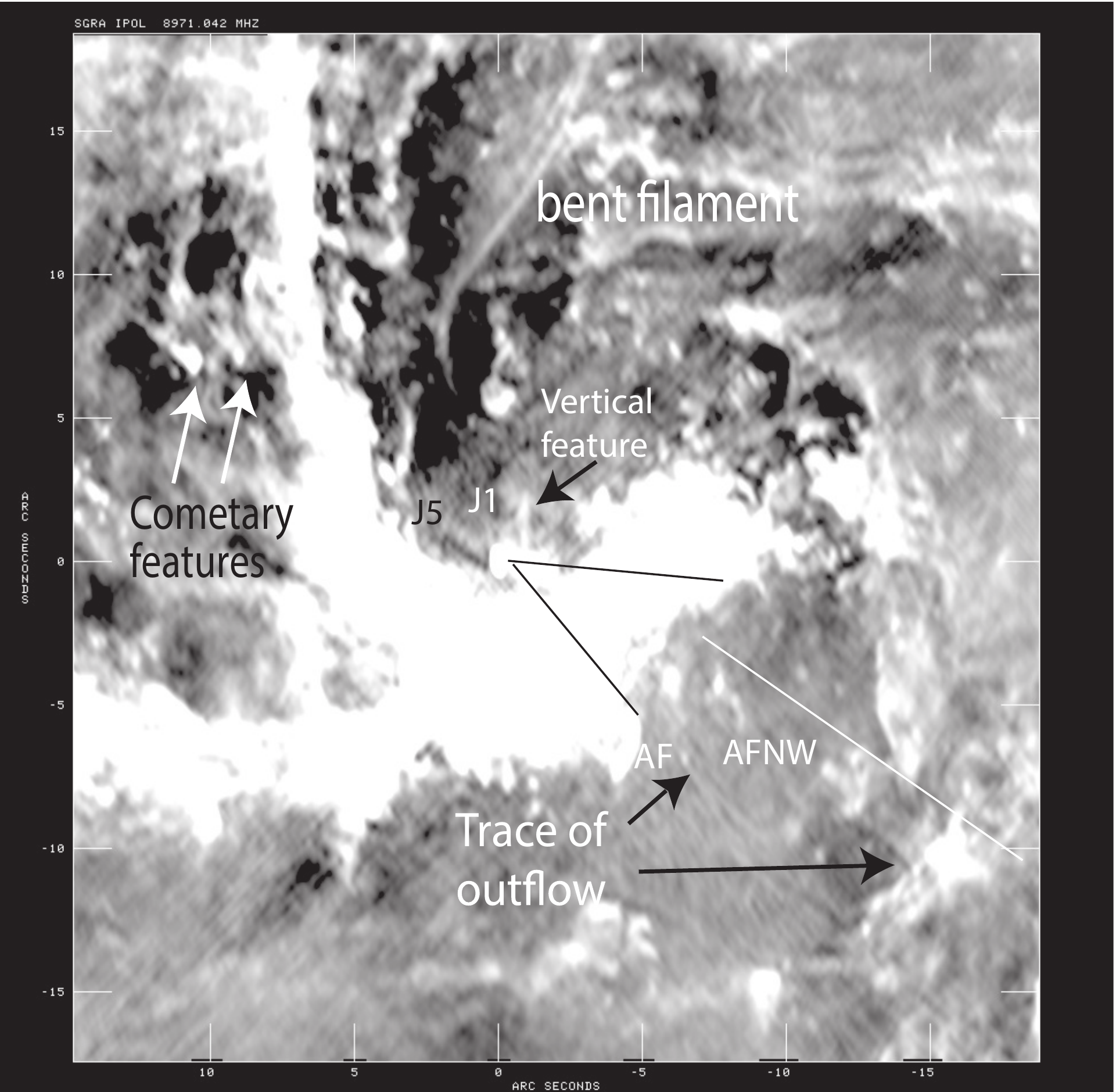}
\caption{
{\it (a, b)} 
Grayscale 9 GHz continuum images of the inner of 20$''$ of Sgr A* with a resolution 
of $0.35''\times0.18''$  (PA=5.61$^\circ$). 
The flux density ranges between -0.6 and 10 mJy beam$^{-1}$ with two different contrasts. 
This data set was taken on August 17, 2019. Black lines show the opening angle of the jet arsing from Sgr A* to the SW.  
The white lines point to a number of sources that lie diagonally along the direction of the outflow.}
\end{figure}

\addtocounter{figure}{-1}
\begin{figure}
\centering
\includegraphics[scale=0.65,angle=0]{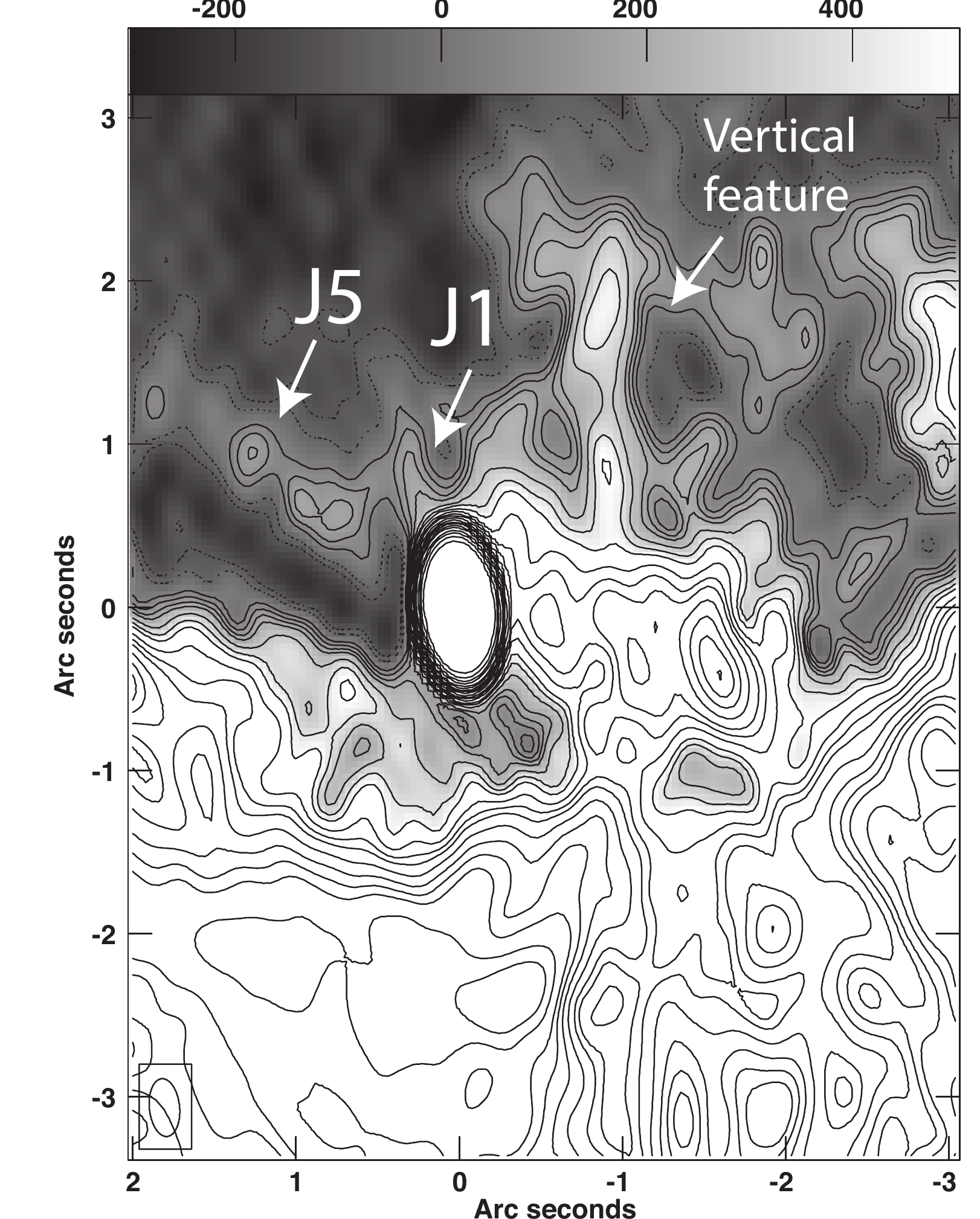}
\includegraphics[scale=0.65,angle=0]{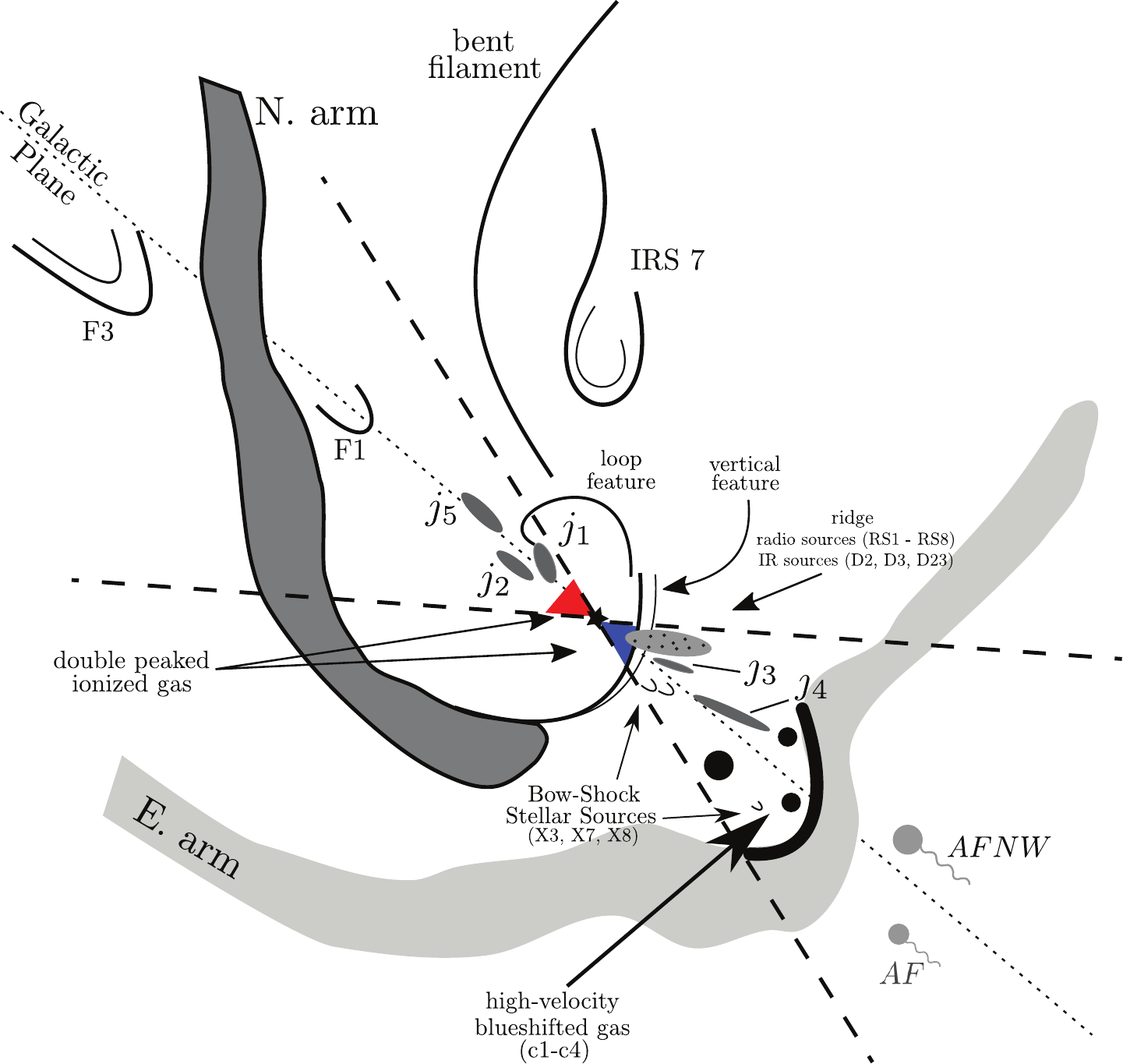}
\caption{
{\it (c)} 
Grayscale contours of 9 GHz emission with levels (-2, -1, 1, 2, 3, 4, 6, 8, 10, 14, 18, 22, 30, 40, 50, 80, 100, 200, 500)
$\times 50\mu$Jy beam$^{-1}$. 
{\it (d)} 
A schematic diagram of prominent features in the inner 20$''$ of Sgr A*.  
The double-peaked  blue and red-shifted mmRL emission associated with Sgr A* is   shown in color. 
The positions of some of the sources, such as bow-shock sources,  indicated on this figure may not be accurate. 
}
\end{figure}

\begin{figure}
\centering
\includegraphics[scale=0.425,angle=0]{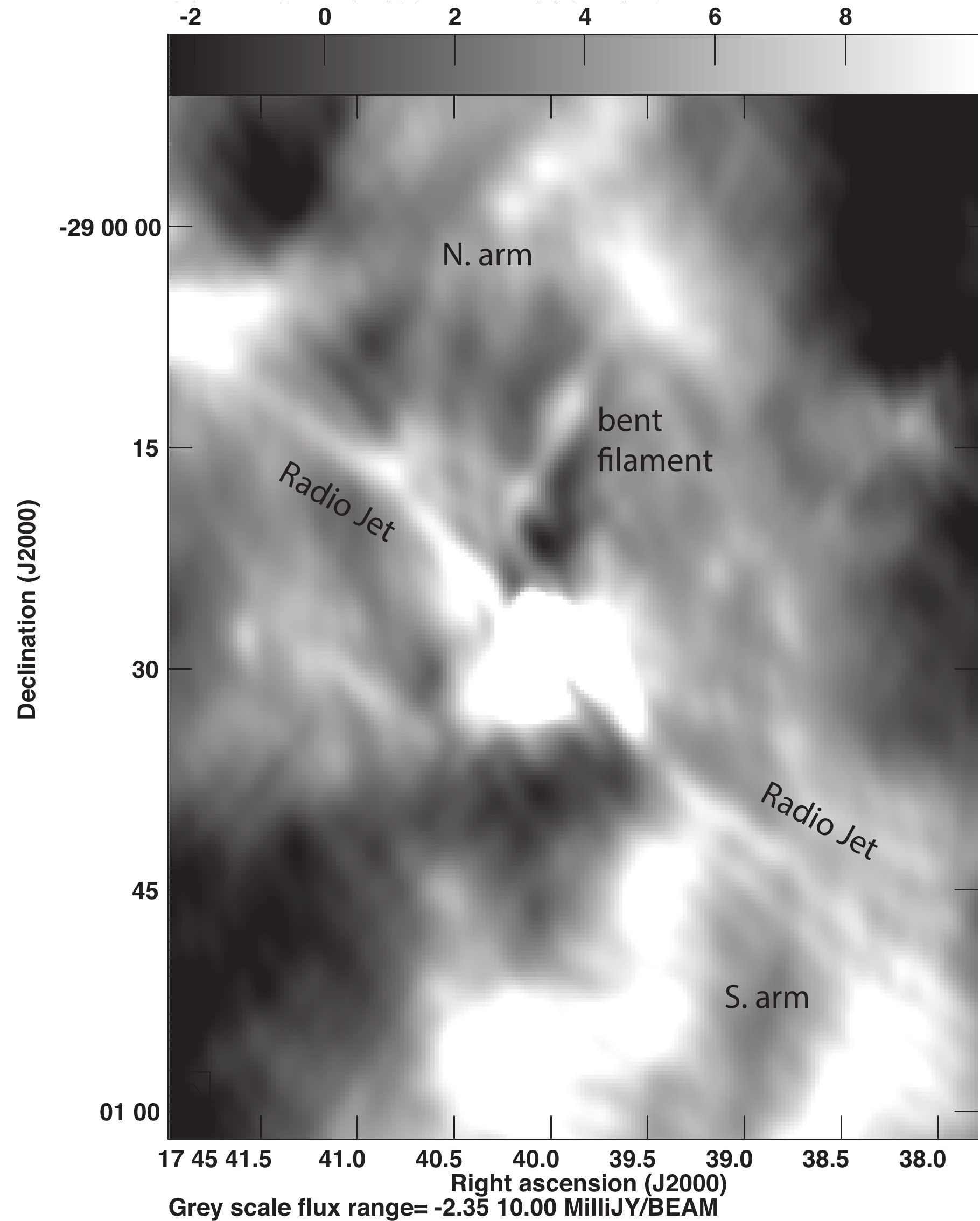}
\includegraphics[scale=0.425,angle=0]{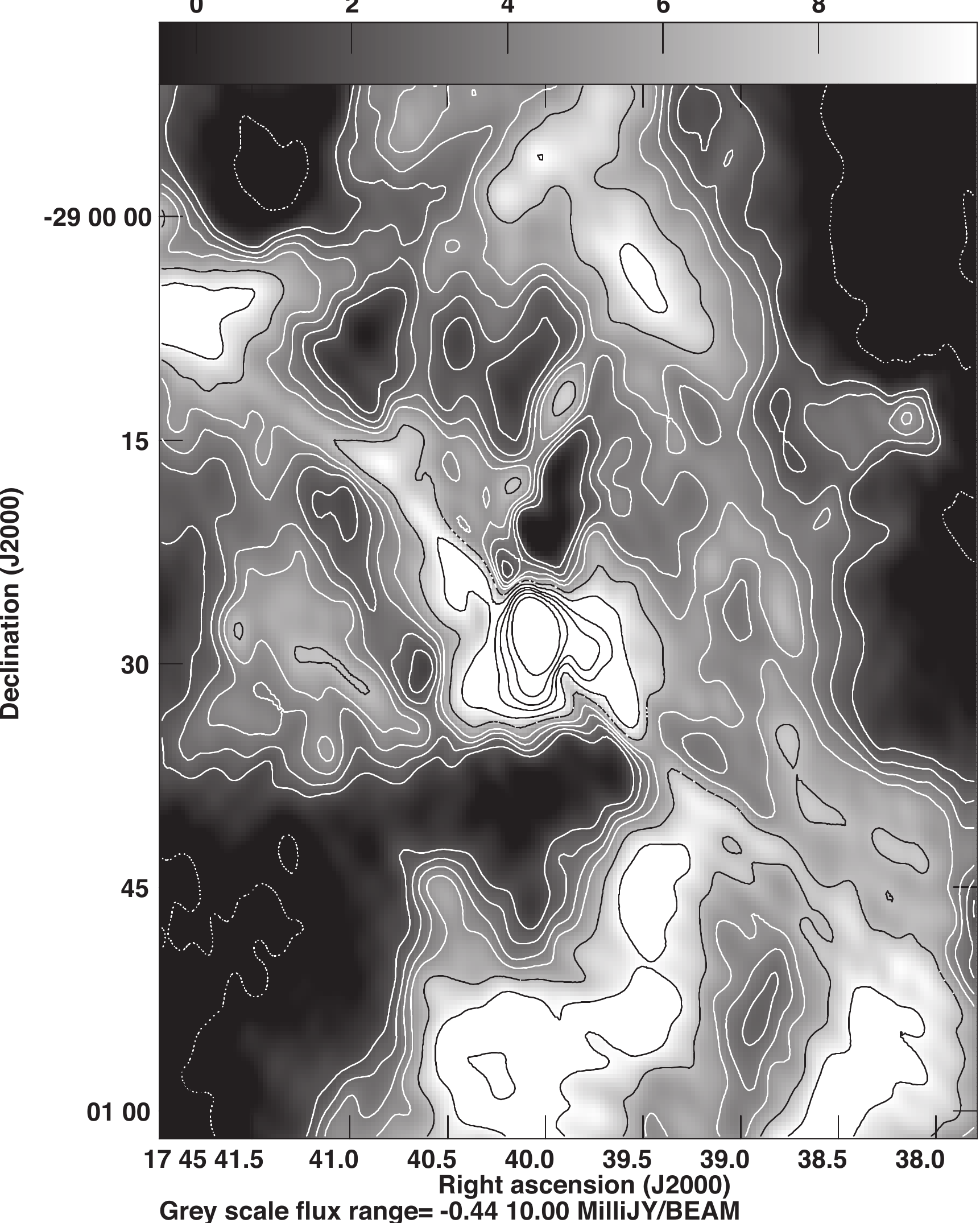}
\includegraphics[scale=0.4,angle=0]{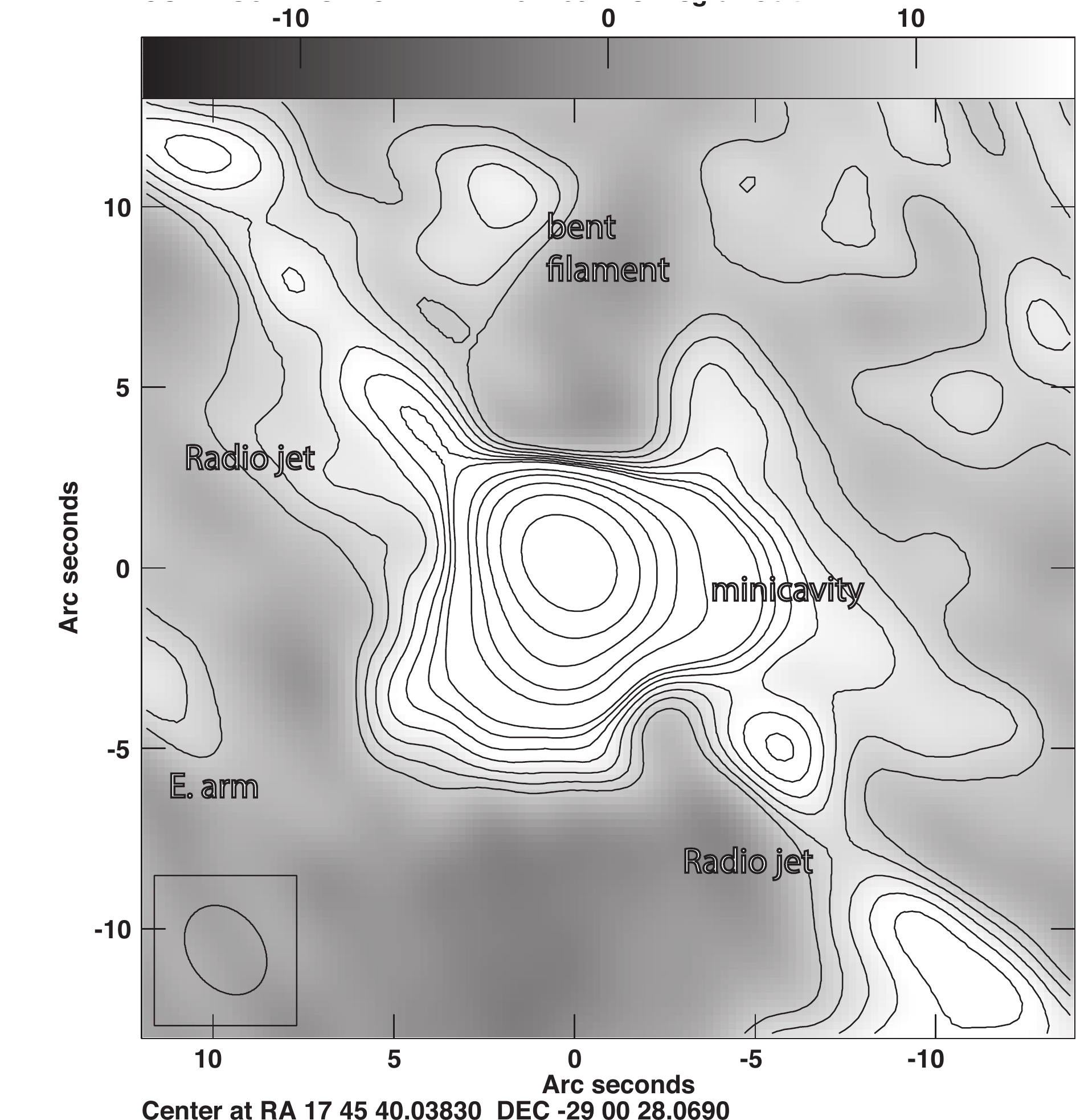}
\includegraphics[scale=0.4,angle=0]{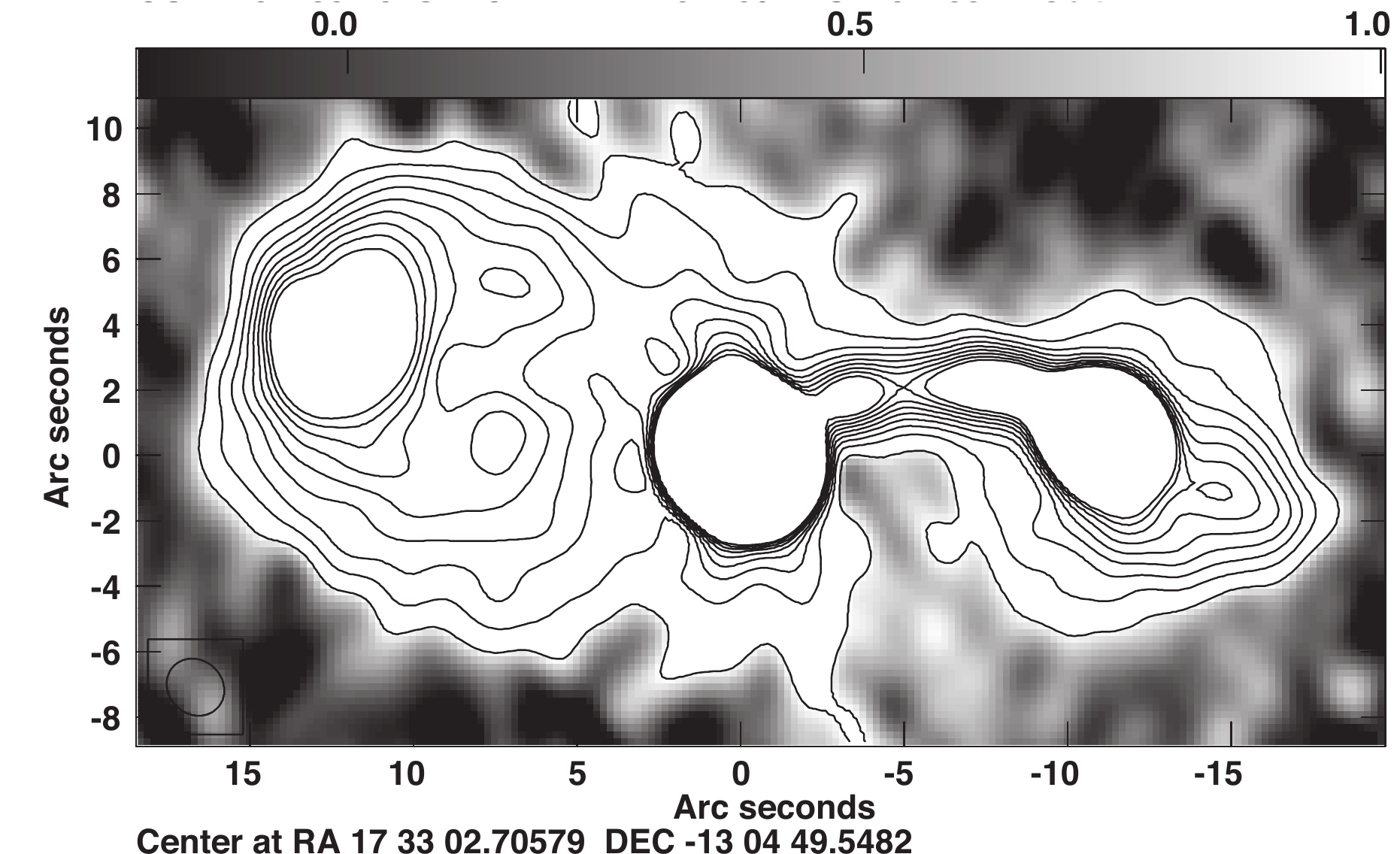}
\caption{
{\it (a)}
1-1.4 GHz image of the mini-spiral constructed by combining eight spectral windows with a resolution of
$2.68''\times2.01''$ (PA=36$^\circ$).
{\it (b)}
Identical to (a) except grayscale contours are presented with levels -2, 2, 3, 4, 5, 7, 10, 15, 20, 30, 50 mJy beam$^{-1}$. 
The coordinates are shown as offsets from the peak position of Sgr A*.
{\it (c)}
Contours of 20cm emission from a narrow spectral window between 1.25 and 1.31 GHz with  levels  at
(-4, 4, 5, 6, 7, 8, 9, 10, 15, 20 30, 50, 100) $\times$ 2 mJy beam$^{-1}$ with a spatial resolution 
$2.94''\times2.17''$ (PA=37.8$^\circ$).
{\it (d)}
Radio emission from the calibrator J1733-1304 corresponding to  the lowest frequency  spectral window at 1.0 GHz with 
a peak flux density of 4.02 Jy beam$^{-1}$ and  contour  levels 
set at (1,  2, 3,....,9)\,  mJy beam$^{-1}$ with a resolution of  
 $1.85''\times1.65''$ (PA=49.92$^\circ$).
}
\end{figure}

\begin{figure}
\centering
\includegraphics[scale=0.45,angle=0]{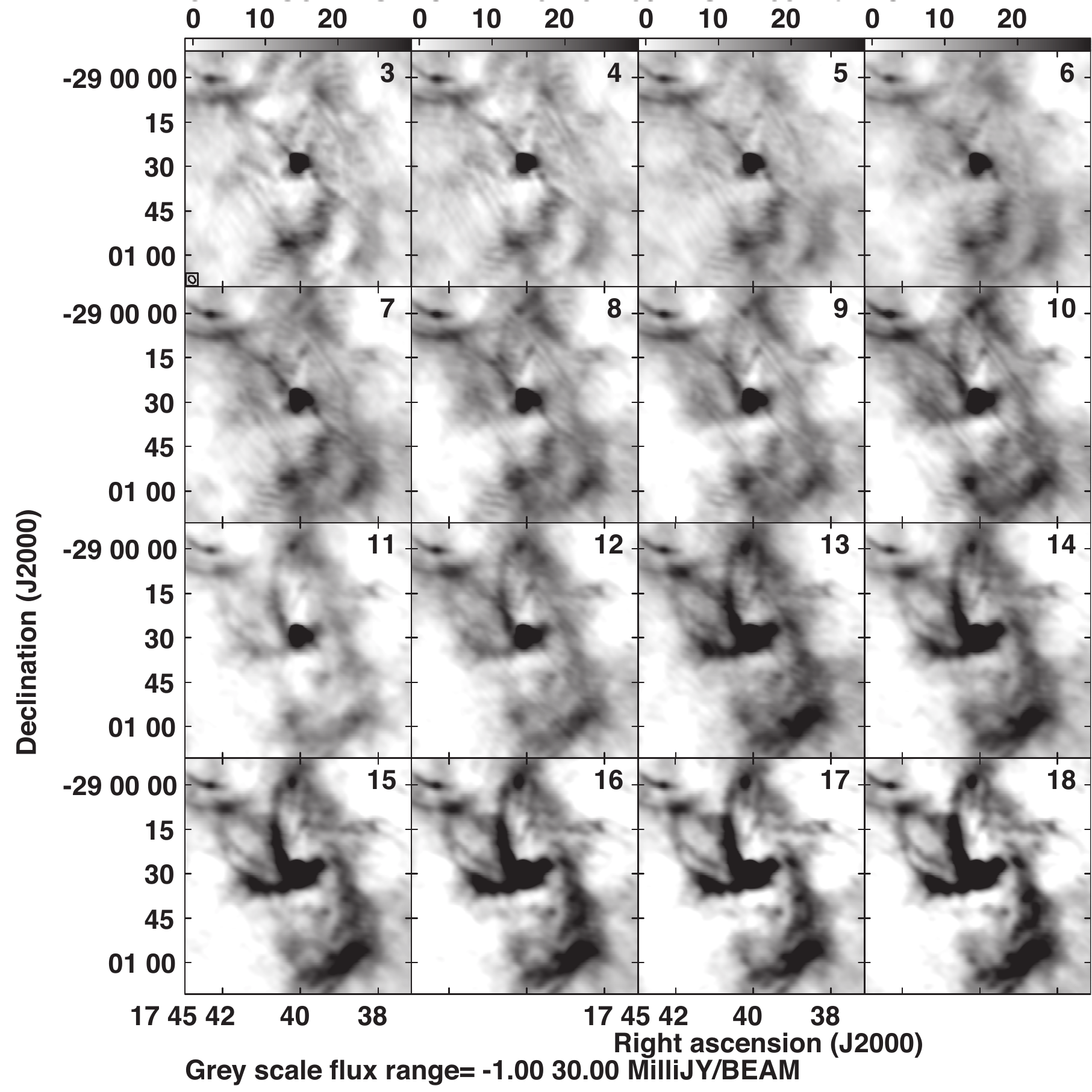}\\
\includegraphics[scale=0.45,angle=0]{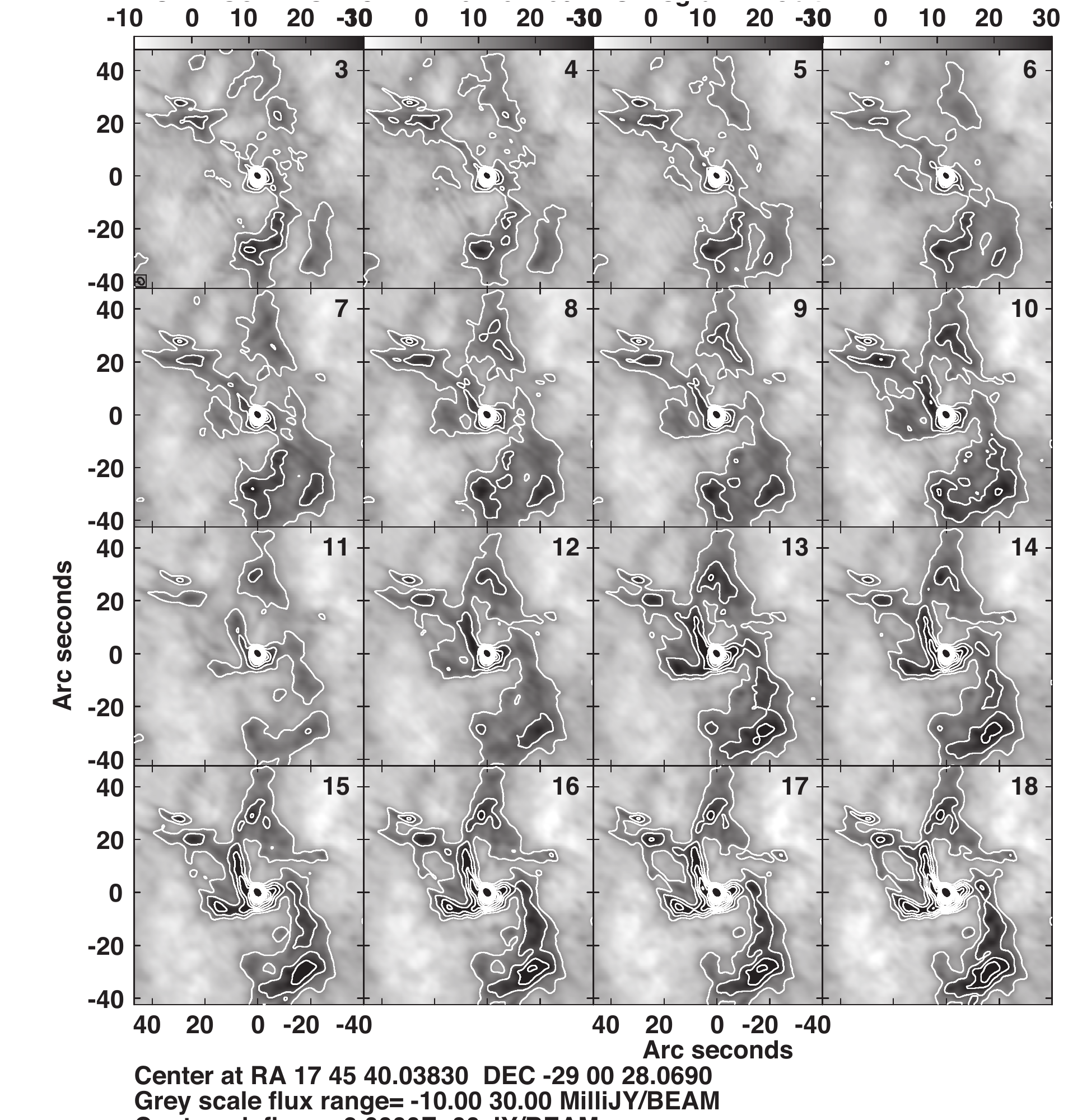}
\caption{
{\it (a)}
Channel maps corresponding to 16 spectral windows  are labeled between 1 and 2 GHz. Black traces
radio emission whereas white in the minispiral is seen in absorption. These channel maps reveal both thermal and nonthermal
sources simultaneous across the 1 GHz bandwidth in the L band.
{\it (b)}
Channel maps corresponding to 16 spectral windows are labeled between 1 and 2 GHz. Black traces
radio emission whereas white is seen in absorption. These channel maps reveal both thermal and nonthermal
sources simultaneous across a 1 GHz bandwidth.
}
\end{figure}

\begin{figure}
\centering
\includegraphics[scale=0.65,angle=0]{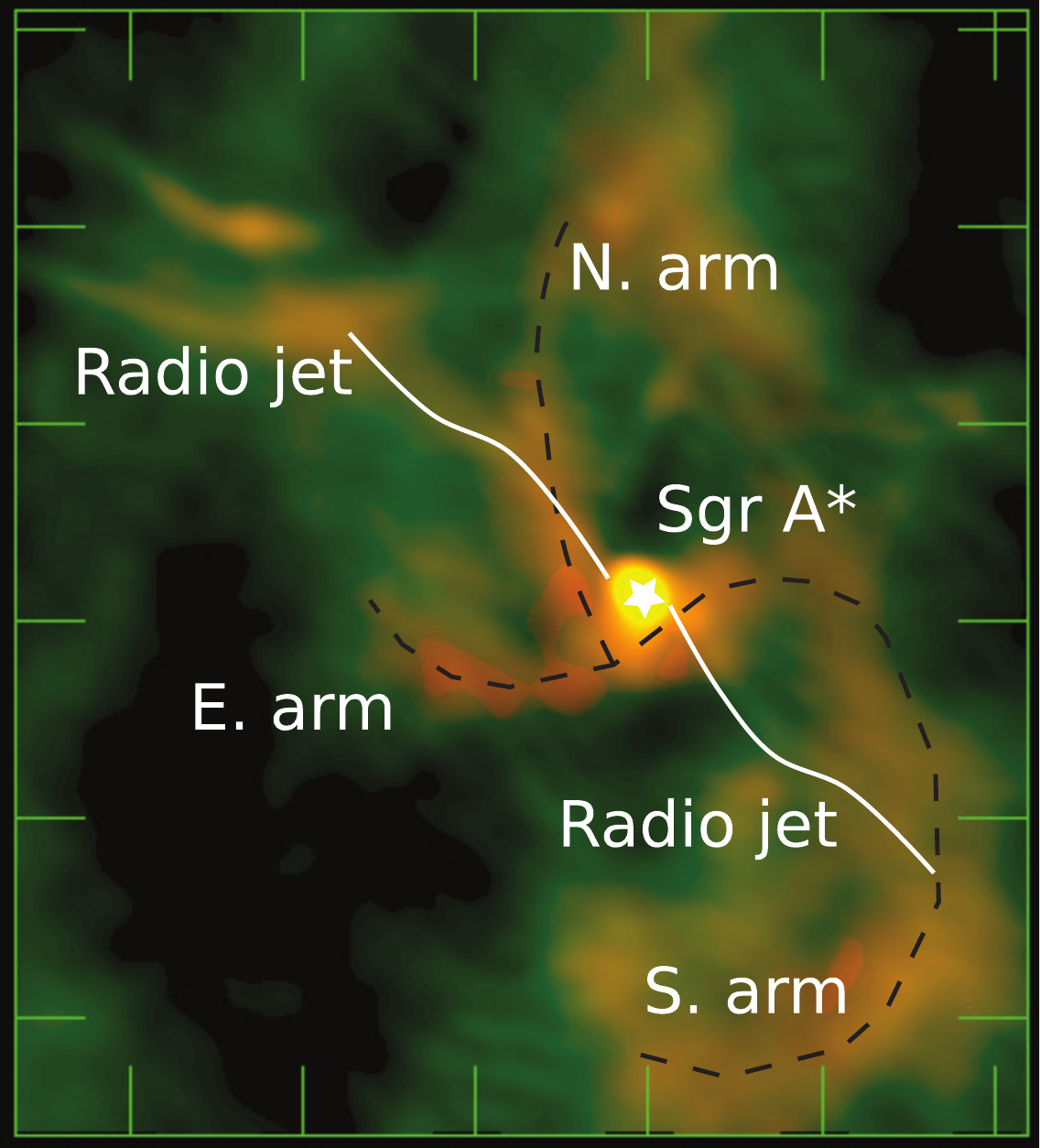}
\includegraphics[scale=0.85,angle=0]{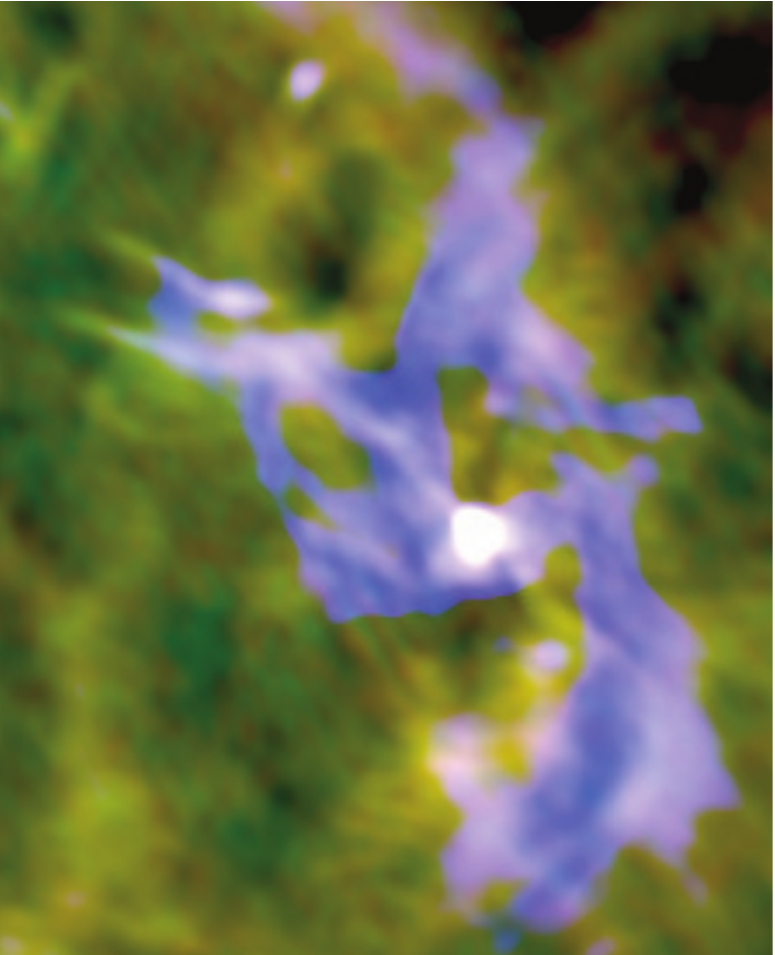}
\caption{
{\it (a)}
A  composite color image of the inner 60$''\times75''$ of Sgr A* between 1-2 GHz.
The mini-spiral thermal feature  and the nonthermal linear features are labeled.
The image is constructed by  using
a narrow spectral window between $\sim1.25$ and $\sim1.31$ GHz  and a broad image taken between 1.6 and 1.9 GHz. 
The mini-spiral thermal feature becomes
optically thick (thin) at low (high) frequencies, thus the jet-feature (the mini-spiral) becomes more visible.
{\it (b)}
Another  composite color image using three different ranges of emission between 1
and 1.9 GHz. The blue represents the integrated intensity due to eight high frequency spectral windows, while the red represents the eight low frequency spectral windows.  The green color represents  the entire bandwidth.
}
\end{figure}

\addtocounter{figure}{-1}
\begin{figure}
\centering
\includegraphics[scale=1.2,angle=0]{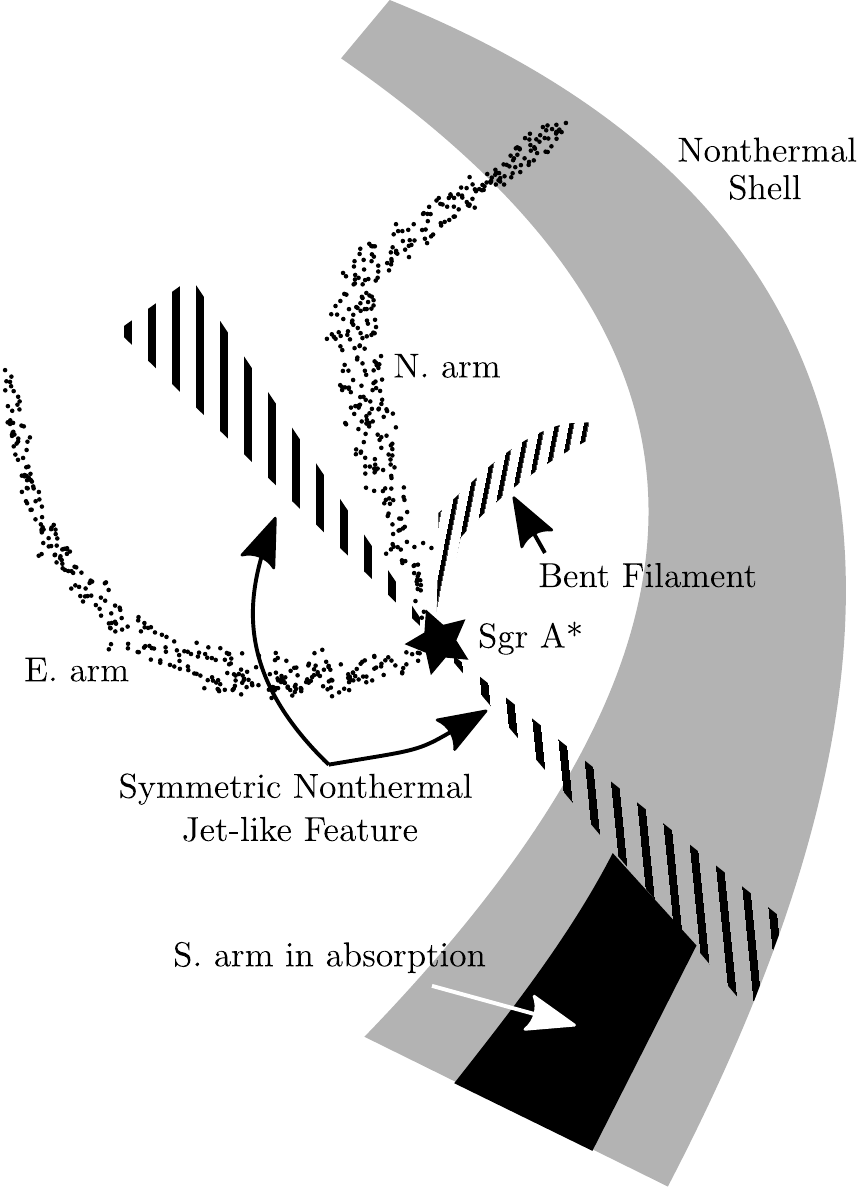}
\caption{
{\it (c)}
A schematic diagram showing prominent nonthermal (hatched) and thermal
(dotted) features at 20cm including a symmetric jet-like feature with respect to Sgr A* and a bent jet-like
feature to the north of Sgr A*.
}
\end{figure}

\begin{figure} 
\centering 
\includegraphics[scale=0.35,angle=0]{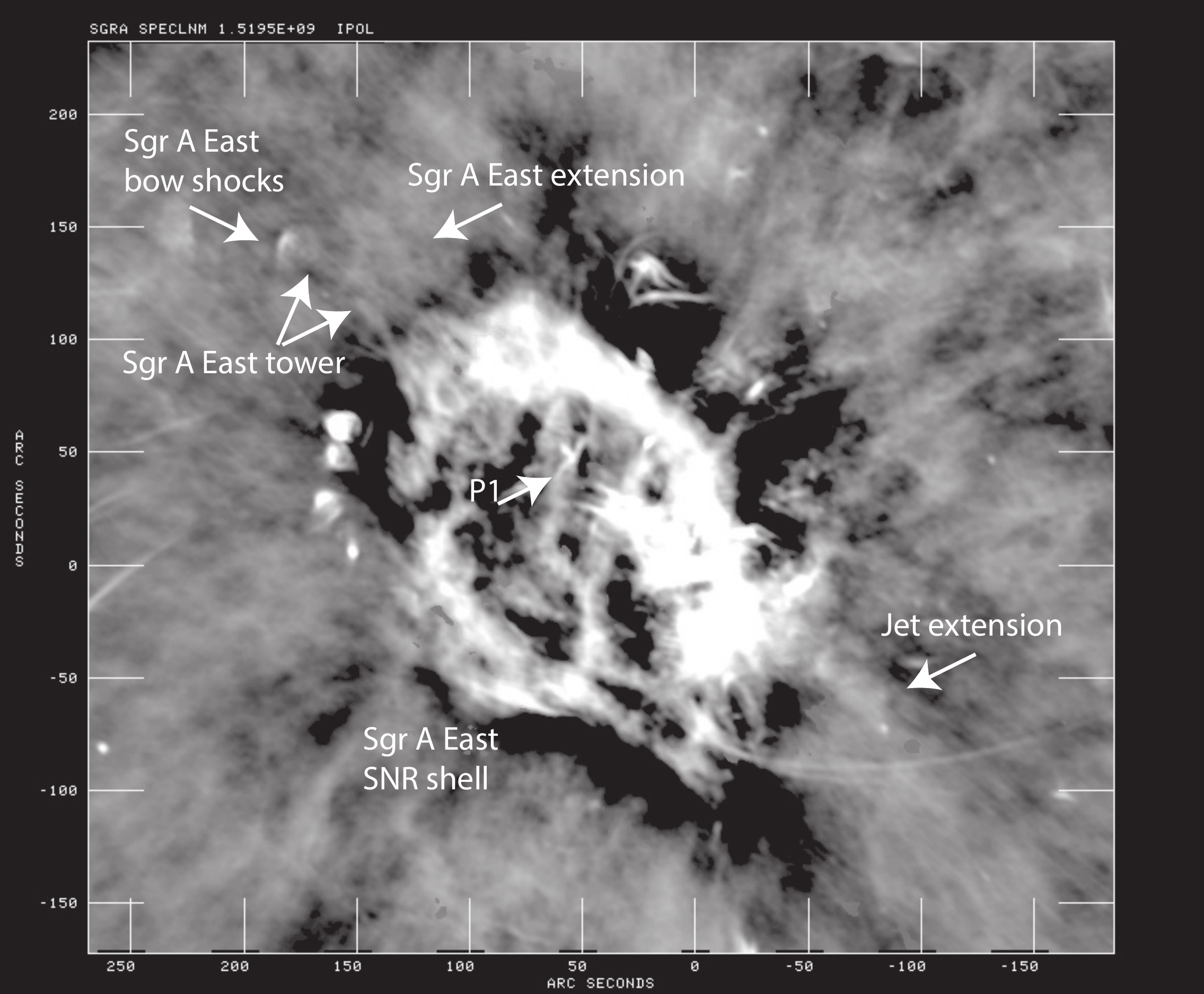} 
\includegraphics[scale=0.35,angle=0]{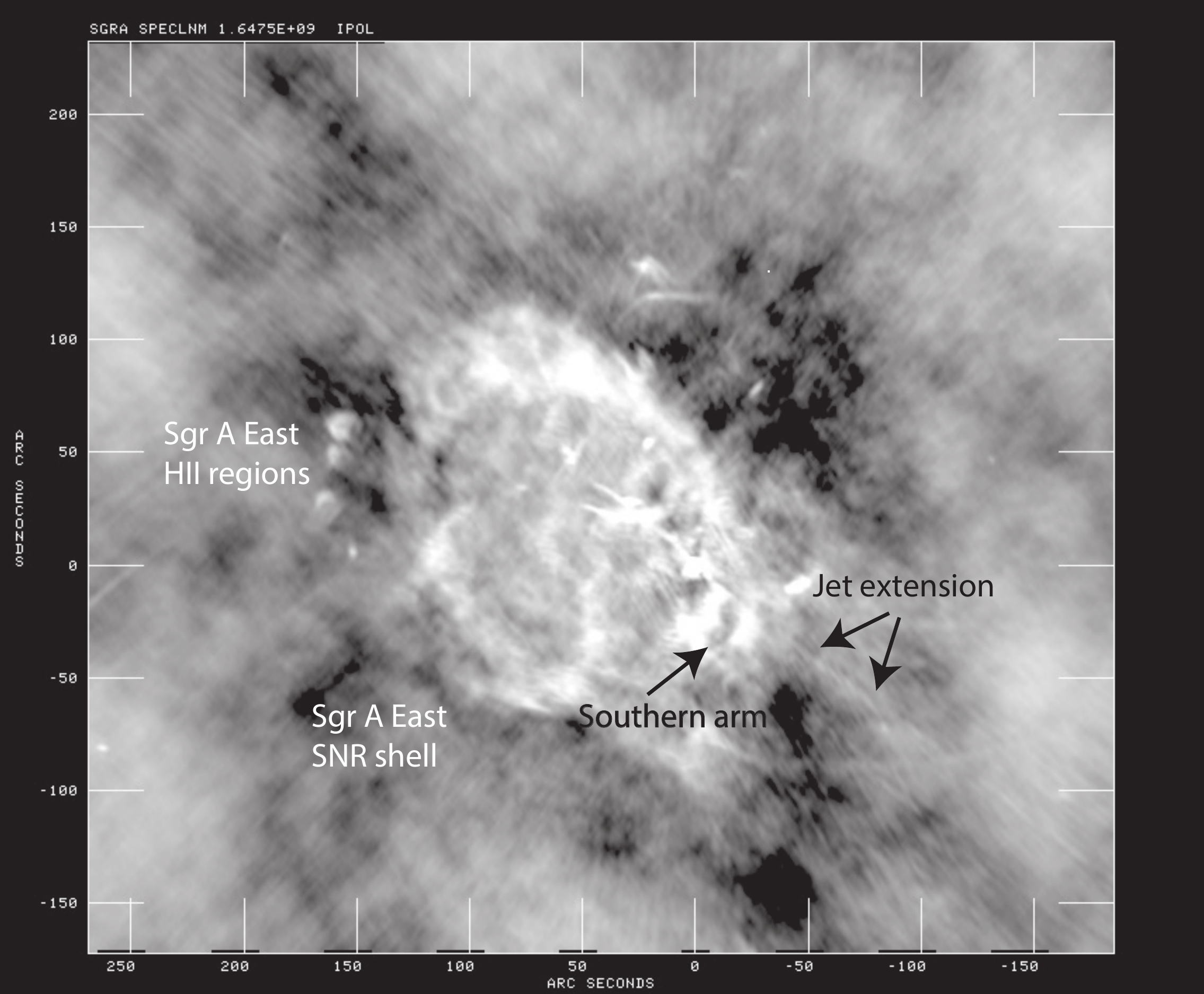} 
\includegraphics[scale=0.5,angle=0]{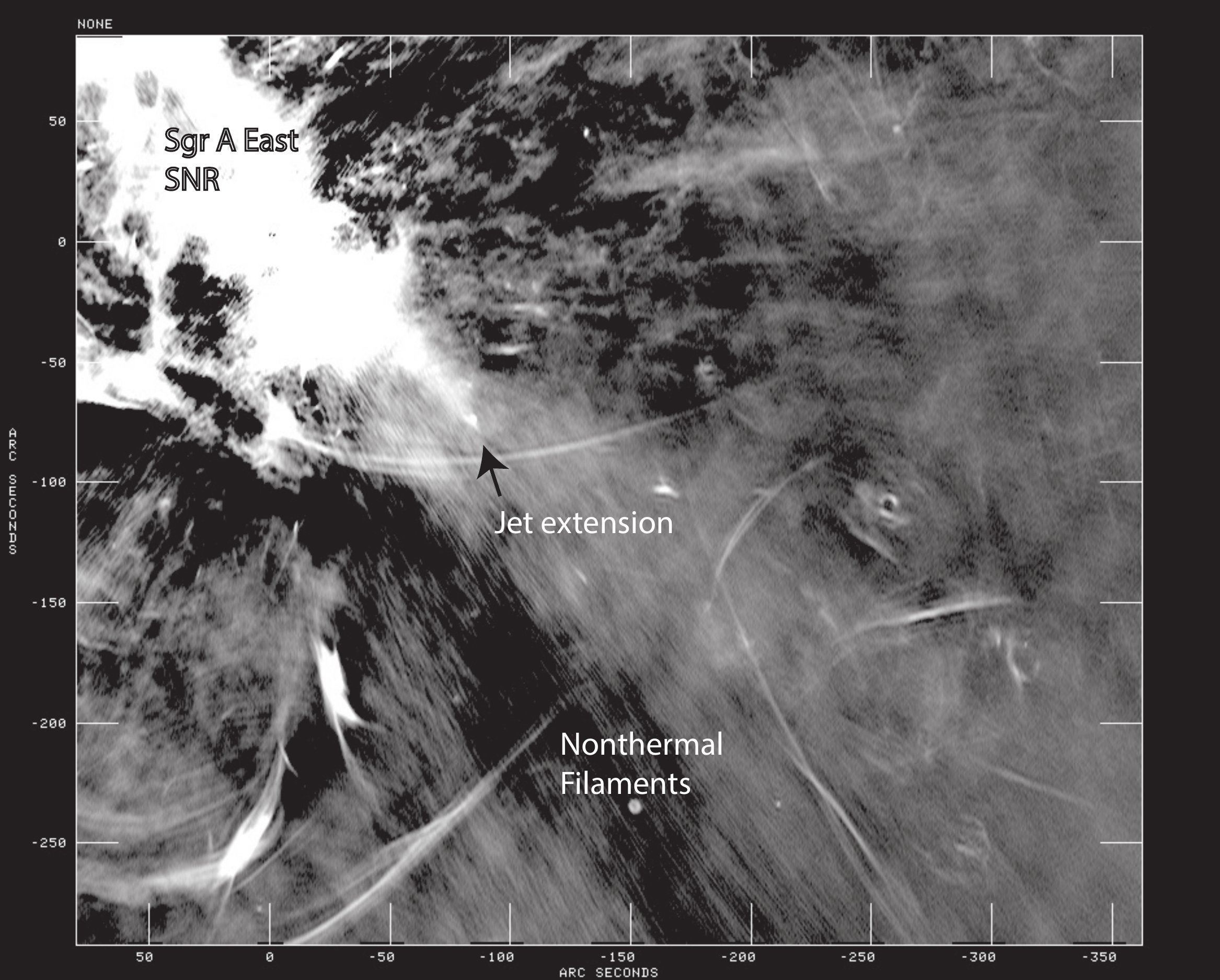} 
\caption{
{\it (a Left)}
Images of the Sgr A East SNR and the mini-spiral are shown 
using the 20cm broad band between  $\sim$1 and $\sim$1.9 GHz
with a resolution of  $2''.68\times2''.03$   (PA=36$^\circ.38$). 
{\it (b  Right)} Similar to (a) except a narrow band  at 1.24 GHz is displayed.
{\it (c Bottom)}
Broad band 20cm VLA image of the region to the SW of Sgr A* with a resolution of 
$1.69''\times0.65''$   (PA=170$^\circ.56$). 
}
\end{figure}

\begin{figure} 
\centering 
\includegraphics[scale=1.2,angle=0]{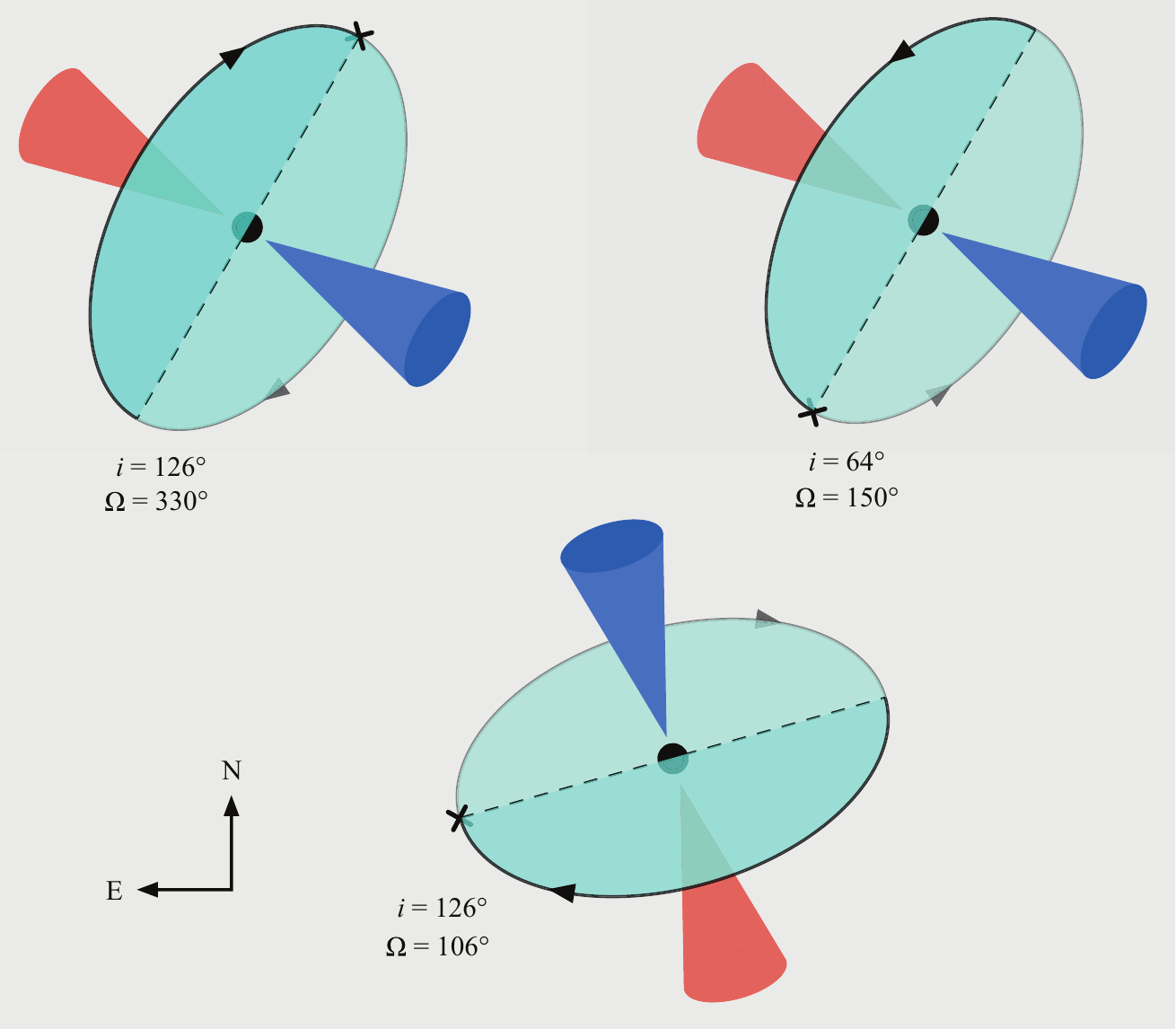} 
\caption{
{\it (a,b top Left and Right)}
Two possible orientations of the accretion disk presumed to be driving the jet from Sgr A*, as 
viewed by an observer.  The red and blue cones represent the red and blue-shifted arms of the jet, which have 
PA 60 and 240 \degr E of N, respectively.  The jet axis is perpendicular to the plane of the disk (cyan), 
which may rotate clockwise (left) or counter clockwise (right). The corresponding PA of the ascending node, 
$\Omega$, is indicated in each case.  
{\it (c Bottom)}
For comparison, we show a jet-disk system with orbital plane 
matching that of the clockwise stellar disk. 
}
\end{figure}



\bsp	
\label{lastpage}
\end{document}